\documentclass[manuscript]{emulateapj}
  
\usepackage{natbib}
\usepackage{psfig}
\usepackage{graphicx}
\usepackage{color}
\usepackage{times}

\def\farcsn{\hbox{$\ \!\!^{\prime\prime}$}}

\def\kms{$\mathrm{km\;s}^{-1}$}
\def\kmsmpc{$\mathrm{km\;s^{-1}\;Mpc^{-1}}$}

\def\msun{\ensuremath{M_{\odot}}}

\def\re{\ensuremath{R_{\rm e}}}

\def\sigmas{\ensuremath{\sigma}}

\def\sigmae{$\sigma_{\rm e}$}
\def\mstar{\ensuremath{M_{\star}}}
\def\mvir{\ensuremath{M_{\rm dyn}}}

\defcitealias{Maraston2009}{M09}
\defcitealias{Maraston2011a}{M11}

\shorttitle{Dynamical properties of SDSS-III/BOSS galaxies} 
\shortauthors{Beifiori et al.}

\begin{document}  
  

\title{Redshift evolution of the dynamical properties of massive galaxies from SDSS-III/BOSS} 


\author{Alessandra Beifiori\altaffilmark{1,2,3}, Daniel
  Thomas\altaffilmark{2,4}, Claudia Maraston\altaffilmark{2}, Oliver
  Steele\altaffilmark{2}, Karen L. Masters\altaffilmark{2,4}, Janine
  Pforr\altaffilmark{5,2}, Roberto P. Saglia\altaffilmark{1,3}, Ralf
  Bender\altaffilmark{1,3}, Rita Tojeiro\altaffilmark{2}, Yan-Mei
  Chen\altaffilmark{6,7}, Adam Bolton\altaffilmark{8}, Joel
  R. Brownstein\altaffilmark{8}, Jonas Johansson\altaffilmark{9,2},
  Alexie Leauthaud\altaffilmark{10}, Robert
  C. Nichol\altaffilmark{2,4}, Donald
  P. Schneider\altaffilmark{11,12}, Robert Senger\altaffilmark{1},
  Ramin Skibba\altaffilmark{13}, David Wake\altaffilmark{6,14}, Kaike
  Pan\altaffilmark{15}, Stephanie Snedden\altaffilmark{15}, Dmitry
  Bizyaev\altaffilmark{15}, Howard Brewington\altaffilmark{15}, Viktor
  Malanushenko\altaffilmark{15}, Elena Malanushenko\altaffilmark{15},
  Daniel Oravetz\altaffilmark{15}, Audrey Simmons\altaffilmark{15},
  Alaina Shelden\altaffilmark{15}, Garrett Ebelke\altaffilmark{15}}
\altaffiltext{1}{Max-Planck-Institut f\"{u}r Extraterrestrische Physik,
Giessenbachstra\ss e, D-85748 Garching, Germany; \email{beifiori@mpe.mpg.de}}
\altaffiltext{2}{Institute of Cosmology and Gravitation, University of
Portsmouth, Dennis Sciama Building, Burnaby Road, Portsmouth, PO1 3FX, UK}
\altaffiltext{3}{Universit\"{a}ts-Sternwarte M\"{u}nchen, Scheinerstrasse 1, D-81679 M\"{u}nchen, Germany}
\altaffiltext{4}{SEPNET, South East Physics Network}
\altaffiltext{5}{NOAO, 950 N. Cherry Ave., Tucson, AZ, 85719, USA}
\altaffiltext{6}{Department of Astronomy, University of Wisconsin-Madison, 475 N. Charter Street, Madison, WI, 53706, USA}
\altaffiltext{7}{Department of Astronomy, Nanjing University, Nanjing 210093, China}
\altaffiltext{8}{Department of Physics and Astronomy, University of Utah, Salt Lake City, UT 84112, USA}
\altaffiltext{9}{Max-Planck Institut f\"{u}r  Astrophysik, Karl-Schwarzschild
Stra\ss e 1, D-85748 Garching, Germany}
\altaffiltext{10}{Institute for the Physics and Mathematics of the Universe
(IPMU), The University of Tokyo, Chiba 277-8582, Japan}
\altaffiltext{11}{Department of Astronomy and Astrophysics, The Pennsylvania State University,
  University Park, PA 16802}
\altaffiltext{12}{Institute for Gravitation and the Cosmos, The Pennsylvania State University,
  University Park, PA 16802}
\altaffiltext{13}{Department of Physics, Center for Astrophysics and Space Sciences, University of California, 9500 Gilman Drive, San Diego, CA 92093}
\altaffiltext{14}{Department of Physical Sciences, The Open University, Milton Keynes, MK7 6AA, UK}
\altaffiltext{15}{Apache Point Observatory, P.O. Box 59, Sunspot, NM 88349-0059, USA}

\begin{abstract}

  We study the redshift evolution of the dynamical properties of $\sim
  180,000$ massive galaxies from SDSS-III/BOSS combined with a local
  early-type galaxy sample from SDSS-II in the redshift range $0.1\leq
  z\leq 0.6$. The typical stellar mass of this sample is
  \mstar$\sim2\times10^{11}$\msun. We analyze the evolution of the
  galaxy parameters effective radius, stellar velocity dispersion, and
  the dynamical to stellar mass ratio with redshift. As the effective
  radii of BOSS galaxies at these redshifts are not well resolved in
  the SDSS imaging we calibrate the SDSS size measurements with
  HST/COSMOS photometry for a sub-sample of galaxies.  We further
  apply a correction for progenitor bias to build a sample which
  consists of a coeval, passively evolving population.  
    Systematic errors due to size correction and the calculation of
    dynamical mass, are assessed through Monte Carlo simulations.  At
  fixed stellar or dynamical mass, we find moderate evolution in
  galaxy size and stellar velocity dispersion, in agreement with
  previous studies.  We show that this results in a decrease of the
  dynamical to stellar mass ratio with redshift at $>2\sigma$
  significance.  By combining our sample with high-redshift literature
  data we find that this evolution of the dynamical to stellar mass
  ratio continues beyond $z\sim0.7$ up to $z>2$ as
  \mvir/\mstar$\sim (1+z)^{-0.30\pm 0.12}$ further strengthening the
  evidence for an increase of \mvir/\mstar\ with cosmic time.  This
  result is in line with recent predictions from galaxy formation
  simulations based on minor merger driven mass growth, in which the
  dark matter fraction within the half-light radius increases with
  cosmic time.

\end{abstract}




\section{Introduction} 
\label{intro}

Early-type galaxies play an important role in observational studies of
galaxy formation and evolution.  Tight empirical correlations between
the observed dynamical and stellar population properties of early-type
galaxies have been derived that set useful constraints to their
formation histories.  These are correlations between size (effective
radius, \re), surface brightness and stellar velocity dispersion
(\sigmae), i.e., the fundamental plane (FP, \citealt{Dressler1987};
\citealt{Djorgovski1987}; \citealt{BBF92,BBF93}), the stellar mass
plane \citep{Hyde2009b, Auger2010}, the fundamental manifold of
galaxies \citep{Zaritsky2008}, as well as correlations between galaxy
color and stellar population age and metal abundance with galaxy mass
\citep[see review by][]{Renzini06}.

Such scaling relations represent a powerful phenomenological tool to
study the co-evolution of baryonic and dark matter in galaxies.  The
latter has been studied extensively for the local galaxy population.
The tightness of the FP has been used to constrain stellar population
variations or dark matter content in galaxies \citep{Renzini1993} or
to study non-homology \citep{Ciotti1996}.
  \citet{Gerhard2001} studied the dynamical properties and dark halo
  scaling relations of giant elliptical galaxies, and found that the
  tilt of the FP is best explained by a stellar population effect and
  not by an increasing dark matter fraction with luminosity.

  \citet{Cappellari2006}, using data from the SAURON survey
  \citep{deZeeuw2002}, suggested that dynamical to stellar
  mass-to-light ratios larger than one are due to dark matter,
  assuming a constant stellar initial mass function (IMF)
  \citep{Kroupa2001} and using self-consistent models \citep[see][for
  a review]{Cappellari2012c}. 
Other studies based on Sloan Digital Sky Survey (SDSS,
\citealt{York2000}) data came to the conclusions that there is an
excess  over the predictions of stellar population models with a fixed
IMF which are luminosity dependent
\citep{Padmanabhan2004,Hyde2009,Hyde2009b}.

These conclusions have recently been revised in
\citet{Cappellari2012b} by means of a large number of axysimmetric
dynamical models including different representations of dark matter
halos reporting a variation of IMF slope with galaxy mass.
In fact, as highlighted by \citet{Thomas2011}, for instance, dark
matter fraction and IMF are highly degenerate.  They studied a sample
of Coma galaxies and their detailed decomposition into luminous and
dark matter reveals that for low-mass galaxies there is a good
agreement between dynamical masses with dark matter halo and lensing
results (galaxies with a $\sigma \sim200$ \kms\ are consistent with a
Kroupa IMF). For higher-mass galaxies ($\sigma >200$ \kms), the
disagreement can be due to either a non constant IMF (Kroupa IMF
under-predicts luminous dynamical masses for galaxies at $\sigma \sim
300$ \kms) or to a dark matter component which follows the light (see
also \citealt{Wegner2012}).

A promising complementary approach to detailed studies of local
galaxies for breaking the degeneracy between dark matter fraction and
IMF is to study the evolution of fundamental plane, dynamical and
stellar population properties of galaxies with look-back time
\citep{Bender1998,vanDokkum1998,Treu2005,Jorgensen2006,Saglia2010,Houghton2012,Bezanson2013b}.
Moreover, dark matter fractions can also be studied in samples of
lensed galaxies at different redshifts (see \citealt{Bolton2012} using
data from both the Sloan Lens ACS sample, SLACS, \citealt{Bolton2006},
\citealt{Bolton2008}, and the BOSS Emission line Lens Survey, BELLS,
\citealt{Brownstein2012}).

A large number of such investigations have been performed in recent
years, analyzing the redshift evolution of galaxy sizes
\citep[e.g.,][]{Daddi2005, Trujillo2006a,
  Trujillo2006b,Trujillo2007,Longhetti2007,Zirm2007,Toft2007,
  vanDokkum2008,Buitrago2008,Cimatti2008,Franx2008,Bernardi2009,Damjanov2009,Saracco2009,Bezanson2009,Mancini2009,Mancini2010,Valentinuzzi2010b,Carrasco2010,Szomoru2012,Newman2012,Saracco2014}
and dynamical properties of galaxies
\citep{vanderWel2005,vanderWel2008,Cenarro2009,Cappellari2009,vanDokkum2009,Newman2010,Onodera2010,Saglia2010,vandeSande2011,Toft2012,Onodera2012,vandeSande2012,Damjanov2013,Belli2014}. The
bottom line is that galaxy sizes appear to decrease (at fixed stellar
mass) and stellar velocity dispersions increase (at fixed stellar
mass) with increasing look-back time. Some of these conclusions are
still controversial, however. \citet{Tiret2011}, for example, argue
that the size evolution disappears when one homogenizes literature
data sets by measuring the stellar mass with the same method and
provides accurate estimates of the sizes to prevent systematic effects
in the \re\ measurements of low $S/N$ high-z compact galaxies
\citep{Mancini2009}.  \citet{Mancini2010} also report evidence for
galaxies as large as local ones at redshift higher than 1.4 suggesting
that not all high-$z$ galaxies are compact.  Moreover, the effect
depends on the photometric band and on whether galaxies have young or
old light-weighted ages and if they reside in clusters or field (see
\citealt{Valentinuzzi2010} for clusters and \citealt{Poggianti2012}
for field studies; see also \citealt{Trujillo2011} for a different
opinion).

So far, galaxies in the distant Universe (look-back times of a few
billion years and above) have not been studied at the same statistical
level as local galaxies. The new data set from the Baryon Oscillation
Spectroscopic Survey (BOSS, \citealt{Dawson2012}) as part of the Sloan
Digital Sky Survey-III \citep[SDSSIII,][]{Eisenstein2011} provides the
opportunity to investigate the dynamical and stellar population
properties of a galaxy sample of unprecedented size up to redshifts
$z\sim 0.7$. The survey is currently obtaining spectroscopic data for
nearly 1.5 million massive galaxies at redshifts between 0.2 and 0.7,
hence up to look-back times of $\sim 6\;$Gyr.  The combination of this
sample with local early-type galaxies from SDSS-I/II allows us to make
a statistically significant link between local galaxy properties and
higher redshift observations.  This is the main aim of the present
paper.

The paper is organized as follows. The galaxy sample is described in
Section~\ref{sec:sample}. Photometric, kinematic, and dynamical
properties are presented in Section~\ref{sec:galaxy_properties}, as
well as the calibration technique for measuring the effective radii
based on a sub-sample of galaxies with HST photometry and the
correction for progenitor bias. The results are presented in
Section~\ref{sec:redshift_ev_all} and discussed in
Section~\ref{sec:discussion}. The paper concludes with
Section~\ref{sec:conclusions}.

Throughout the paper we assume the following cosmology with $H_0=71.9$
\kmsmpc, $\Omega_{\rm m}=0.258$, and $\Omega_{\Lambda}=0.742$
following the cosmology used for the stellar mass determination of
BOSS galaxies in DR9 \citep{Maraston2012}.

\section[]{Data}
\label{sec:sample}
We use the galaxy sample from SDSS-III/BOSS covering a redshift range
$0.2\la z\la 0.7$. To leverage our study on the redshift evolution we
combine this sample with a local sample of massive galaxies at $z\sim
0.1$ drawn from SDSS-II.

\subsection{Main galaxy sample from SDSS-III/BOSS}
Data are taken from the SDSS-III/BOSS Data Release Nine (DR9,
\citealt{SDSSDR9}).  BOSS \citep{Dawson2012,Smee2012}, one of the four
surveys of SDSS-III \citep{Eisenstein2011}, is taking spectra for 1.5
million luminous massive galaxies with the aim of measuring the cosmic
distance scale and the expansion rate of the Universe using the
Baryonic Acoustic Oscillations (BAO) scale \citep{Anderson12}.
 Data are taken with an upgraded version of the multi-object
  spectrograph on the SDSS telescope \citep{Gunn2006}.
  BOSS galaxy targets are selected from the SDSS {\it ugriz} imaging
  \citep{Fukugita1996,Gunn1998,Stoughton2002}, including new imaging
  part of DR8 mapping the southern Galactic hemisphere.  A series of
  color cuts have been used to select targets for BOSS spectroscopy
  (Padmanabhan et al. 2014, in preparation).  The selection criteria
  are designed to identify a sample of luminous and massive galaxies
  with an approximately uniform distribution of stellar masses
  following the Luminous Red Galaxy (LRG, \citealt{Eisenstein2001})
  models of \citet{Maraston2009}.
  The galaxy sample is composed of two populations: the higher
  redshift Constant Mass Sample (CMASS, $0.4<z<0.7$) and the
  Low-Redshift Sample (LOWZ, $0.2<z<0.4$). A fraction (around 1/3) of
  those LOWZ galaxies derived with those cuts have been already
  observed in SDSS-I/II and are included in the BOSS sample, but they
  are not re-observed if they had reliable redshifts.  These two
  population are well separated in the $(g-r) $ and $(r-i)$ colors
  diagram \citep{Masters2011}.

  To extract a working set of objects from the entire BOSS sample we
  matched galaxies from different catalogs of stellar velocity
  dispersions and stellar masses described in the following Sections
  using the keywords {\tt PLATE, MJD, FIBERID} that uniquely determine
  a single observation of a single object.  The final merged catalog
  comprises $491,954$ galaxies that are included in DR9.
We selected objects from the LOWZ and CMASS samples ({\tt
  BOSS\_TARGET1 = Target flags}) with a good platequality ({\tt
  PLATEQUALITY='good'}), with object class 'galaxy' ({\tt
  class\_noqso=GALAXY}), with high-confidence redshifts ({\tt
  ZWARNING\_NQSO=0}), and for which we have a unique set of objects in
the case of duplicate observations ({\tt SPECPRIMARY=1}).

The final sample we will analyze comprises $\sim 180,000$ objects (37\% of
the original sample) obtained after applying some additional redshift
cuts and quality cuts to stellar velocity dispersions and stellar
masses that we will discuss in Section~\ref{sec:galaxy_properties}.

\subsection{Local galaxy sample from SDSS-II}

In order to connect our BOSS galaxies to the local galaxy population,
we use a sample of galaxies from the SDSS Data Release Seventh
\citep[DR7,][]{Abazajian2009}.  We select early-types galaxies
following \citet{Hyde2009}. Galaxies had to be well fitted by a de
Vaucouleurs profile in the $g$ and $r$-bands ({\tt fracDeV\_g}=1 {\tt
  fracDeV\_r}=1), with an early-type like spectrum ({\tt eClass} $<$
0), extinction-corrected $r$-band de Vaucouleurs magnitudes in the
range $14.5<${\tt deVMag\_r}$< 17.5 $ (this results in a tighter limit
compared to that of the SDSS Main Galaxy sample, see details in
\citealt{Hyde2009}), measured stellar velocity dispersion in DR7 {\tt
  velDisp} $>0$, and with an axis ratio in $r$-band of $b/a >0.6$, to
be more likely pressure supported. As described in \citet{Hyde2009}
the DR7 $b/a$ distribution shows two distinct populations separated by
this axis-ratio value with a 20\% of low-luminosity objects at
$b/a<0.6$.\footnote{We do not apply this cut in our BOSS sample
  because  $b/a$ from the SDSS imaging could be highly
    unreliable.  In fact, we do not find  the same clear
  separation in the $b/a$ distribution of BOSS galaxies. Also,
    the BOSS $b/a$ distribution looks different probably due to the
    large uncertainties on $b/a$ from SDSS imaging.  However, we
    assess the typical percentage of galaxies which should have a
    $b/a<0.6$ by using the sub-sample of BOSS galaxies with COSMOS
    photometry (see Section~\ref{subsec:size_cosmos} for details) and
    find that $\sim 22$\% of galaxies have a $b/a<0.6$, which is
    consistent with \citet{Masters2011} findings. Finally, the choice
    of this cut does not affect appreciably our results.}  This
retains $\sim 123,500$ DR7 objects.

  \citet{Bernardi2010} did a detailed comparison of different methods
  to select early-types in the literature (morphologically based,
  colors, or structural parameters-based methods), showing that
  \citet{Hyde2009} cuts give similar results to other methods but is
  more efficient in discriminating elliptical galaxies from spirals,
  which is important in our analysis.
  
We select galaxies with {\tt zWarning}$=0$ and apply some constraints
on the errors of the parameters: errors on
$\re<$70\%, errors on the axis ratios $0<err_{b/a}<1$, errors
in $\sigmas <$30\%. Only galaxies with stellar
velocity dispersion of 70$<$\sigmas$<$550\kms\ were selected,
following the BOSS cuts.

\section[]{Galaxy properties}
\label{sec:galaxy_properties}

The primary aim of this work is to study the redshift evolution of the
dynamical properties of BOSS galaxies.  Stellar masses are taken from
\citet{Maraston2012} and stellar velocity dispersions from
\citet{ThomasD2012}. Both of these quantities are included in the DR9
data release.  We used a sub-sample of 240 galaxies with additional
HST/COSMOS photometry and BOSS spectra \citep{Masters2011} to derive a
calibration for galaxy sizes from DR8.  In the following sections we
describe the galaxy parameters in more detail.  In
Section~\ref{subsec:galaxy_mass} we describe how the effective radii
and stellar velocity dispersion are combined to derive virial masses.

The redshifts used in our analysis are those extracted from DR9 as
{\tt z\_noqso} with formal $1\sigma$ error given by {\tt z\_err\_noqso}
(outputs of the BOSS pipeline as described in \citealt{Bolton2012b}
and Schlegel et al. 2014, in preparation).
Redshifts are successfully determined for $\sim 98\%$ of CMASS
galaxies \citep[][their Table~1]{Anderson12}.  Errors in the measured
redshift are less than about 0.0002 ($\sim 60$ \kms).  BOSS observed a
few galaxies at $z>0.7$ and $z<0.2$. In our analysis we focus on the
redshift range $0.2\leq z\leq 0.7$ and therefore excluded all the
galaxies outside this redshift range. This cut retains 90\% of the
galaxies ($444,118$ objects).

\subsection[]{Stellar mass}
\label{subsec:photo_masses}

We use the stellar masses \mstar\ from \citet{Maraston2012} as
published in the DR9 data release\footnote{Stellar masses have been
  derived for galaxies with non-zero photometry in i-band {\tt
    modelmag\_i $>$ 0.0}, {\tt z\_err\_noqso $\geq$ 0} and {\tt
    z\_noqso $>$ z\_err\_noqso}.}. These masses are derived through
broad-band spectral energy distribution (SED) fitting of model stellar
populations on SDSS $u,g,r,i,z$ {\tt modelMag} magnitudes from DR8,
scaled to the $i$-band {\tt cmodelMag} magnitude. The BOSS
spectroscopic redshift is used to constrain the fits. We note that the
DR9 data release also provides alternative mass estimates from
\citet{Chen2012} \citep[see][for discussion]{Maraston2012}.

\citet{Maraston2012} use different types of templates to derive
stellar masses; passive models, star-forming models or a mix, chosen
so as to match the galaxy expected galaxy type, based on a color cut,
in BOSS\footnote{This assignment of passive or star-forming models
  following a color cut is strictly valid at $z>0.4$.  However, since
  LOWZ galaxies are generally red, \citet{Maraston2012} assumed the
  criterion valid over the full BOSS range.}.  In this paper we used
stellar masses obtained with the passive template that is a mix of old
populations with a spread in metallicity and which was found to
reproduce well the colors of luminous red galaxies at redshift 0.4 to
0.7 \citep{Maraston2009}. This maximally old and passive LRG template
minimizes the risk of underestimating \mstar\ which is the typical
effect that occurs when one determines stellar masses from light
\citep{Maraston2010,Pforr2012}.

We adopt the stellar masses derived from the median of the probability
distribution function (PDF). Typical uncertainties on \mstar\ are
$<0.1\;$dex, independently of redshift \citep[see][for
details]{Maraston2012}. 

The Portsmouth stellar mass pipeline described in
  \citet{Maraston2012} provides stellar masses using various types of
  configurations, i.e. \citet{Kroupa2001} or \citet{Salpeter1955} IMF,
  passive \citep{Maraston2009} or star-forming \citep{Maraston2006}
  models, and considering or not the mass-loss in the stellar
  evolution.
The subset of calculations used here adopt a Kroupa IMF.  As is well
known, a Salpeter IMF produces systematically larger stellar masses by
about a factor 1.6 \citep{Maraston2005,Bolzonella2010}. Systematic
uncertainties are mainly due to the choice of the IMF ($\sim
0.2\;$dex).

We removed galaxies for which \mstar\ was not properly determined due
to unreliable values of redshift and photometry issues.  This cut
retains 387,590 galaxies (79\% of the full sample).

\subsection[]{Size}
\label{sec:size}

The photometric data used in this paper were derived using the
SDSS-III DR8 pipeline \citep{Aihara2011}. One of the main updates
performed in DR8 compared to \citet{Stoughton2002} is the correction
for sky levels (in particular for extended galaxies) which past
studies found to be overestimated \citep{Bernardi2007b,
  Lauer2007b,Guo2009}.  This issue has also been addressed in detail
in \citet{Blanton2011}.  SDSS effective radii are adopted from the DR8
catalog. In this section we describe this measurement and its
calibration with COSMOS/HST imaging.

\subsubsection[]{SDSS DR8 effective radii}
\label{subsec:size_sdss}
Effective radii are estimated using
seeing-corrected de Vaucouleurs effective radii ({\tt deVRad}) and
the associated errors ({\tt deVRadErr}, of the order of $\sim 5-25$ \%
depending on redshift). Those values correspond to the effective
radius along the semi-major axis derived on elliptical aperture.

BOSS galaxies are often not well resolved in SDSS imaging. The average
seeing of the SDSS survey is 1\farcs05 from the {\tt PSF\_FWHM} in the
$i$-band (lower and upper percentile of 0\farcs88 to 1\farcs24), which
is better than in previous data releases due to the repeated images
taken during the survey \citep[see][]{Ross2011,Masters2011}.  The
median effective radius of BOSS galaxies (after having applied the
previous cuts on the sample) is 1\farcs24 (upper and lower percentile
2\farcs12 and 0\farcs72, respectively).  As a consequence, seeing
effects will affect size measurements (see \citealt{Saglia1993b},
\citealt{Saglia1997}, \citealt{Bernardi2003a}), which we need to
correct for.  To develop a seeing correction we compare SDSS galaxy
radii with measurements based on high-resolution HST/COSMOS imaging
(Section~\ref{subsec:size_calibration}).

The surface brightness models used by pipeline to obtain galaxy sizes
are relatively simple (single parameter fits, i.e. exponential or de
Vaucouleurs profiles); a more complex model is not feasible for this
analysis because of the limitation of the image resolution.  We tested
this conclusion and found that performing S\'ersic fits on BOSS images
yields strong degeneracies between effective radius and S\'ersic
index, preventing robust estimate of effective radii.  We estimate
SDSS circularized radii as $R_{\rm e,circ}=R_{\rm e}\times q^{1/2}$
where $R_{\rm e}$ is the semi-major axis of the half-light ellipse and
$q=b/a$ is the axis ratio which is part of the DR8 catalog.

The effective radii should, in principle, be referred to a fixed
rest-frame wavelength to account for the fact that early-types have
color gradients, so on average their optical radii are larger in
bluer bands and at higher $z$ this effect will make the sizes larger
\citep[see][ for discussions]{Bernardi2003a, Hyde2009}.  We did not
apply any correction for this trend, because uncertainties of the size
calibration ($\sim$ 10-25\%) are larger than the typical variation in
size measured in different filters (from 4\% to 10\%,
\citealt{Bernardi2003a}, \citealt{Hyde2009}).  In
Section~\ref{subsec:mass_size} we describe the average effect this
could have on the size evolution studies.

Effective radii were converted to physical radii using the code of
\citet{Hogg1999} which presents the scale conversion between arcsec
and kpc for our given cosmology.  Hereafter, we will use the notation
\emph{pipeline \re} to represent SDSS effective radii in kpc. We further
remove from the catalogs galaxies with unreliable values of \re\ and
their errors.  This cut retains $\sim370,000$ objects.

\subsubsection[]{COSMOS effective radii}
\label{subsec:size_cosmos}

\begin{figure*}
\begin{center}
\includegraphics[width=0.7\textwidth]{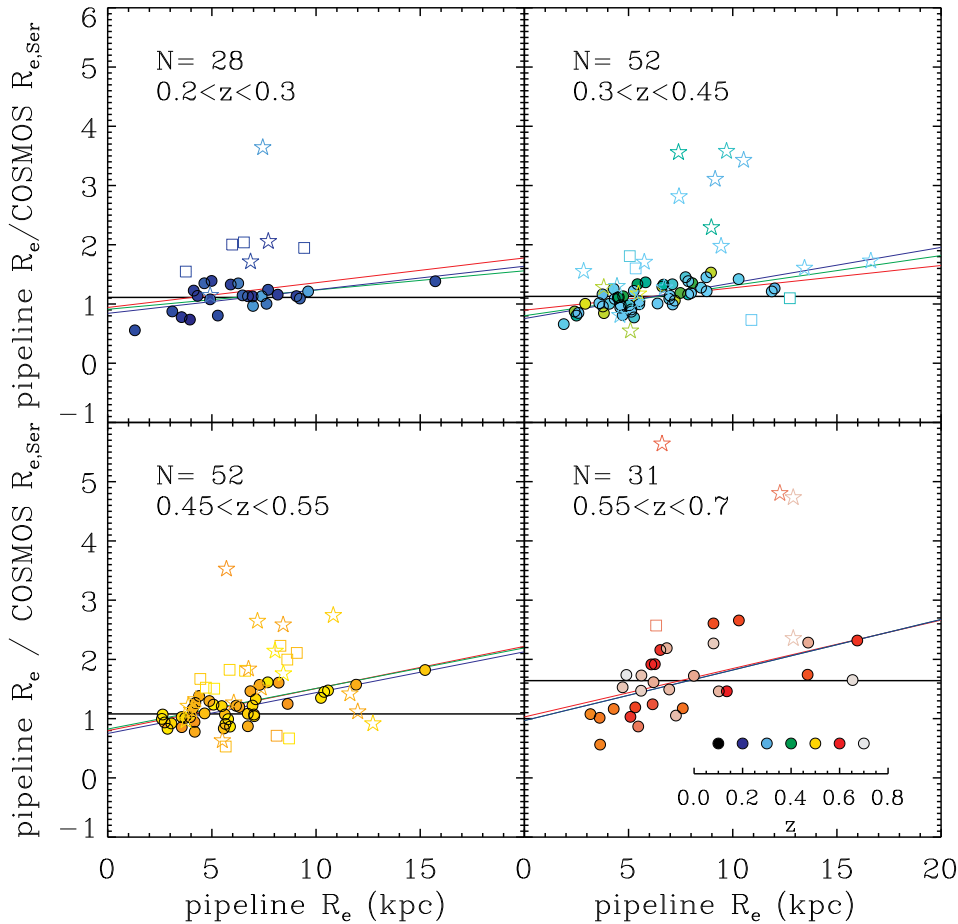}
\end{center}
\caption{Ratio between pipeline \re\ from SDSS DR8 and COSMOS
  \ensuremath{R_{\rm e, Ser}} as a function of pipeline \re. Points
  are coded as a function of redshift ($0.2\leq z\leq0.3$, $0.3<z\leq
  0.45$, $0.45<z<0.55$, and $0.55\leq z \leq 0.7$). Different lines
  represent different fitting procedures: the red line is a linear
  fit, the green line is a linear fit with one $2\sigma$ clip, the
  blue line is iterative $2\sigma$ clipping, the horizontal black line
  is the single offset. Circles are the points that have been used in
  the fit and open squares are points discarded in the iterative sigma
  clipping. Open stars are multiple systems in HST imaging which are
  unresolved in the SDSS images \citep{Masters2011}. Labels in each
  panel give the number of galaxies used in the fit including objects
  discarded in the iterative $2\sigma$ clipping (multiple systems
    are not included in this number).}
\label{fig:sdss_re_cosmos_re_vs_sdss}
\end{figure*}

\citet{Masters2011} constructed a sub-sample of 240 BOSS galaxies with
HST/COSMOS imaging \citep{Koekemoer2007,Leauthaud2007}. This sample
was used to calibrate the SDSS DR8 radii part of the CMASS sample.  We
adopted the effective radii from the public {\tt Zurich Structure
  \& Morphology Catalog v1.0}\footnote{Available at \\
  http://irsa.ipac.caltech.edu/data/COSMOS/datasets.html}
\citep{Scarlata2007,Sargent2007} derived from $I$-band (F814W) ACS
images. Effective radii are available from this catalog for 224 of
the 240 objects.

The catalog contains galaxy structural parameters derived from a
two-dimensional decomposition using the {\tt GIM2D} code \citep[Galaxy
Image 2D,][]{Simard1998} on the ACS/HST images 
as deep as $I_{\rm AB}\sim22.5$ \citep{Sargent2007}. We
used the seeing-corrected effective radii {\tt R\_GIM2D} resulted from
the one component S\'ersic fit, of the form

\begin{equation}
I(r)=I_{\rm e} 10^{[-b_n((r/R_{\rm e,"})^{1/n}-1)]} 
\end{equation}

\noindent
where $I_{\rm e}$ is the effective intensity, and the constant $b_n$
is defined in terms of the shape parameter {\it n} and is chosen so
that $R_{\rm e,"}$ encloses half of the total luminosity and it is
measured in arcsec.  The quantity $b_n$ can be well approximated by
$b_n=0.868n-0.142$ \citep{Caon1993}.
The {\tt Zurich Structure \& Morphology Catalog v1.0} contains two
values of the effective radius from {\tt GIM2D}, {\tt R\_GIM2D} and
{\tt R\_0P5\_GIM2D} which correspond to the PSF-corrected effective
radii. The choice of one of the two is purely arbitrary and they show
a negligible median difference of 0\farcs001 (with {\tt R\_GIM2D}
being smaller than {\tt R\_0P5\_GIM2D}).
The catalog gives the uncertainties on effective radii as 99\%
confidence lower and upper error on {\tt R\_GIM2D} ({\tt
  LE\_R\_GIM2D}, {\tt UE\_R\_GIM2D}, respectively). Those errors are
$<1$\% of {\tt R\_GIM2D}, much smaller than the errors in \re\ from
SDSS photometry ($\sim 15$ \% in the COSMOS/BOSS sub-sample).

Sizes are converted to circularized effective radii $R_{\rm e, serc}$
following the procedure described
in \citet{Saglia2010}\footnote{The code is available at \\
  http://www.mpe.mpg.de/$\sim$saglia/rps\_software.html} which calculates
the half-light radius obtained from the classical curve of growth
analysis of the intrinsic S\'ersic profile.  The procedure requires
{\tt R\_{GIM2D}}, S\'ersic index {\tt SERSIC\_N\_GIM2D}, and axis
ratio $q=b/a$, where $a$, $b$ are the semi-major and minor axis of the
half-light ellipse, and $q=1-\epsilon$, where $\epsilon$ is the
ellipticity of the object {\tt ELL\_GIM2D}.
  Here we use a different parametrization than in
  Section~\ref{subsec:size_sdss} because we have additional
  information from the S\'ersic fit (the two parametrization would
  give consistent results for a wide range of $q$).

  We did not apply any correction to account for the fact that
  rest-frame wavelengths are different at different redshifts using
  only $i$-band data because this sample is only used for calibration
  purposes, and uncertainties due to the size calibration we derive
  using COSMOS radii are much larger than the wavelength variation at
  the redshifts considered (see Section~\ref{subsec:size_sdss}).
  Moreover COSMOS has just one filter available for those
  galaxies. COSMOS radii were converted to kpc
  in the same way as SDSS radii using the code of \citet{Hogg1999}.
  Hereafter, we will use the notation \emph{COSMOS
      \ensuremath{R_{\rm e, Ser}}} to represent COSMOS effective
  radii in kpc.

\subsubsection[]{Comparison SDSS vs COSMOS}
\label{subsec:size_calibration}

A significant fraction (23 \%) of BOSS galaxies with HST imaging are
unresolved multiple systems in SDSS imaging \citep{Masters2011}. To
derive an accurate calibration we excluded these unresolved multiple
systems from our analysis (44 galaxies) using the public catalog of
\citet{Masters2011} \footnote{Available at
  http://www.icg.port.ac.uk/$\sim$mastersk/BOSSmorphologies/}.
The redshift range we wish to study is $0.2\leq z\leq0.7$, and we
discarded 4 additional galaxies at higher redshifts and 13 at
$z<0.2$. This leaves us a sample of 163 galaxies.  COSMOS redshifts
are adopted, as not all of these objects (158 out of 163) have BOSS
redshifts (this will not change our results, because the median
difference between COSMOS and BOSS redshifts is negligible,
7.21$\times 10^{-6}$). We do not correct for PSF variations in the
SDSS imaging, because the PSF is reasonably stable and the effect is
negligible compared to the overall correction derived here.

\begin{figure*}
\begin{center}
\includegraphics[width=0.33\textwidth]{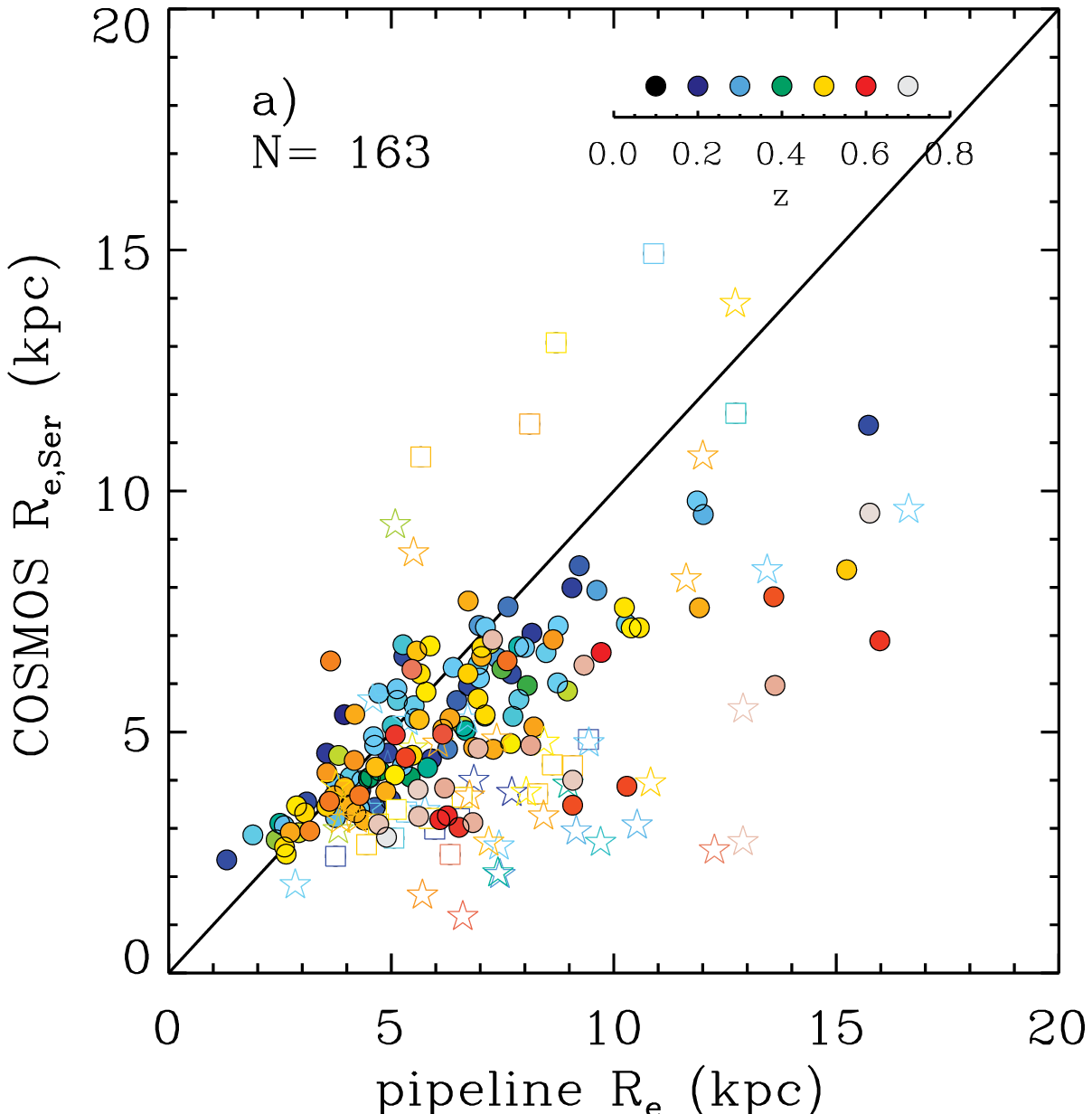}
\includegraphics[width=0.33\textwidth]{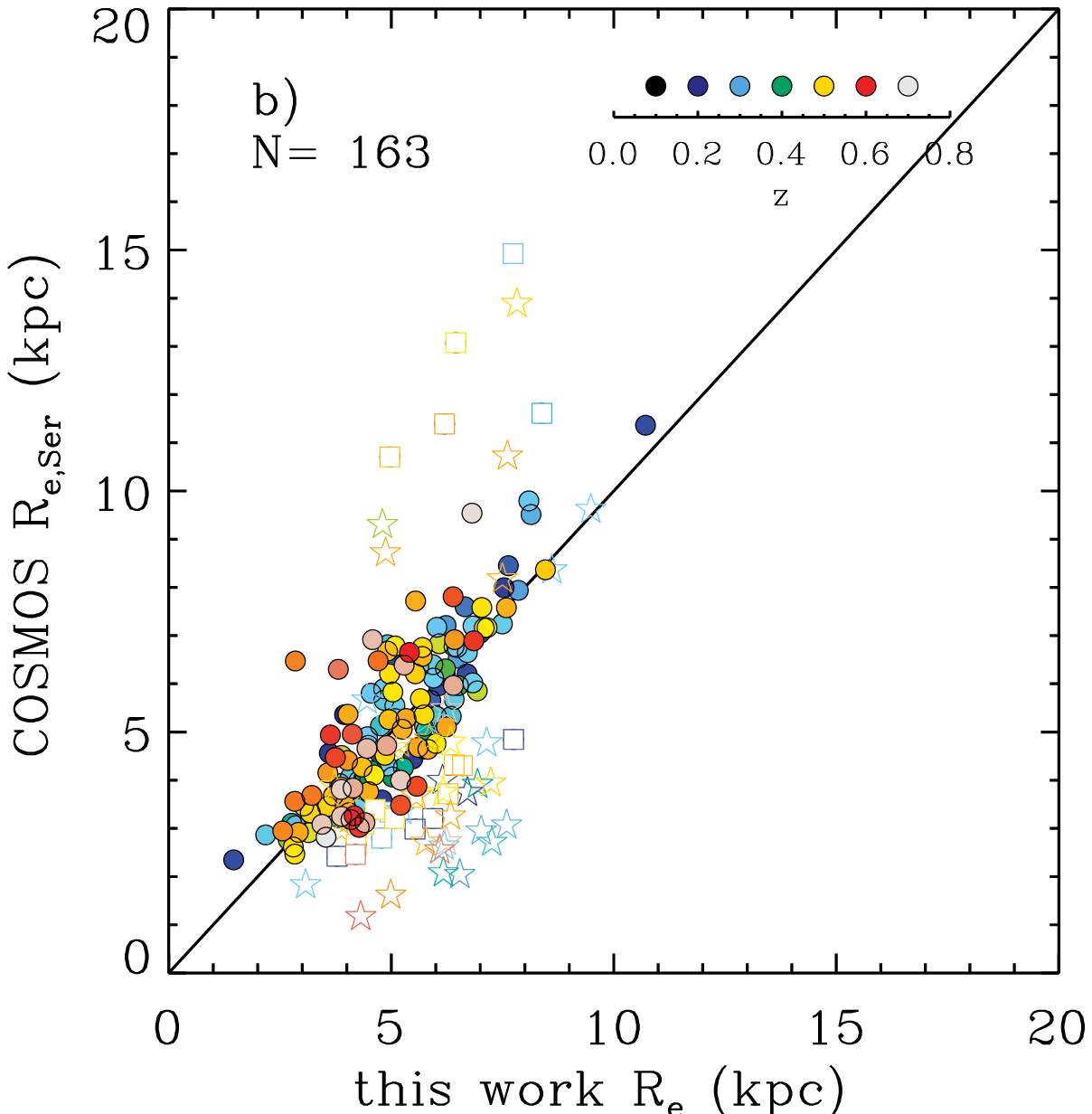}
\includegraphics[width=0.33\textwidth]{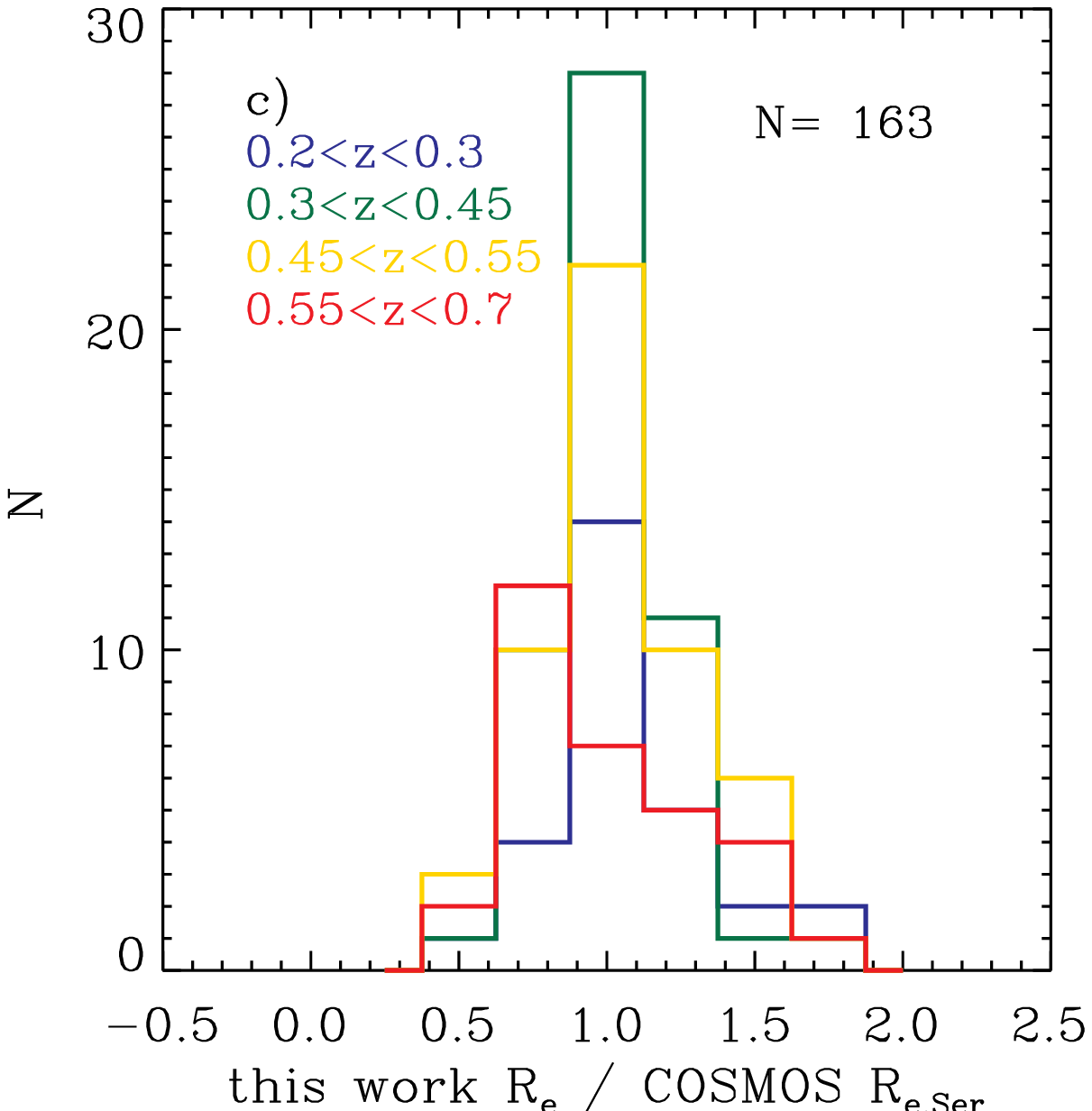}
\end{center}
\caption{{\em Left panel:} COSMOS $R_{\rm e, Ser}$ as a function of
  pipeline SDSS \re\ coded in terms of redshift. {\em Central panel:}
  COSMOS $R_{\rm e, Ser}$ as a function of rescaled SDSS \re\ (used in
  the present work) coded in terms of redshift. Symbols are as in
  Figure~\ref{fig:sdss_re_cosmos_re_vs_sdss}. {\em Right panel:}
  Distribution of the ratio between SDSS rescaled \re\ and COSMOS
  $R_{\rm e, Ser}$ for the four redshift bins in
  Figure~\ref{fig:sdss_re_cosmos_re_vs_sdss}. Histograms contain also
  discarded objects but not multiple systems.  Legends in each
    panel give the number of galaxies used in the derivation of the
    size correction including objects discarded in the iterative
    $2\sigma$ clipping (multiple systems are not included in this
    number).}
\label{fig:sdss_re_cosmos_re_vs_sdss_re_correction}
\end{figure*}

Figure~\ref{fig:sdss_re_cosmos_re_vs_sdss} shows the ratio between SDSS
and COSMOS effective radii as a function of SDSS effective radius for
four redshift bins. The ratio is around $1.1$ and SDSS radii can be
overestimated by up to a factor two. The discrepancy between SDSS and
COSMOS radii increases with increasing SDSS radius \citep[see
also][]{Masters2011}.
\citet{Masters2011}\footnote{\citet{Masters2011} estimated the size
  correction only for the CMASS sample and used major axis radii.}
found that a single offset was reproducing their data in which they
compared the ratio SDSS over COSMOS radii as a function of COSMOS
radii. In this work we compare the ratio of SDSS over COSMOS radii as
a function of SDSS radii to derive a correction for the full BOSS
sample.  As is to be expected, there is also some redshift dependence,
in the sense that SDSS radii overestimate most the true radii at
higher redshifts. Also, as expected, the scatter of the relationships
increases with redshift owing to the decrease in SDSS imaging quality.

The size calibration could be affected by the larger uncertainties in
the SDSS effective radii due to the higher than typical  sky background 
($\geq$ 60-70\%) of SDSS images in the COSMOS field
\citep{Masters2011,Mandelbaum2012} (on the other side seeing is
smaller than typical of 10-15\%) which could give relations between
SDSS and COSMOS radii not universal for the full BOSS sample.

We performed fits to these relationships in the four redshift bins
independently. We tested for linear correlations applying different
levels of sigma clipping in the linear fits: no sigma clipping, red
line in Figure~\ref{fig:sdss_re_cosmos_re_vs_sdss}; just one $2\sigma$
clipping green line in Figure~\ref{fig:sdss_re_cosmos_re_vs_sdss}; and
an iterative $2\sigma$ clipping blue line in
Figure~\ref{fig:sdss_re_cosmos_re_vs_sdss} with a maximum of three
iterations.

We fitted a linear relation of the form $R_{\rm e, pipeline}/R_{\rm e,
  COSMOS, Ser}=a +b (R_{\rm e,pipeline})$.  The best-fit quantities
$a$ and $b$, the number of galaxies used in the fit after the sigma
clipping, the scatter of the relations $c$  (which include
  objects discarded by the $2\sigma$-clipping) and their associated
errors obtained as $1\sigma$ uncertainties for each redshift bin are
listed in Table~\ref{tab:sizecorr}. {\rm The least-square fits were
  performed using the {\tt MPFIT} algorithm \citep{Markwardt2009}
  under the {\tt IDL}\footnote{Interactive Data Language is
    distributed by Exelis Visual Information Solutions. It is
    available from
    http://www.exelisvis.com/ProductsServices/IDL.aspx.}
  environment}. Fits with and without sigma clipping are consistent
within the errors.

The slope of the relation increases slightly with redshift as to be
expected.  The scatter about the relation is comparable in the first
three redshift bins, while the last redshift bin shows a considerably
higher scatter (see Table~\ref{tab:sizecorr}, $c=0.49\pm0.06$). For
this reason, we will only use the first three redshift bins in the
analysis.

We tested the significance of the fits through an F-test by comparing
the resulting $\chi^2$ values for free and fixed slope fits accounting
for the number of degrees of freedom, and find
a maximum probability of no relation to be $\sim$2 \%, which
confirms the statistical validity of our fits. The final fits we adopt
for the radius correction in each redshift bin are the linear fits
with iterative $2\sigma$ clipping (blue lines)  because they give
  corrections with a smaller scatter compared to other
  fits (of 6-30\%).  The open squares in
Figure~\ref{fig:sdss_re_cosmos_re_vs_sdss} are those galaxies that were
discarded in the sigma clipping. Open stars represent unresolved
multiple systems not considered in the fits.

We additionally searched for correlations of the effective radius with
several other DR8 structural parameters like axis ratio {\tt
  b/a}, {\tt fracdev}, and the difference between {\tt fiber2mag}
and {\tt modelmag} with the aim at finding the best parameter space
for the radius correction. None of these parameters helped improving
the radius correction.

 The size correction we derive here accounts also for the fact
  that galaxies in our sample could have been better described by a
  S\'ersic profile rather than a de Vaucouleurs, therefore we should
  consider our calibrated sizes as "S\'ersic-like".

\begin{table}
\label{tab:correction}
\begin{scriptsize}
\begin{center}
\caption{Size correction for the four redshift bins.}
\begin{tabular}{l c c c c}
\hline
\hline
\noalign{\smallskip}
\multicolumn{1}{c}{$z$ range} &
\multicolumn{1}{c}{$N$} &
\multicolumn{1}{c}{$a$}  &
\multicolumn{1}{c}{$b$}  &
\multicolumn{1}{c}{$c$}  \\
\hline
$0.2 \leq z \leq 0.3$          &24    &0.84  $\pm$0.11   &0.04  $\pm$0.01 & 0.29$\pm$ 0.04\\
$0.3 < z \leq 0.45$            &48    &0.75  $\pm$0.06   &0.06  $\pm$0.01 & 0.19$\pm$ 0.02 \\
$0.45 < z < 0.55$              &41    &0.75  $\pm$0.04   &0.07  $\pm$0.01 & 0.32$\pm$ 0.03 \\
$0.55 \leq z \leq0.7$         & 30   & 0.97 $\pm$0.17   &0.08  $\pm$0.02 & 0.49$\pm$ 0.06\\
           
\hline
\noalign{\smallskip}
\label{tab:sizecorr}
\end{tabular}

\begin{minipage}{\columnwidth}

{\sc Notes.} ---
A correlation of the form $R_{\rm e, pipeline}/R_{\rm e, COSMOS, Ser}=a
+b(R_{\rm e,pipeline})$ is assumed. $N$ is the number of points
  used in the fit after  iterative $2\sigma$ clipping. Uncertainties on each
  parameter are 1$\sigma$ errors. The rms scatter $c$ is derived as
  deviation of the data about the fits  considering also objects
    discarded by the $2\sigma$-clipping.
\end{minipage}
\end{center}
\end{scriptsize}
\end{table}

\subsubsection{Radius calibration}
\label{subsec:radiicalib}

\begin{figure*}
\begin{center}
\includegraphics[width=0.9\textwidth]{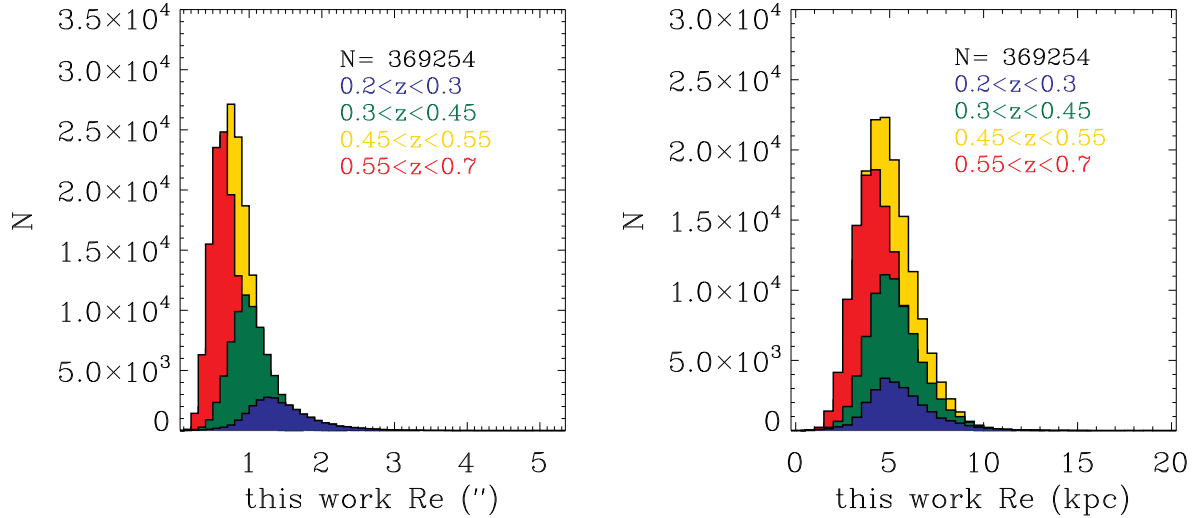}
\end{center}
\caption{Histogram of the effective radii derived in this work (after
  size correction), in arcseconds {\em (left panel)} and kpc {\em
    (right panel)} for various redshift bins as color-coded in
  Figure~\ref{fig:sdss_re_cosmos_re_vs_sdss_re_correction}.}
\label{fig:re_distrib}
\end{figure*}

Figure~\ref{fig:sdss_re_cosmos_re_vs_sdss_re_correction} middle panel b)
shows the final corrected radii that are obtained using the fits of
Figure~\ref{fig:sdss_re_cosmos_re_vs_sdss}. For comparison the left a)
panel shows the uncorrected radii.  Panel c) presents the distribution
of the ratio between the SDSS and COSMOS radii in various redshift
bins after the correction. The radii agree well at all redshifts after
the correction has been applied. More specifically, the median ratio
between our rescaled \re\ and COSMOS $R_{\rm e, Ser}$  is 1.02 (upper and lower
quartile 1.34, 0.86 and mean 1.10) for $0.2\leq z\leq0.3$, 0.98 (upper
and lower quartile 1.15 and 0.85 and mean 1.01) for $0.3<z\leq0.45$,
1.01 (upper and lower quartile 1.31 and 0.84 and mean 1.05) for
$0.45<z<0.55$, and 0.99 (upper and lower quartile 1.41 and 0.73 and
mean 1.02) for $0.55\leq z\leq 0.7$. Median values of the
distributions in each redshift are compatible within $\pm 1
\sigma/N^{1/2}$, where $N$ is the number of objects.  Typical errors
on rescaled radii range from 0.7 kpc at $z\sim 0.2$ to 1.0 kpc at
$z\sim 0.7$ and median radii range from 5.34 to 4.28 kpc, mode 4.73 to
3.48 kpc, in this redshift range. If we did not apply the size
correction we would have larger radii (median sizes range from 5.72
kpc at $z\sim0.25$ to 5.40 kpc at $z\sim0.55$, 6.51 kpc at
$z\sim0.65$, mode from 5.00 to 4.00 kpc and 5.00 at $z\sim0.65$).

Filled circles in
Figure~\ref{fig:sdss_re_cosmos_re_vs_sdss_re_correction} are those
galaxies that were used in the previous section to derive the
calibration. Open squares are those objects that were discarded in the
iterative $2\sigma$ clipping.  Open stars are the unresolved multiple
systems discarded from the calibration. Most of the multiples are
strong outliers in these plots and they would have been discarded
during the sigma clipping fit.

Figure~\ref{fig:re_distrib} shows the resulting distributions of
galaxy effective radii (in both arcseconds and kpc) for the final
sample of 369,254 galaxies in the various redshift bins. The size
distribution can be described by a log-normal function (as the typical
size distribution at low redshift, \citealt{Shen2003,Bernardi2003a})
but with different peaks of the distributions suggesting a variation
of typical sizes with $z$. In Section~\ref{sec:redshift_ev_all} we
present our results using both the corrected SDSS size and pipeline
sizes (which we circularized using SDSS axis ratio for this purpose).

\subsubsection{Systematic errors}

The systematics in the error budget have been assessed through Monte
Carlo simulations which account for uncertainties in both parameters
$a$ and $b$ of the fit. The average errors vary with redshift in a
non-linear fashion. The errors are $\sim 0.6$ kpc at $z\sim 0.25$,
$\sim 0.2$ kpc at $z\sim0.55$, and $\sim 0.6$ kpc at $z\sim0.65$.

 We additionally include in our Monte Carlo simulations the impact
  of unresolved multiple systems, which have systematically
  overestimated sizes.  By using the COSMOS/BOSS sub-sample we can
  estimate that those correspond to the $\sim$ 6\% of the galaxies in
  this sample (see left panel of Figure~\ref{fig:multiples_re_histo}).
  The sizes of the two components which are resolved in the COSMOS
  imaging (and unresolved in the SDSS imaging) allow us to assess the
  contribution of unresolved multiple systems in our analysis, which
  seems to be negligible compared to other systematic uncertainties
  (see Appendix~\ref{sec:multiples} for details).

More detail on the Monte Carlo simulations are given also in
Section~\ref{sec:redshift_ev_all}, where we discuss the impact on the
final science analysis.

\subsection[]{Stellar velocity dispersion}
\label{subsec:sigma}

Stellar velocity dispersions (\sigmas) are taken from the Portsmouth
Spectroscopic pipeline described in \citet{ThomasD2012}, also
available in DR9. Briefly, stellar kinematics are derived by means of
the Penalized Pixel-Fitting method pPXF \citep{Cappellari2004} in
spectra in which emission lines are fitted with Gaussian templates by
using the GANDALF code \citep{Sarzi2006}.  The stellar population
models of \citet{Maraston2011a} have been adopted to fit the stellar
continuum. These are based on a hybrid model between MILES stellar
library \citep{SanchezBlazquez2006} and theoretical spectra at bluer
wavelengths from UVBLUE \citep{Rodriguez05}. Stellar population models
based on the MILES library have a resolution of 2.54 \AA\ FWHM
\citep{Beifiori2011b}, and therefore needed to be only slightly
downgraded to match the BOSS spectral resolution ($R\sim 1800-2000$ at
$5000\;$\AA, 2.78 \AA\ $-$2.50 \AA\ FWHM). Stellar velocity
dispersions have been measured in the typical rest-frame wavelength
range $4500-6500$ \AA\ most suitable for stellar kinematics analysis
due to the presence of strong absorption features
\citep{Bender1990,Bender1994}.

Stellar velocity
dispersions from the Portsmouth Spectroscopic pipeline agree
  within a few percent with other DR9 measurements of \sigmas\
by \citet{Bolton2012b} and \citet{Chen2012} \citep[see][{for a
  detailed comparison of the systematic offsets between
  methods}]{ThomasD2012}. \citet{ThomasD2012} show that the typical
error distribution on the \sigmas\ measurements for BOSS galaxies
peaks at 14 \%, and 93 \% of the measurements have an error below 30
\%. We therefore selected objects with an error in \sigmas\ below 30
\% for the present study to be as inclusive as possible while still
maintaining an acceptable accuracy in velocity dispersion (large
errors are due to the low signal-to-noise, {\it S/N}, of BOSS spectra,
mean $\sim 4.4$ from {\tt S\_N median}, which is sufficient to measure
velocity dispersions, \citealt{ThomasD2012}).  This cut is not as
tight as is generally applied but it allows us not to be affected by
biases due to sample selection (for example, a common tighter cut with
a relative error $<$ 10\% would discard most of the low \sigmas\
galaxies at high redshift).  \citet{ThomasD2012} also show that
  \sigmas\ determinations show no bias with $S/N$. Errors on
  \sigmae\ slightly vary with redshift, from 12 \kms\ at $z\sim 0.25$
  to 39 \kms\ at $z\sim 0.65$.

  Besides the cut in relative error below 30\%\ we further restrict
  our sample to values of $70 \leq \sigma \leq 550$ \kms. We discard
  velocity dispersions below 70 \kms\ because of the limit in
  instrumental resolution of the BOSS spectrograph, and velocity
  dispersions above 550 \kms\ to exclude contamination by potential
  multiple systems \citep{Bernardi2003a,Bernardi2006,Bernardi2008}.
The final number of galaxies that survive these additional cuts in
velocity dispersion is $\sim 370,000$, which is 75 \% of the original
sample. 

The stellar velocity dispersions from BOSS spectroscopy ($\sigma_{\rm
  ap}$) are measurements within the 2\farcsn\ diameter aperture of the
BOSS fiber. Therefore, we applied an aperture correction to translate
the BOSS velocity dispersions to the aperture corresponding to the
effective radius using the relation of \citet{Cappellari2006} derived
from the integral field data of the SAURON sample

\begin{equation}
\sigma_{\rm e}=\sigma_{\rm ap} \times(r_{\rm  ap}/R_{\rm e,"})^{0.066}
\end{equation}

\noindent
in which $\sigma_{\rm e}$ is the stellar velocity dispersion within
$R_{\rm e,"}$, and $r_{\rm ap}=1$\farcsn\ is the radius of the BOSS
fiber. $R_{\rm e,"}$ is taken from the rescaled effective radii
converted to arcsecond. The relation of \citet{Cappellari2006} is
consistent with that of \citet[][slope=0.06]{Mehlert2003} and slightly
steeper but in agreement within the errors with older determinations
by \citet[][slope=0.04]{Jorgensen1995}.

Aperture corrections depend on galaxy profile and systematic evolution
in the light profile of galaxies could affect the stellar velocity
dispersion, as well as this rescaling factor could change from local
SAURON galaxies to the higher redshift BOSS galaxies.  However, we
expect this effect to be negligible as the aperture corrections are
small (maximum 3\% at higher redshift) because the fiber diameter is
close to the typical effective radius of galaxies at the redshifts
studied here (see Figure~\ref{fig:re_distrib}).  Typical uncertainties
after the aperture correction range from 5 to 16\% of \sigmae\ (13 to
39 \kms).
\subsubsection{Systematic errors}
We performed Monte Carlo simulations to estimate systematic errors on
the aperture correction due to the size calibration (see
Section~\ref{subsec:radiicalib}), and have found them to be small. On
average \sigmae\ changes by $\sim 1.5$\kms, which is well below the
measurements errors.

\begin{figure*}
\begin{center}
\includegraphics[width=0.45\textwidth]{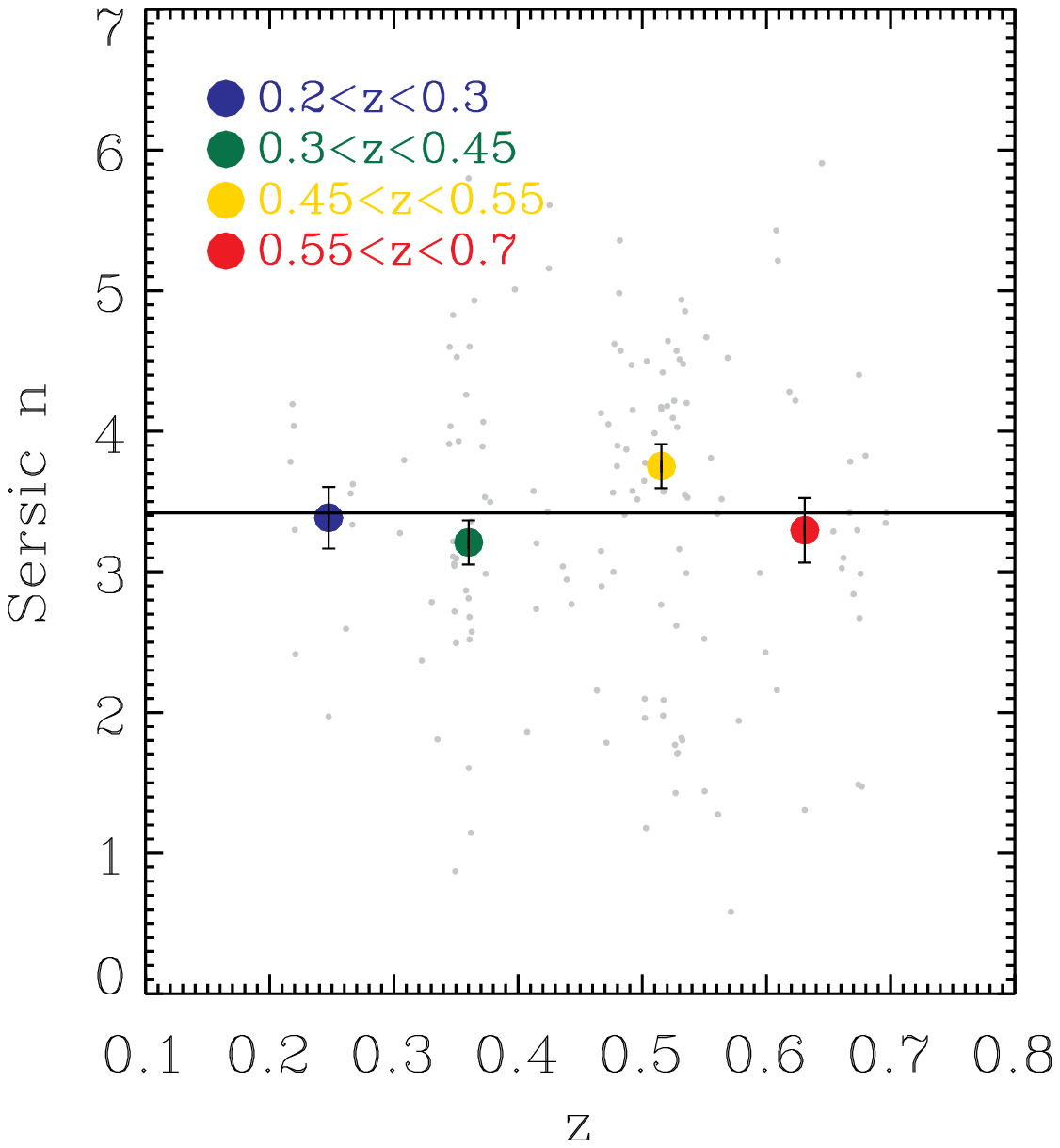}
\includegraphics[width=0.45\textwidth]{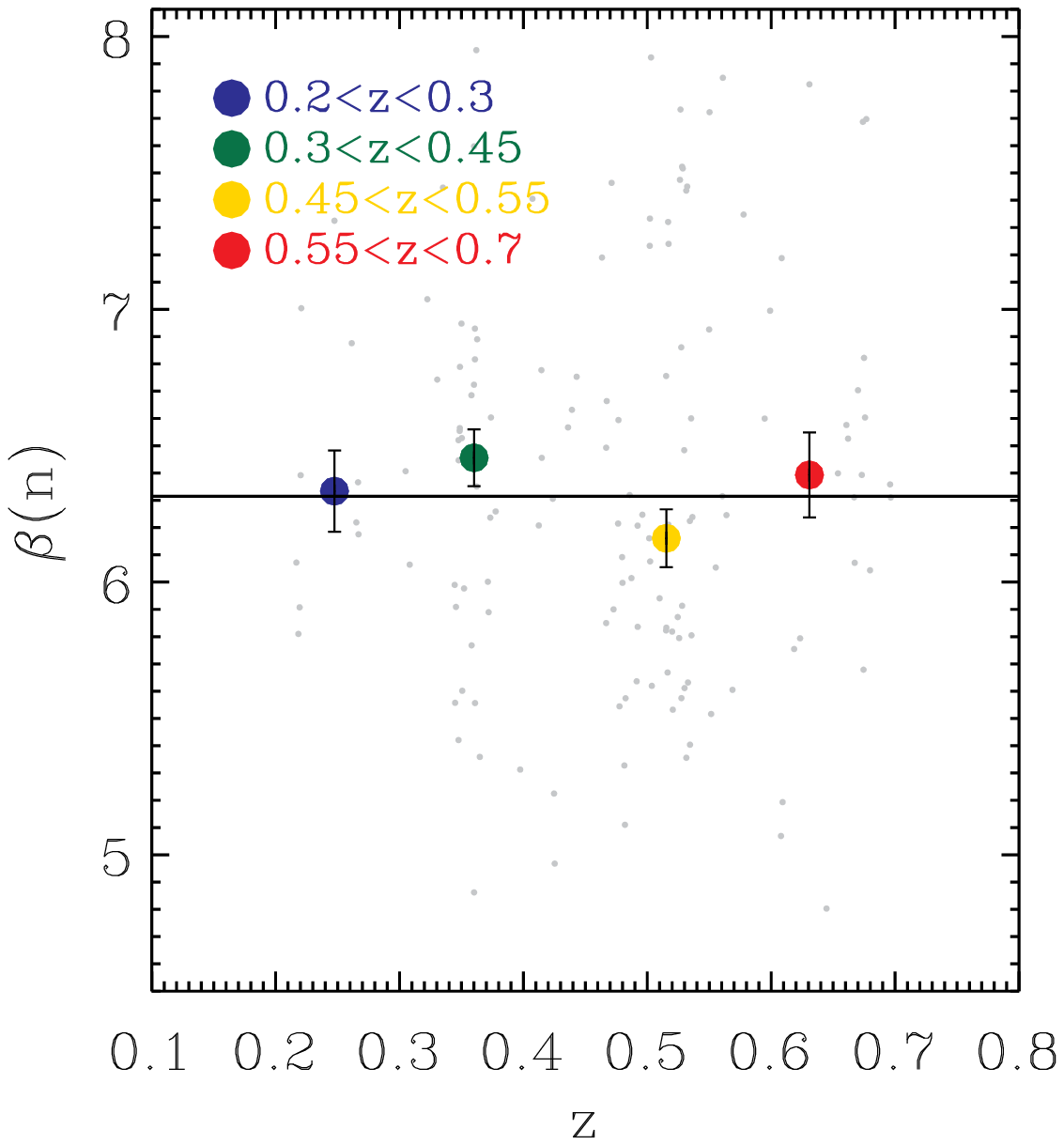}
\end{center}
\caption{S\'ersic indices ({\em left panel}) and $\beta$ parameters
  ({\em right panel}, see Equation~\ref{eq:vir}) of the COSMOS/HST
  sample for various redshift bins as color-coded in
  Figure~\ref{fig:sdss_re_cosmos_re_vs_sdss_re_correction}. Colored
  symbols with error bars are the medians. The continuous line is the
  median value.  The total number of objects is 150.}
\label{fig:alpha_z}
\end{figure*}

\subsection[]{Dynamical mass}
\label{subsec:galaxy_mass}

Following \citet{Beifiori2012}, we estimate dynamical galaxy mass from
the effective radius and velocity dispersion within the effective
radius using the virial mass estimator as

\begin{equation}
M_{\rm dyn}=\beta(n) \; R_{\rm e} \sigma_{\rm e}^2/G 
\label{eq:vir}
\end{equation}

\noindent
where $G$ is the gravitational constant and $\beta$ is a dimensionless
constant that depends on galaxy structure, often adopted as
$\beta=5.0\pm0.1$ for local galaxies, see \citet{Cappellari2006}.

Even though based on measurements within the effective radius, this
virial estimator is designed to capture the total dynamical mass of a
galaxy. A caveat is that this might only be true as long as total mass
traces light. \citet{Thomas2011} found that this assumption might not
be consistent with lensing studies. They suggest that
Equation~\ref{eq:vir} may only yield about 86 \% of the true total
dynamical mass. However, this will only affect the absolute scale of
the dynamical to stellar mass ratios that we derive, while their
evolution with redshift will remain the same. As we are mostly
interested in the latter, the main conclusions of this work will not
be affected.  As an additional check we therefore compared the
\mvir/\mstar\ derived here with those measured for local galaxies from
more sophisticated dynamical modeling in and find good agreement (see
Section~\ref{subsec:masses_boss}). Still, it should be emphasized that
any change in dynamical mass found here reflects a change of dynamical
mass within $1\re$.

\subsubsection{Dependence on structural parameters}
\label{subsecstructural_cosmos}

The appropriate value of $\beta$ is actually a function of the
S\'ersic shape index {\it n} \citep{Trujillo2004, Cappellari2006}.
\citet{Taylor2010} showed that dynamical masses and stellar masses
correlate well when the structure of the galaxy is taken into account
(see also Section~\ref{subsec:calib_masses_cosmos}).  They find that
dynamical masses estimated with the homology assumption exhibit
residual trends with galaxy structure properties, so they introduce a
structure-corrected dynamical mass adopting a constant $\beta$ that
is S\'ersic index dependent \citep{Bertin2002}.  Note that the virial
mass estimator of \citet{Cappellari2006} (Equation~\ref{eq:vir}) has
been calibrated on dynamical masses from Schwarzschild modeling where
no assumption about homology is made.

For our sample of BOSS galaxies we cannot make any statements in this
respect since SDSS images do not have the necessary angular resolution
to perform S\'ersic fits. However, we can expect this effect to be
negligible, as the BOSS galaxy sample is restricted to massive
galaxies in a relatively narrow mass range \citep{Maraston2012} and
limited redshift range so that variations of the S\'ersic index will
be minimal.  Moreover, the fact that our size calibration is based
  on S\'ersic \re\ from COSMOS, allows us to account for possible
  differences between de Vaucouleurs profiles and S\'ersic profiles
  resulting in a ``S\'ersic-like'' calibrated radii. 

  We verify this assumption with the COSMOS
  sub-sample for which S\'ersic indices are available.  The {\tt
    Zurich Structure \& Morphology Catalog v1.0} also contains values
  of galaxy S\'ersic index, $n$. This allows us, for this sub-sample,
  to account for the variation of the parameter $\beta$ with $n$ and
  encapsulate the effects of galaxy structure on \mvir\ (by assuming a
  constant $\beta$ for all galaxies Equation~\ref{eq:vir} implicitly
  assumes that all galaxies are homologous).
 
  We estimate $\beta(n)$ following the analytic expression
    between $\beta(n)$ and the S\'ersic index \citep[Equation 20
    of][]{Cappellari2006}, which is theoretically derived for spherical
    and isotropic models with a S\'ersic profile for different values of
    $n$ (\citealt{Bertin2002}, see also \citealt{Taylor2010} for a
    discussion of its importance on the SDSS sample).

    Figure~\ref{fig:alpha_z}, right panel, shows the dependence of the
    $\beta(n)$ parameter on redshift for each galaxy in the sub-sample
    (gray points). Colored circles are the median $\beta(n)$ for each
    redshift bin. We find that the median $\beta(n)$ is $\sim 6.3$ for
    all redshifts bins (see continuous line in
    Figure~\ref{fig:alpha_z}, right panel).
  This is larger than the local values of 5 generally adopted,
    and yields systematically higher masses by $\sim$20\%. The
    reason is the relatively low S\'ersic indices (between 3.38 and
    3.30 at $z\sim0.25$ or $z\sim0.6$, as shown in 
    Figure~\ref{fig:alpha_z}, left panel) for the COSMOS sample,
    compared to typical S\'ersic indices for local galaxies.

    The key point illustrated by Figure~\ref{fig:alpha_z}, however, is
    that both $n$ and $\beta(n)$ do not evolve with redshift. As we
    focus in the redshift evolution and not absolute values for
    dynamical mass, the present study is not affected by a systematic
    offset in $\beta$. We will use \mvir\ derived using a median
    $\beta=6.3$, which is the median value $\beta$ derived using the
    BOSS/COSMOS photometry.

\subsubsection{Dependence on aperture} 

The dynamical mass obtained using the virial mass estimator (see
Section~\ref{subsec:galaxy_mass}) is based on stellar kinematics
within an aperture of 1 effective radius and scaled to {\em total}
dynamical mass via equation~\ref{eq:vir}. This quantity is compared
with the {\em total} stellar mass from \citet{Maraston2012} based on
{\tt cmodelMag} magnitudes. Hence both dynamical and stellar masses
are {\em total} masses, which ensures a consistent comparison.

Still, the total dynamical mass is derived from observations within
the effective radius, while the stellar mass comes from the total
stellar light. We explore therefore the possible presence of a
systematic effect from the different apertures in which kinematics and
stellar populations have been measured. To this end we compare \mstar\
derived from {\tt modelmag} (rescaled to $i$-band {\tt cmodelmag}) and
\mstar\ from aperture magnitudes within \re\ (rescaled to $i$-band
{\tt cmodelmag}). This test is presented in
Appendix~\ref{sec:ap_stellar_masses}.

In brief, the difference between the two sets of masses is $\sim 0.08$
dex. The stellar masses measured from SED fitting within 1 \re\ are
higher by this amount, because of the higher $M/L$ ratio within
$1\re$. We emphasize, however, that this quantity is an {\em
  overestimate of the true total mass}. Nevertheless, it is reassuring
to verify that this systematic difference is relatively small. Most
importantly, the offset is independent of redshift (see
Appendix~\ref{sec:ap_stellar_masses}). Hence the science analysis of
this work is not affected, because we study redshift dependence and do
not focus on absolute ratios between dynamical and stellar masses. We
also note that the dynamical to stellar mass ratio is always larger
than 0.08 dex in our redshift range, hence \mstar\ does never exceed
\mvir\ ensuring physically meaningful solutions throughout.

\subsubsection{Dependence on rotation}
\label{subsecgal_unres_rot}

The possible presence of unresolved rotation is another complication
that could affect our mass estimates from
Equation~\ref{eq:vir}. \citet{vanderMarel2007} have measured increased
rotational support at $z\sim0.5$ and argue that data at different
redshifts can be affected by rotation, with a stronger impact on
low-\sigmas\ galaxies which are more rotationally supported than
galaxies at high \sigmas. As our BOSS sample consists of massive
galaxies in a relatively small mass range \citep{Maraston2012},
however, we expect this effect to be negligible.

\begin{figure*}
\begin{center}
\vbox{
\hbox{
\includegraphics[width=0.33\textwidth]{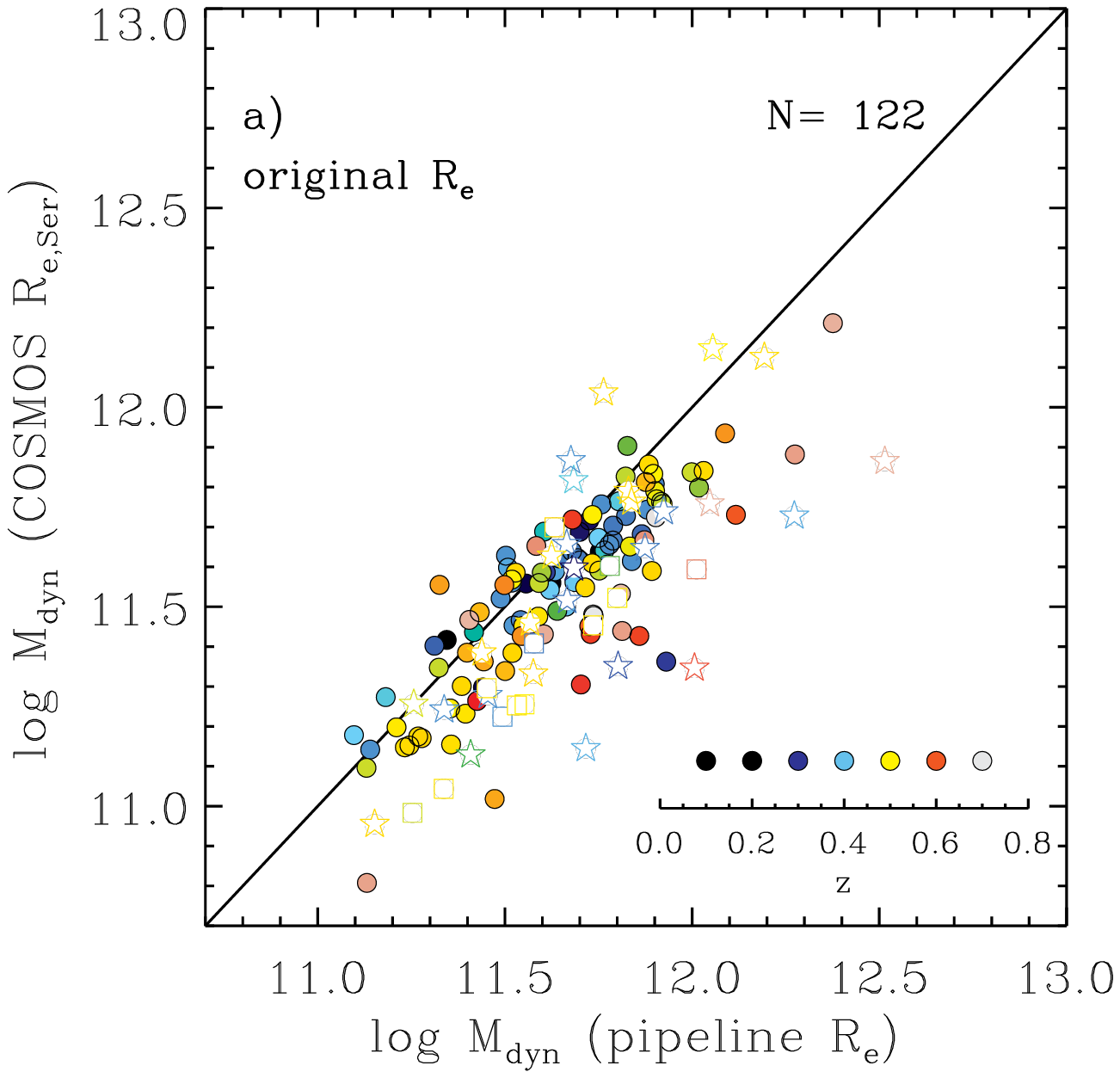} 
\includegraphics[width=0.33\textwidth]{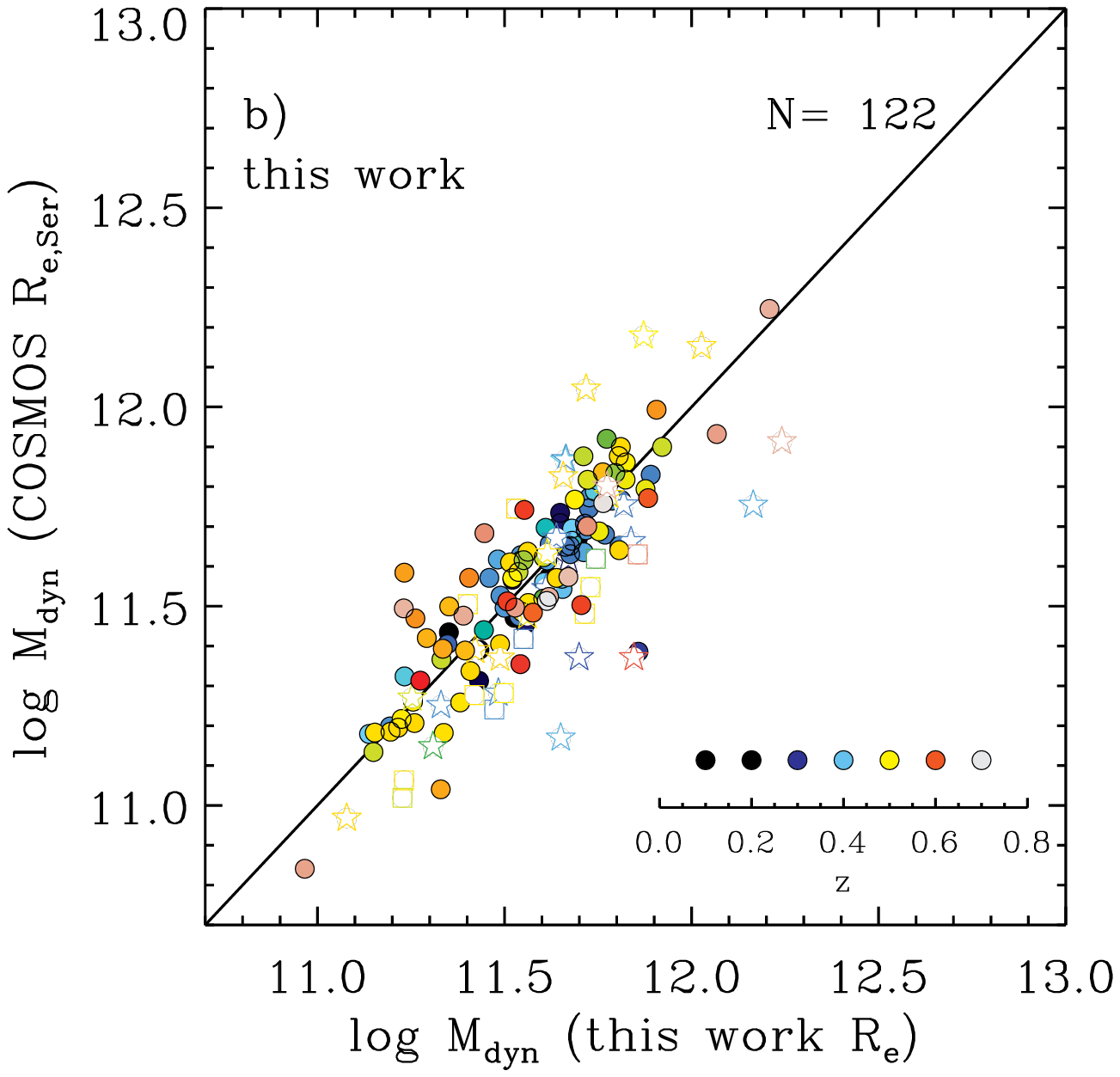}
\includegraphics[width=0.33\textwidth]{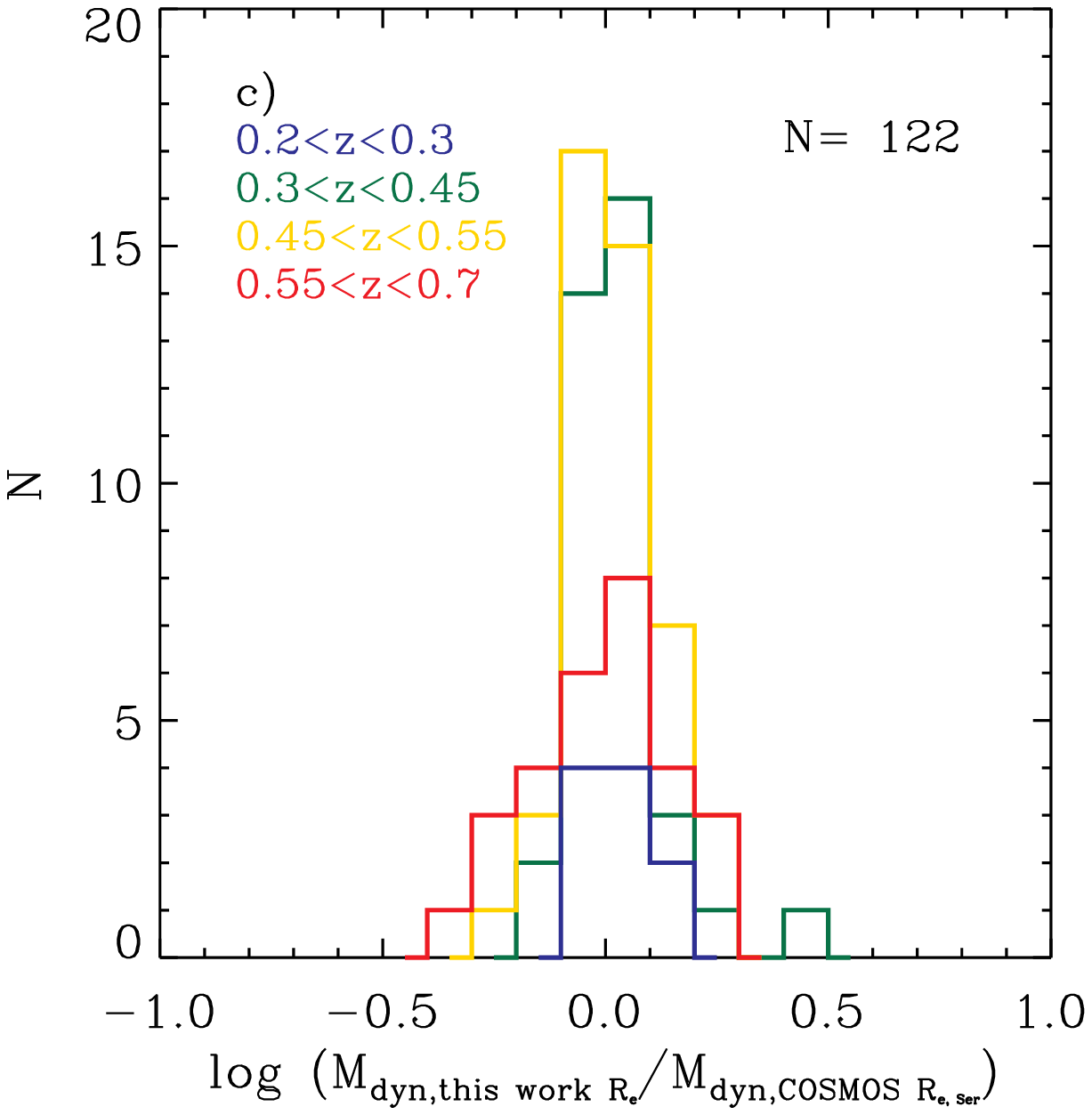} 
}}
\end{center}
\caption{{\em Left panel}: Dynamical masses derived using the original
  SDSS \re\ assuming a constant $\beta=6.3$ versus dynamical masses
  derived using COSMOS $R_{\rm e, Ser}$ adopting a variable $\beta$
  based on the S\'ersic index. Symbols are as in
  Figure~\ref{fig:sdss_re_cosmos_re_vs_sdss}. {\em Central panel:}
  Same as left panel but using rescaled SDSS \re\ adopted in the
  present study. {\em Right panel:} Distribution of the ratio of
  dynamical masses from rescaled SDSS \re\ and COSMOS $R_{\rm e, Ser}$
  for the four redshift bins of
  Figure~\ref{fig:sdss_re_cosmos_re_vs_sdss}.  Legends in each
    panel give the number of galaxies used in the derivation of the
    size correction including objects discarded in the iterative
    $2\sigma$ clipping (multiple systems are not included in this
    number).}
\label{fig:sdss_mvir_cosmos_mvir_corrected}
\end{figure*}

\subsubsection{Dependence on galaxy type}
\label{subsecgal_type}

Finally, in deriving \mvir\ with Equation~\ref{eq:vir}, we implicitly
assume that the measured value of \sigmae\ is dominated by the bulge
component. For late-type galaxies we expect that the disc contribution
to \sigmae\ results in a broader distribution of \mvir, since the
\sigmae\ may not represent the actual dynamical state of those
galaxies which is dominated by rotation (see
Section~\ref{subsecgal_unres_rot}) as well as the $\beta$ parameter
we used could not be appropriate for late-type galaxies with low
S\'ersic index.  As shown in \citet{Masters2011}, however, the
majority of BOSS galaxies ($74\pm6$\%) have early-type morphology and
the remaining later types are bulge dominated, hence this effect will
be negligible. We tested this assumption by only considering
early-type galaxies for the CMASS sample using the morphological cut
$(g-i)>2.35$ by \citet{Masters2011}.  We compared dynamical masses
derived with COSMOS \re\ and adopting $\beta$ based on the S\'ersic
index and dynamical masses with \re\ derived in this work and found a
good agreement between early-types and the full COSMOS/BOSS
sub-samples, with a scatter around the one-to-one relation consistent
within the errors ($\sim$0.14 dex).

\subsubsection[]{Calibrated virial masses for the COSMOS sub-sample}
\label{subsec:calib_masses_cosmos}
 
As an additional test we compare our virial mass estimates based on
the re-scaled effective radii with virial masses derived directly from
the COSMOS effective radii, the result is shown in
Figure~\ref{fig:sdss_mvir_cosmos_mvir_corrected}. The left-hand panel
shows the comparison between virial masses derived using COSMOS \re\
and the {\em uncorrected} SDSS \re. As expected, there is a clear
offset to higher virial masses from SDSS imaging because of the
overestimation of galaxy radii.

The re-scaled radius of this work remedies this problem. The central
panel of Figure~\ref{fig:sdss_mvir_cosmos_mvir_corrected} shows the
comparison between virial masses derived using COSMOS \re, and
adopting a variable $\beta$ based on the S\'ersic index (see
Section~\ref{subsecstructural_cosmos}) and the {\em corrected} SDSS
\re\ of the present work (by using a constant $\beta=6.3$ as described
in Section~\ref{subsecstructural_cosmos}). Mass estimates agree well
at all redshifts with a scatter of $\sim 0.14$ dex, which is well
within the errors. The right-hand panel presents the distribution of
the logarithmic ratio between COSMOS and SDSS masses after correction.
The distribution is symmetric around zero for all redshift bins with a
maximum deviation of $\sim$0.5 dex.

\subsubsection{Random and systematic errors}

The final errors in \mvir\ are a combination of uncertainties in
\sigmae\ (which account for the aperture scaling of \sigmas), \re, the
statistical uncertainties due to the rescaling factor of \re, and
$\beta$.  This results in median random errors of $\sim 0.15$ dex
depending on redshift (from 0.08 dex at $z=0.25$ to 0.18 dex at
$z=0.65$).
Based on Monte Carlo simulations we estimate median systematic errors
due to the size calibration (see Section~\ref{subsec:radiicalib}) and
uncertainty of $\beta(n)$ to be $\sim$ 0.04 dex.
  
\subsection{Local SDSS-II early-type galaxy sample}
\label{sec:localETGs}
We combine the SDSS-III/BOSS sample described above with a local
sample of massive galaxies at $z\sim 0.1$ drawn from SDSS-II. The
galaxy properties of this local sample are presented in the following
sections.

\subsubsection{Stellar mass}

Stellar masses and ages were estimated from the SED fitting of the $u,
g, r, i, z$-band photometry following the same prescription of
\citet{Maraston2012} with passive templates (the LRG model by
\citealt{Maraston2009} mentioned earlier).  We homogenize the stellar
mass distribution by selecting a sub-sample that matches the mass
distribution of the BOSS sample. We constructed this sub-sample using
the stellar mass distribution in the lowest BOSS redshift bin
($0.2<z<0.3$) as reference. For each stellar mass bin we randomly
selected from the local galaxy distribution a number of galaxies equal
to the number of galaxies in the low-$z$ BOSS one.  This cut on the
local early-types population retains $12,089$ galaxies. A discussion
on the impact of the science analysis in this paper from this
homogenization is given in Appendix~\ref{sec:mass_distrib_local}.

\subsubsection{Size}
We collect photometry and effective radii \re\ from DR8 in which the
correction for the sky over-subtraction of previous releases is
already implemented (see discussion in Section~\ref{subsec:size_sdss})
and no further correction to sizes (see \citealt{Hyde2009} for
details) has been applied. This is motivated by the fact that we
selected galaxies at redshift $z<0.2$ that are resolved in the SDSS
imaging with \re$> $ FWHM of the PSF (retaining 96\% of the objects).

\subsubsection{Stellar velocity dispersion}
We collect redshift and stellar velocity dispersions from the DR7
catalogs. \citet{ThomasD2012} show that their DR7 \sigmas\ are
consistent with SDSS pipeline \sigmas\ at the few percent level (see
their Figure~1).  The median offset across all the stellar velocity
dispersion is $\sim$ 1\%. However, this offset increases
  towards high stellar velocity dispersions.  We can quantify the
correct offset to apply to DR7 \sigmas\ looking at Figure 4 of
\citet{ThomasD2012} where their \sigmas\ are compared to
\citet{Bolton2012} \sigmas\ within BOSS, which is the relevant mass
range.\footnote{\citet{Bolton2012} is the same code that produced the
  SDSS \sigmas.}. The offset is 4\%, which we correct for in the
SDSS-II sample.

We further rescaled stellar velocity dispersions to the value at \re,
following the procedure described in Section~\ref{subsec:sigma},
accounting for the fact that DR7 galaxies were observed with a
3\farcsn\ aperture.  The variation in \sigmae\ for the aperture
correction in local SDSS galaxies is ~2\% (\sigmae\ within Re on average
2\% smaller than the SDSS ones, and median ratio between aperture size
and \re\ is $\sim$0.72).

\subsubsection{Dynamical mass}
\label{subsecstructural_local}

Dynamical mass is derived from stellar velocity dispersion and size in
the same way as for the BOSS sample as described in
Section~\ref{subsec:galaxy_mass}. To ensure internal consistency we
use the same redshift independent parameter $\beta=6.3$ as for the
BOSS sample derived from the BOSS/COSMOS photometry (see
Section~\ref{subsecstructural_cosmos}).

\subsection{Correction for progenitor bias}
\label{sec:progenitor}

BOSS target selection was designed to obtain a nearly uniform stellar
mass distribution across the redshift range $0.2\leq z\leq 0.7$.
Still, the sample needs to be corrected for effects from progenitor
biases \citep[e.g.,][and references therein]{Valentinuzzi2010,
  Saglia2010,Cimatti2012}, as higher-$z$ galaxies in the sample are
not necessarily progenitors of the lower-$z$ galaxies in the sample
\citep[see also][]{Tojeiro2012}.

\begin{figure}
\begin{center}
\includegraphics[width=\columnwidth]{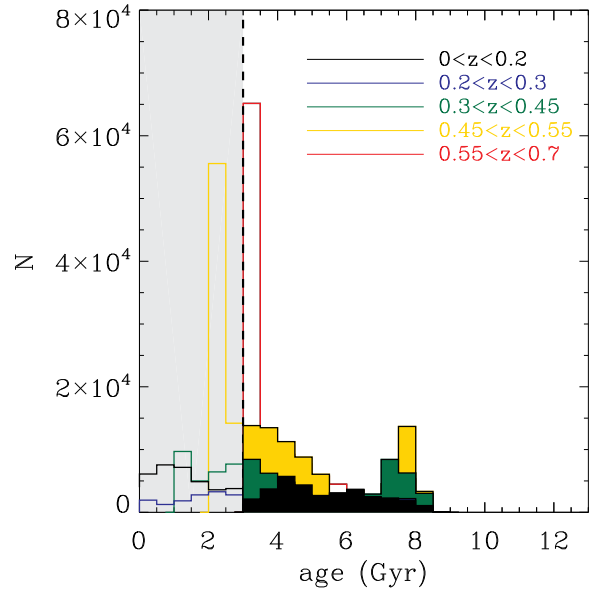}
\end{center}
\caption{Histograms of the ages for different redshift bins, 
    evolved to the highest redshift bin by subtracting the look-back
    time, which are used to apply the progenitor-bias cut (see text
  for details). Galaxies in the shaded region, i.e., with an age below
  $3\;$Gyr, have been discarded.}
\label{fig:progenitor_bias}
\end{figure}

To correct for the progenitor bias we compare -- in each redshift bin
-- the galaxy ages from \citet{Maraston2012} (one of the products of
the SED fit, see Section~\ref{subsec:photo_masses}) and remove those
galaxies from the low-$z$ sample whose ages (evolved to the highest
redshift bin by subtracting the look-back time) would be lower than a
given age threshold which is the time needed for a typical galaxy to
become passive.  For each redshift bin we select galaxies such
  that their age follows

\begin{equation}
t_{\rm g}(z)-(t_{\rm u}(z)-t_{\rm u}(z=0.65)) > 3 Gyr
\end{equation}

\noindent
where $t_{\rm g}$ is the age of a galaxy at a give redshift, $t_{\rm
  u}$ is the age of the universe at the same redshift and $t_{\rm
  u}(z=0.65)$ is the age of the universe at the median redshift of the
highest redshift bin.  Histograms of the evolved ages for
different redshift bins are shown in
Figure~\ref{fig:progenitor_bias}. Galaxies in the shaded region have
been discarded.  As the age threshold we chose 3 Gyr, adopting the age
limit used in \citet{Maraston2012}\footnote{\citet{Maraston2012} set a
  minimum age of 3 Gyr for the mass calculation using the passive
  template in order to minimize the chance to underestimate the mass
  by underestimate the galaxy age. This age limit translates into the
  assumption of a high-formation epoch for the massive and passive
  galaxies in CMASS.} for calculating stellar masses (see their
Section 3.1 for discussion). This threshold is only slightly larger
than the 1.5 Gyr suggested by \citet{vanDokkum2001a}.

We considered the highest redshift bin as a reference and we evolved
all other redshift bins including the local SDSS early-type sample. By
discarding galaxies with age$\; <3\;$Gyr, we retain 268,938 galaxies,
which corresponds to the $\sim 65\%$ of the initial local and BOSS
samples as shown in Figure~\ref{fig:progenitor_bias}.

We obtain similar results using the tighter selection criteria
described in \citet{Cimatti2012}, which select in each redshift bin
the galaxies with ages within $\pm 1\sigma$ of the age distribution
for each redshift bin accounting for the cosmic time elapsed from
one bin to the other. This selection also discards objects at older
ages and provides a sample size that is $\sim$54\% of the initial one.

\citet{Poggianti2012} found that galaxy sizes are correlated to
luminosity-weighted ages such that older galaxies will show a stronger
size evolution, with a stronger effect in clusters than in the
field. Our progenitor bias correction minimizes that effect.

\begin{figure*}
\begin{center}
\includegraphics[width=0.90\textwidth]{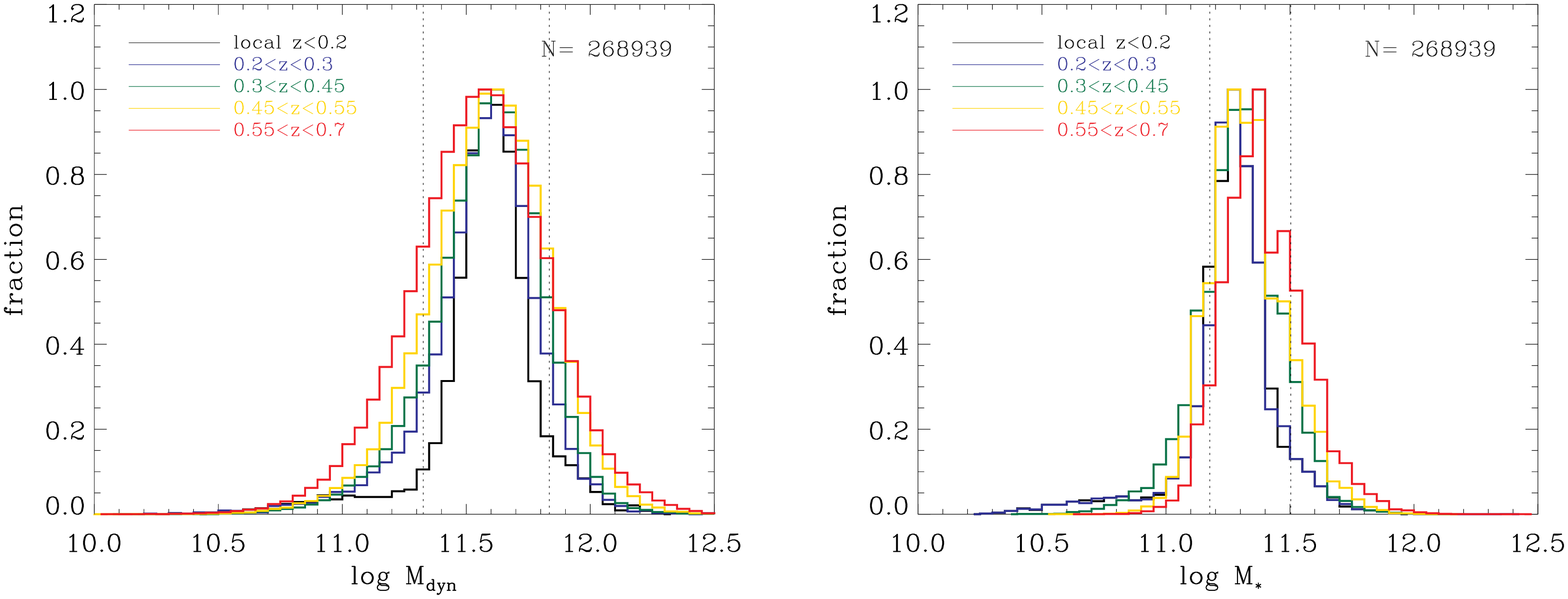}
\includegraphics[width=0.45\textwidth]{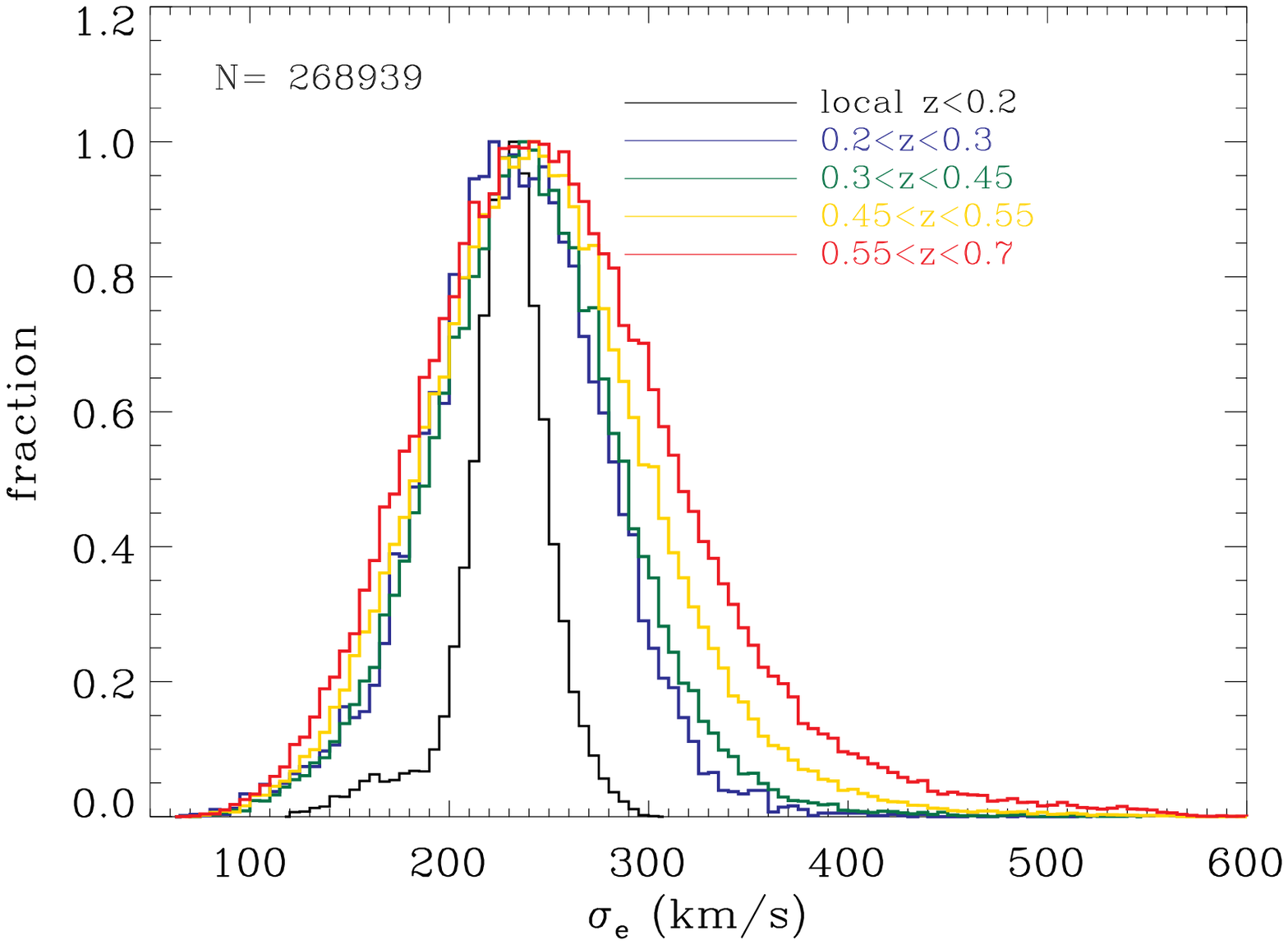}
\includegraphics[width=0.45\textwidth]{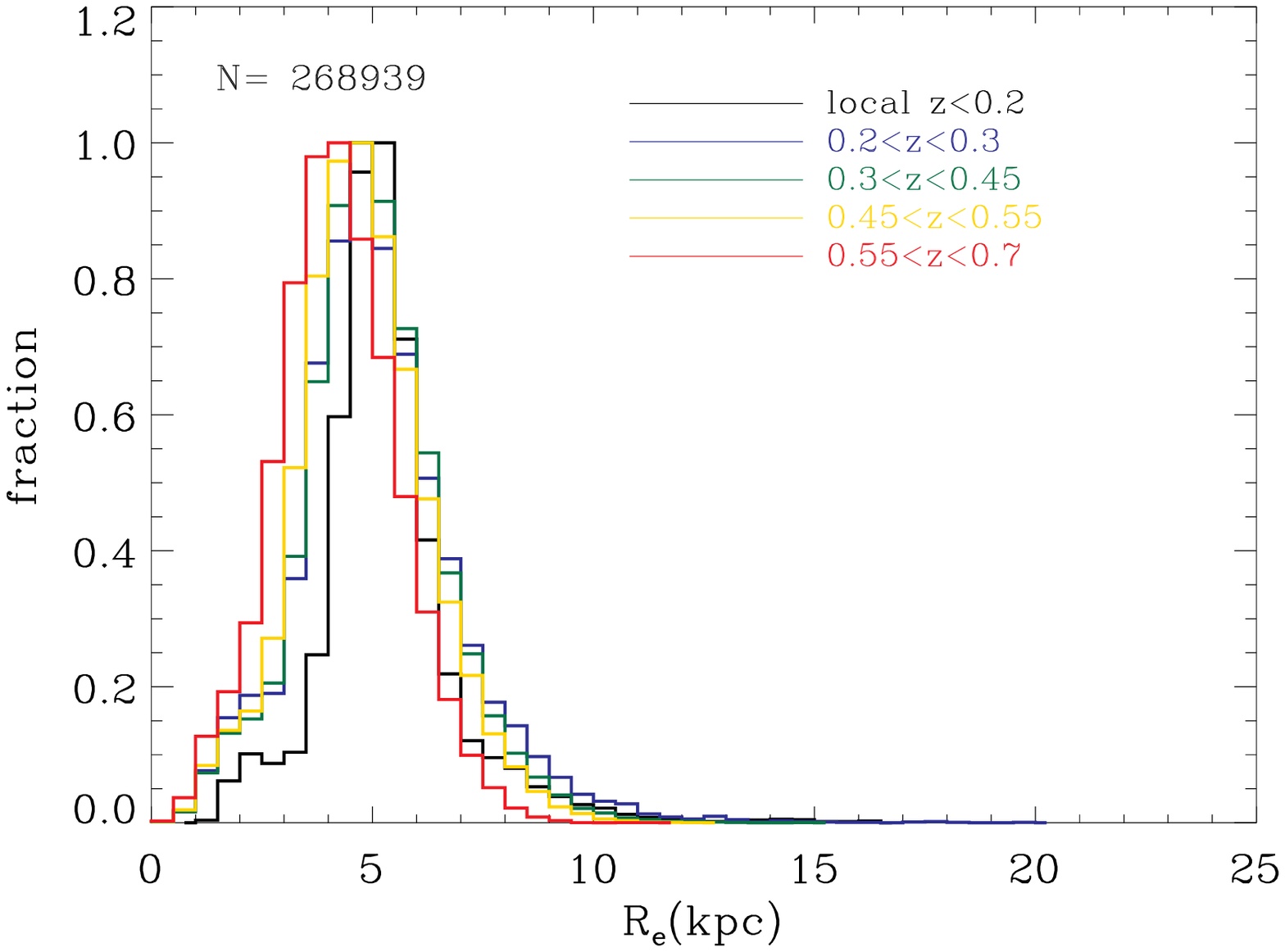}
\end{center}
\caption{{\em Top panels:} Distribution of \mvir\ (with $\beta=6.3$,
  {\em left panel)} and \mstar\ {\em (right panel)} for various
  redshift bins, normalized to the peak value in each bin. The BOSS
  mass distributions are fairly uniform over the redshift range under
  analysis (see also Figure~10 in \citealt{Maraston2012}).  Local
  early-types from SDSS-II are selected to have the same stellar mass
  distribution of the lowest BOSS redshift bin. Dotted-black lines
  indicate the $\pm 1\sigma$ of the mass distributions adopted for the
  present analysis. {\em Bottom left and right panels}: distributions
  in stellar velocity dispersion distribution and effective radius. The
  progenitor-bias correction has been applied in all cases.}
\label{fig:progenitor_bias_mass_distribution}
\end{figure*}

The distributions of \mstar, \mvir, \sigmae, and \re\ of the final
sample after correction for progenitor bias are shown in
Figure~\ref{fig:progenitor_bias_mass_distribution} for various
redshift bins. The typical median stellar mass is around
$\log$\mstar$\sim 11.28$ dex, the median \re$\sim 5.2$ kpc, and
$\sigma_{\rm e}\sim 231$ \kms.

To study the effect of the progenitor bias correction on the redshift
evolution of these quantities, we have performed a re-analysis for a
sample without progenitor bias correction presented in
Appendix~\ref{sec:noPB_show}. It can be seen that generally results
are consistent. Most importantly, the evolution of \mvir/\mstar\ is
fairly stable against the progenitor-bias correction, hence the main
conclusions of these paper do not critically depend on the progenitor
bias correction.

\begin{table*}
\begin{scriptsize}
\begin{center}
\begin{minipage}{0.5\textwidth}
\caption{Fitting parameters for the redshift evolution of galaxy parameters
  between $0.1\leq z\leq0.55$.}
\begin{tabular}{c r r r r}
\hline
\hline
\noalign{\smallskip}
\multicolumn{1}{c}{} &
\multicolumn{2}{c}{\mstar} &
\multicolumn{2}{c}{\mvir}  \\
\hline
\noalign{\smallskip}
\multicolumn{1}{c}{Parameter} &
\multicolumn{1}{c}{slope} &
\multicolumn{1}{c}{zero point} &
\multicolumn{1}{c}{slope} &
\multicolumn{1}{c}{zero point} \\
\hline
\re                              &$-0.49\pm 0.26$ & $0.76\pm 0.04$  &$-0.37\pm 0.20$ &$0.73\pm 0.03$  \\
\sigmae                      &$ 0.12\pm 0.02$ & $2.36\pm 0.01$ &$ 0.18\pm 0.06$ & $2.35\pm 0.01$ \\
$M_{\rm dyn}/M_{\star} $ &$-0.48\pm 0.23$ & $0.35\pm 0.03$   &$-0.55\pm 0.17$ & $0.36\pm 0.02$\\ 

\hline
\noalign{\smallskip}
\label{tab:fits}
\end{tabular}
\end{minipage}
\begin{minipage}{\textwidth}
  {\sc Notes.} --- Uncertainties on each parameter are $1\sigma$
  errors derived from Monte Carlo simulations.  The relation we fitted
  for \re\ is $\log R_{\rm e} =\log R_{\rm e,0} +\beta (1+z) $, for
  \sigmae\ is $\log \sigma_{\rm e} =\log \sigma_{\rm e,0} +\gamma
  (1+z)$, and for $M_{\rm dyn}/M_{\star}$ is $\log (M_{\rm
    dyn}/M_{\star}) = \log (M_{\rm dyn}/M_{\star})_0+\delta(1+z)$.
\end{minipage}
\end{center}
\end{scriptsize}
\end{table*}

\section[]{Results}
\label{sec:redshift_ev_all}

\begin{figure*}
\vbox{
\hbox{
\includegraphics[width=0.33\textwidth]{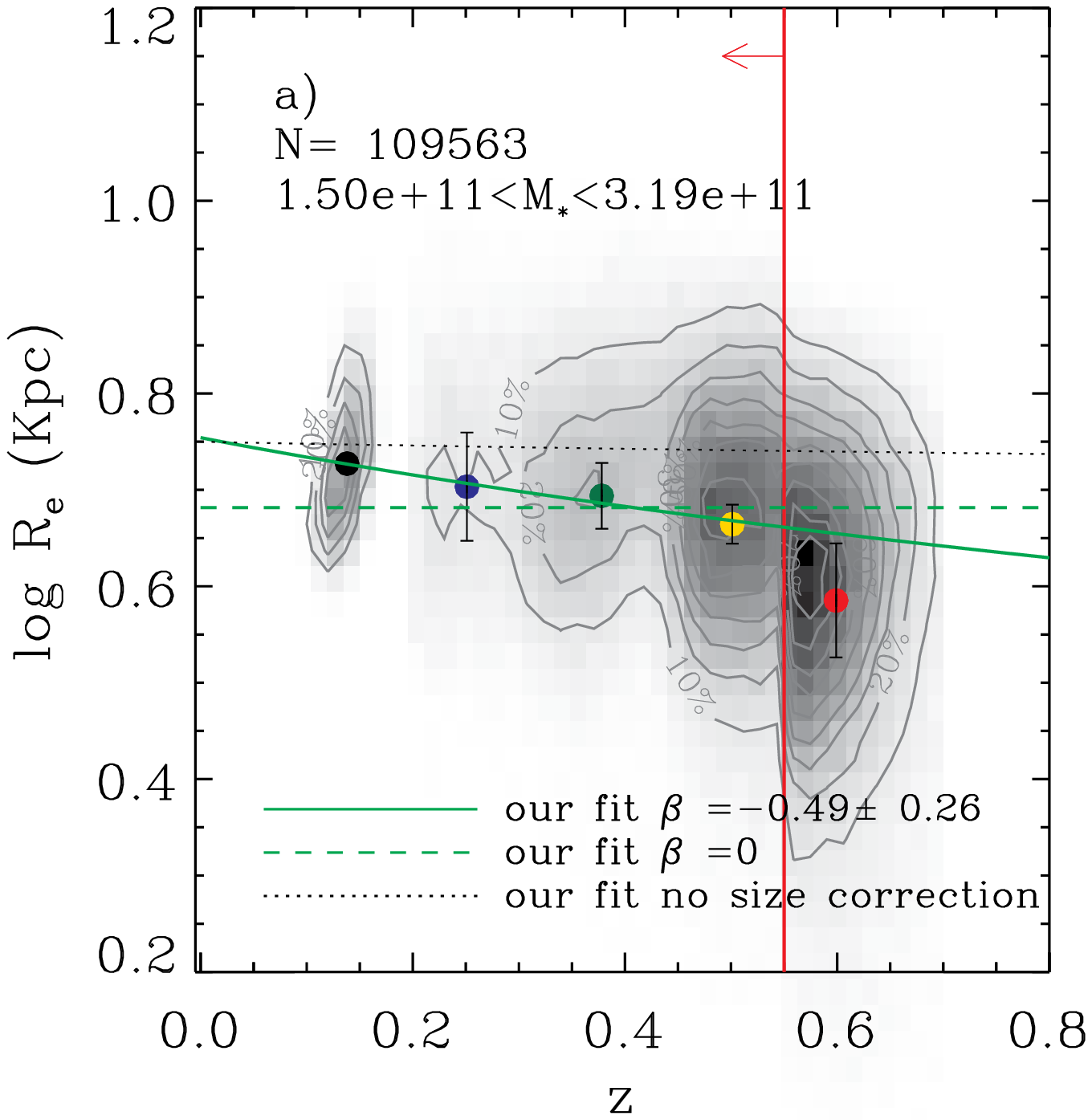}
\includegraphics[width=0.33\textwidth]{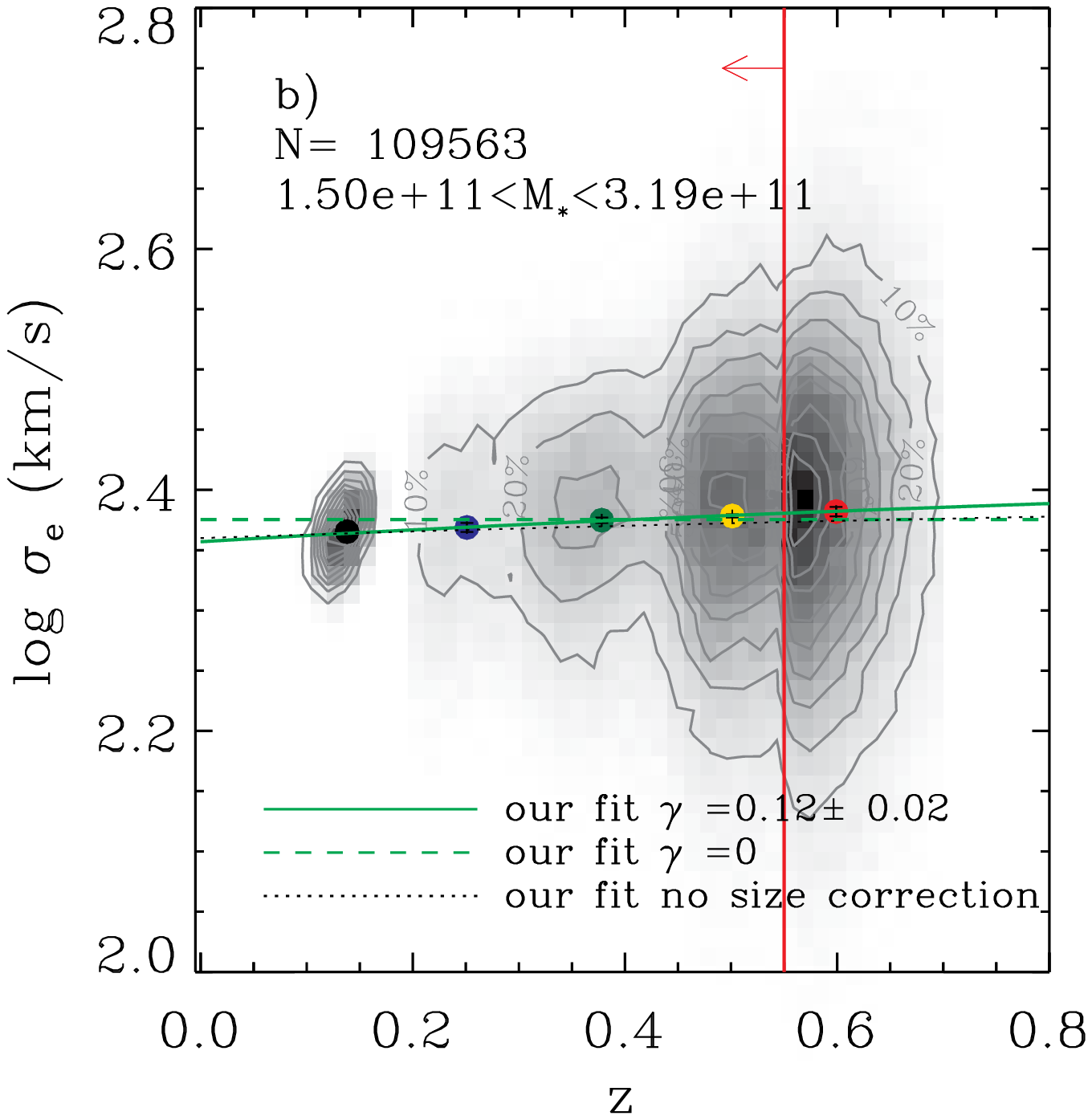} 
\includegraphics[width=0.33\textwidth]{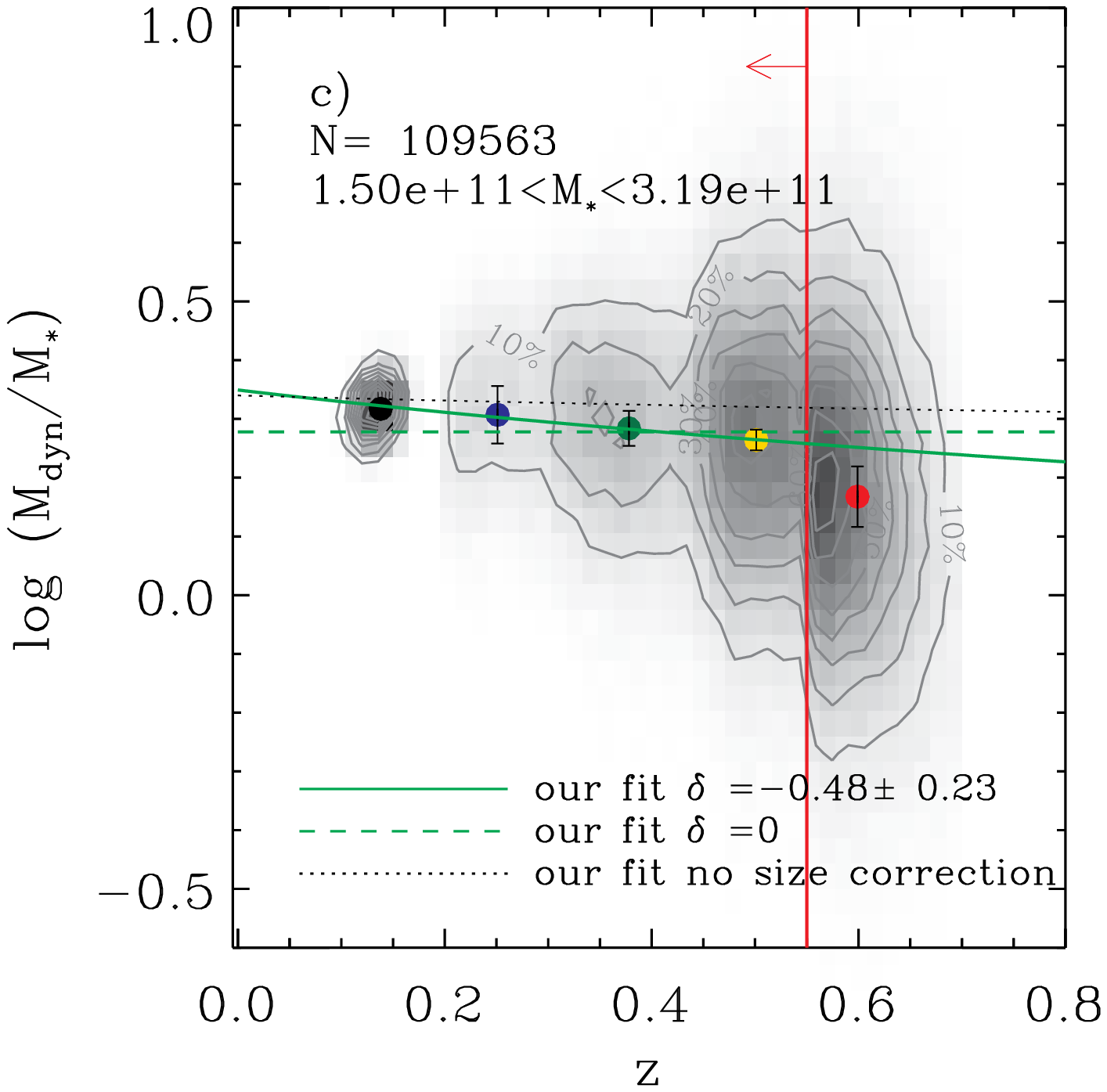}}}
\vbox{
\hbox{
\includegraphics[trim= 0 0 0 30, clip,width=0.33\textwidth]{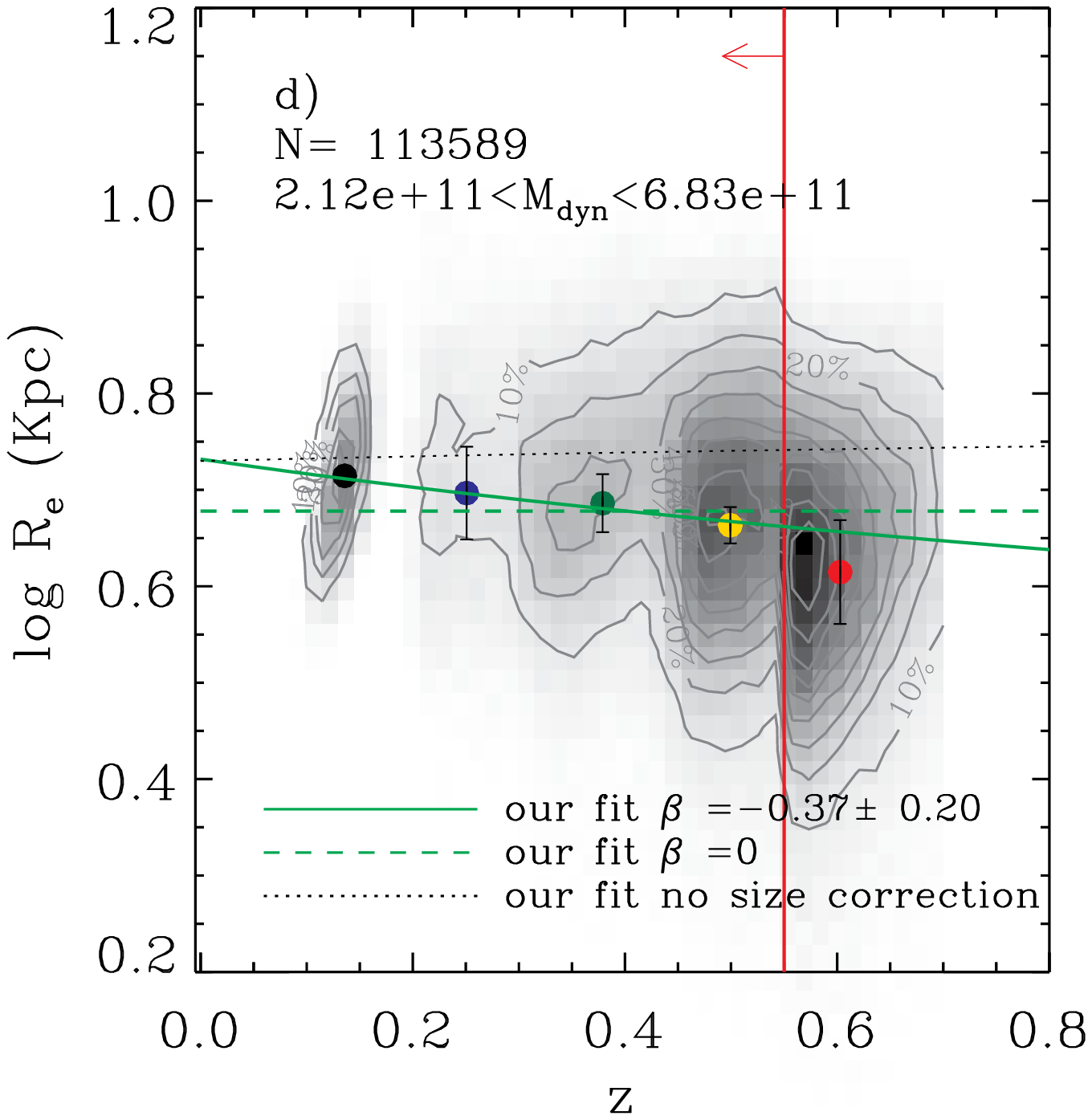}
\includegraphics[width=0.33\textwidth]{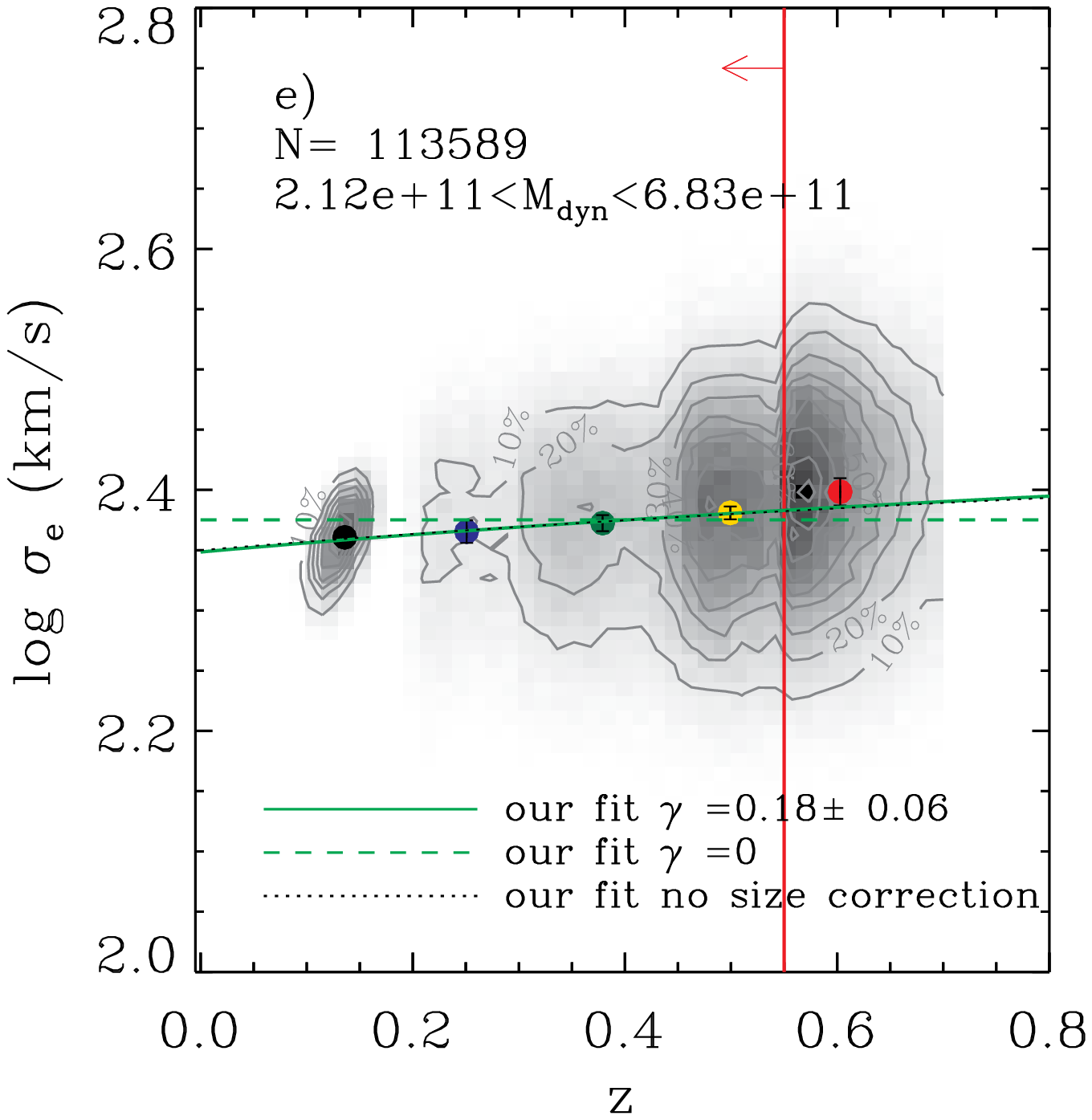}
\includegraphics[trim= 0 0 0 30, clip ,width=0.33\textwidth]{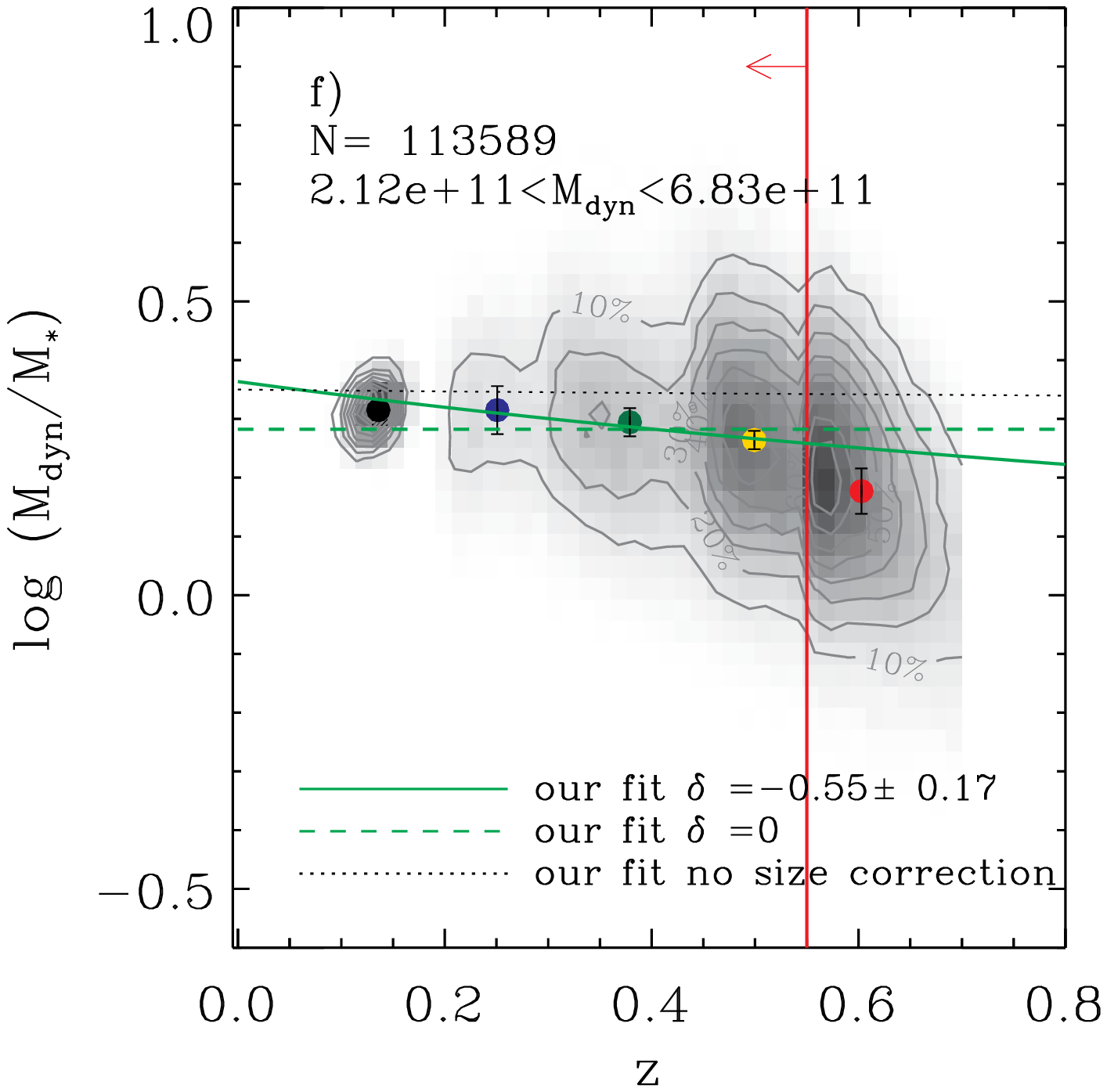}}}
\caption{{\em Left panels:} Effective radius \re\ as a function of
  redshift. {\em Central panels:} Stellar velocity dispersion \sigmae\
  as a function of redshift. {\em Right panels:} Ratio between
  dynamical and stellar mass \mvir/\mstar\ as a function of redshift.
  {\em Top and bottom panels} are for galaxies selected using \mstar\
  and \mvir, respectively (within $\pm 1 \sigma$ of the mass
  distributions, total number of galaxies given by the labels).  The
  shaded contour region indicates the full sample after correction for
  progenitor bias. Contours show 10 equally-spaced density levels
  showing the percentage of galaxies compared to the peak value of
  each plot. The colored filled circles are the median values in four
  redshift bins.  The green solid line is a linear fit, the green
  dashed line is a fit with zero slope for comparison. The black
  dotted line is a linear fit to the sample if no size correction is
  applied.  Error bars of the median points indicate the $1\sigma$
  uncertainty derived from Monte Carlo simulations. The red-continuous
  lines and arrows indicate the range where we fit our data.}
\label{fig:mvir_mstar_sigma_re}
\end{figure*}

\begin{figure*}
\vbox{
\hbox{
\includegraphics[width=0.33\textwidth]{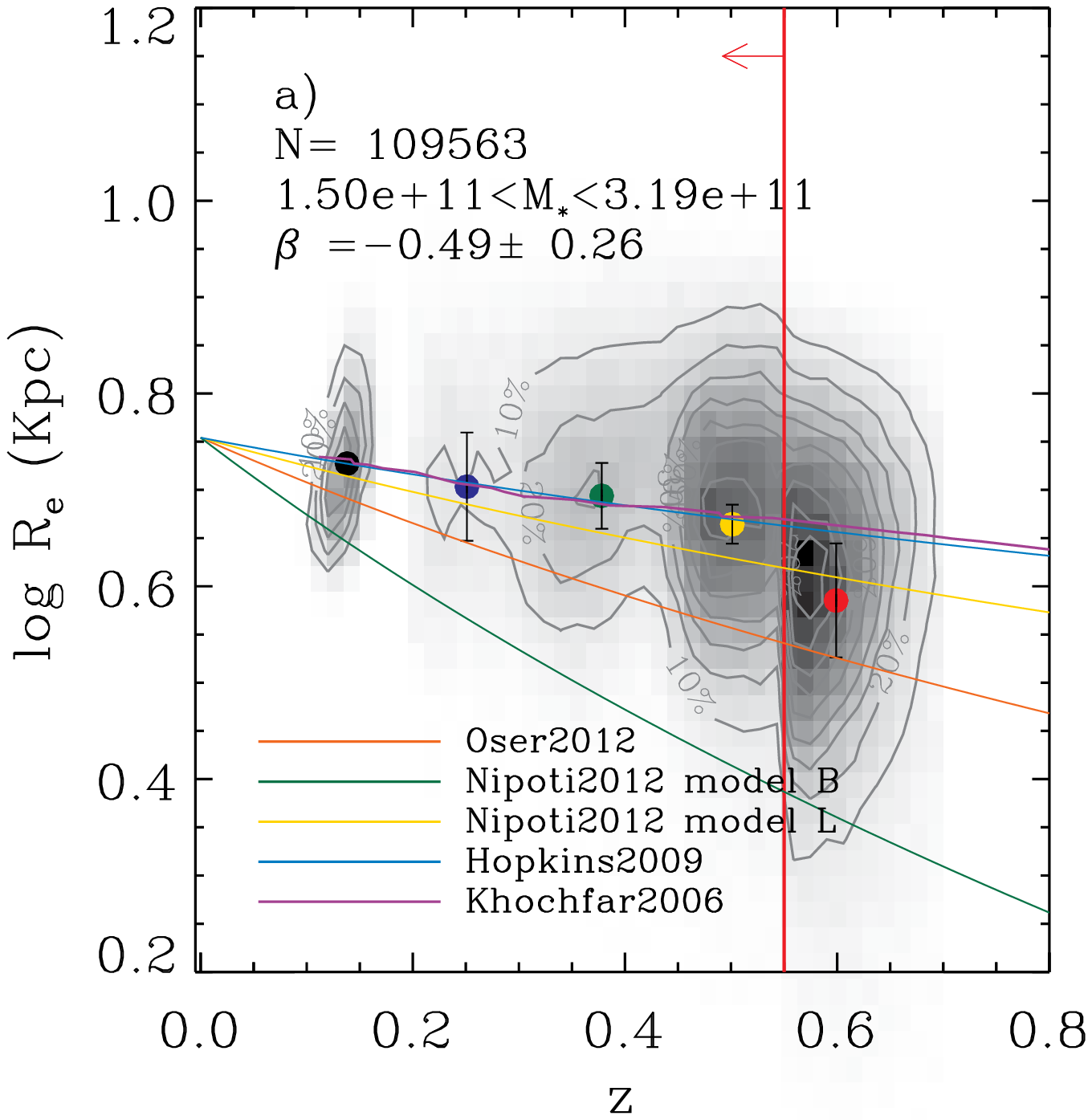}
\includegraphics[width=0.33\textwidth]{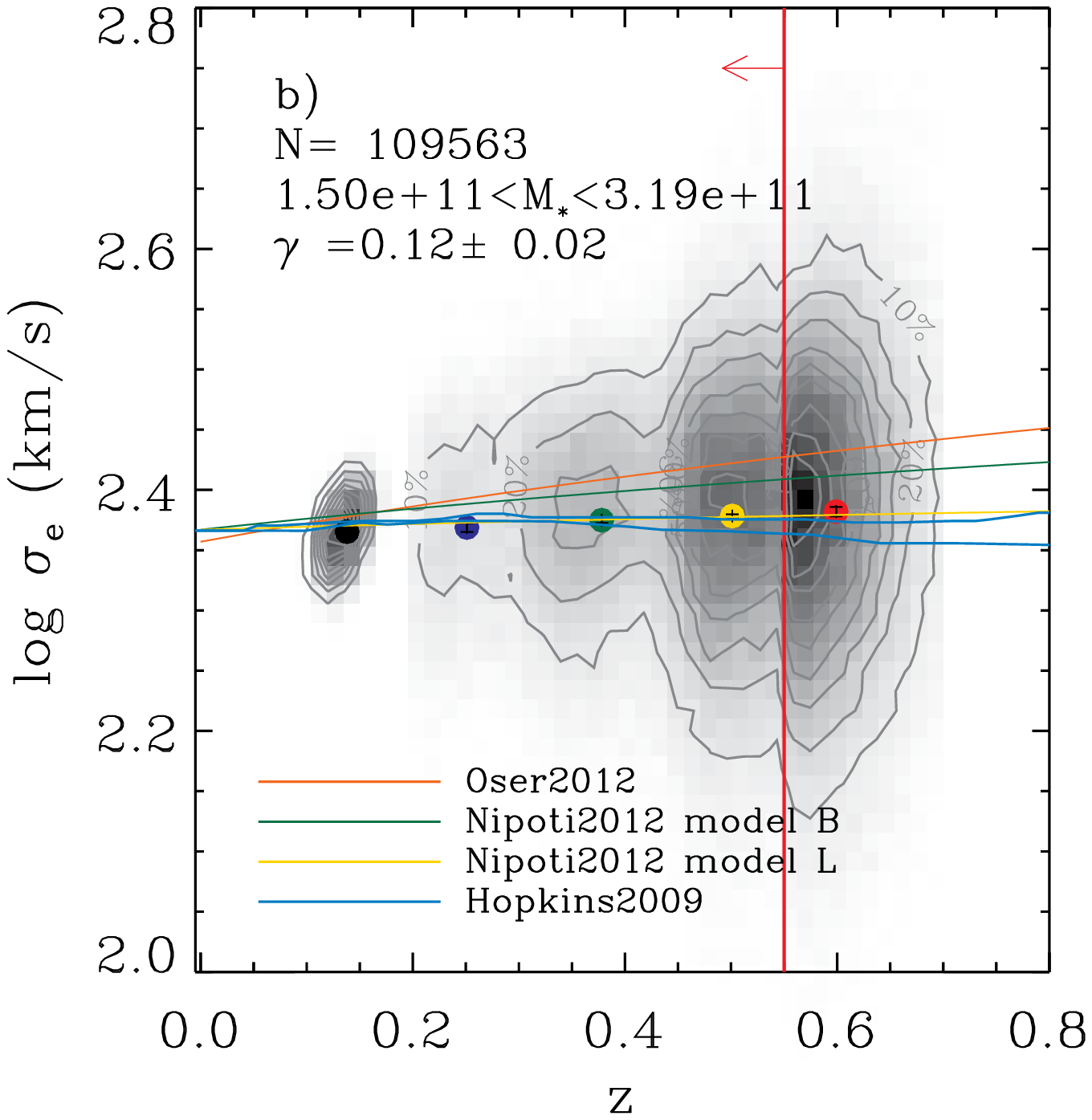}
\includegraphics[width=0.33\textwidth]{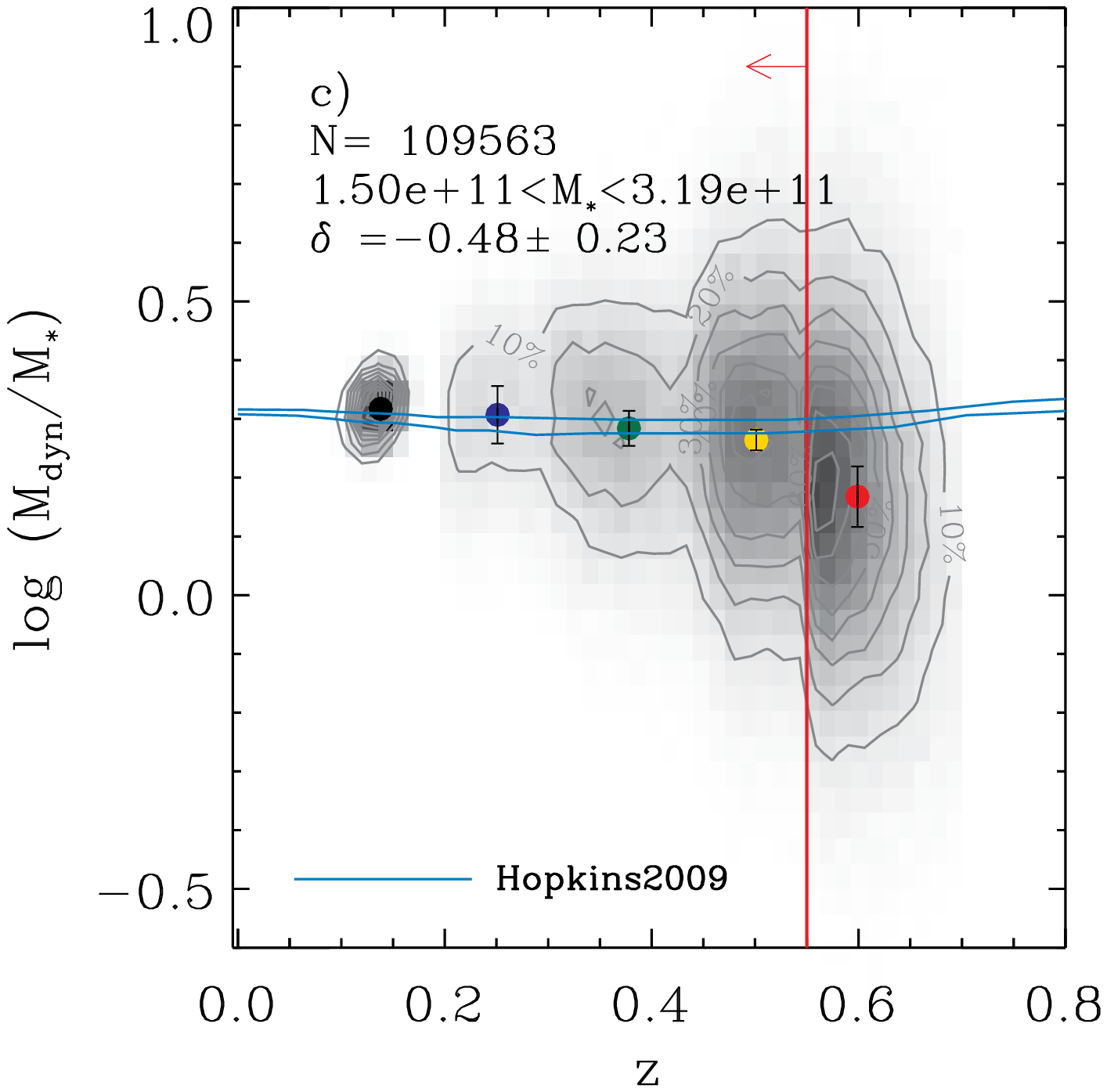}}}
\vbox{
\hbox{
\includegraphics[trim= 0 0 0 30, clip, width=0.33\textwidth]{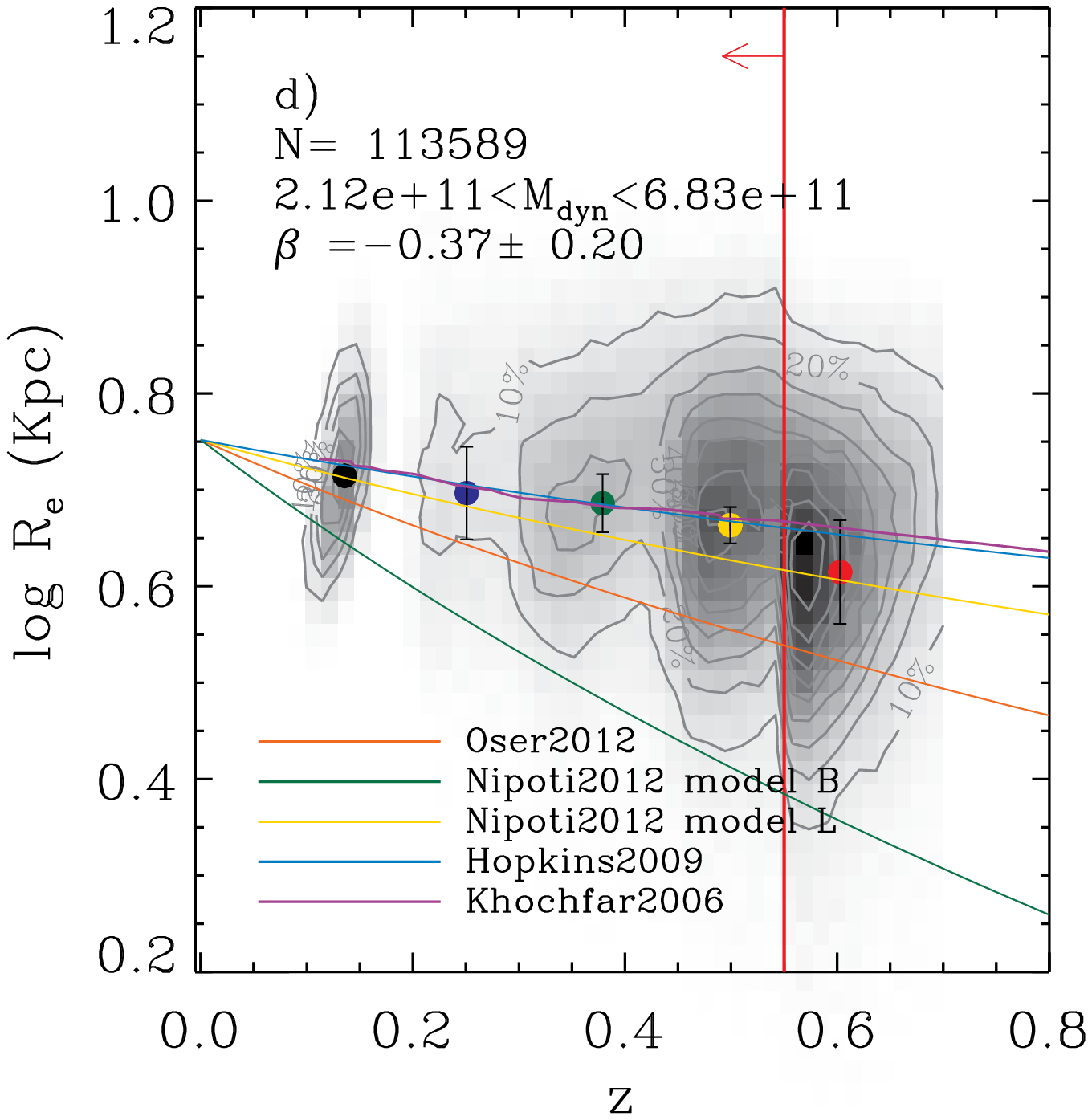}
\includegraphics[width=0.33\textwidth]{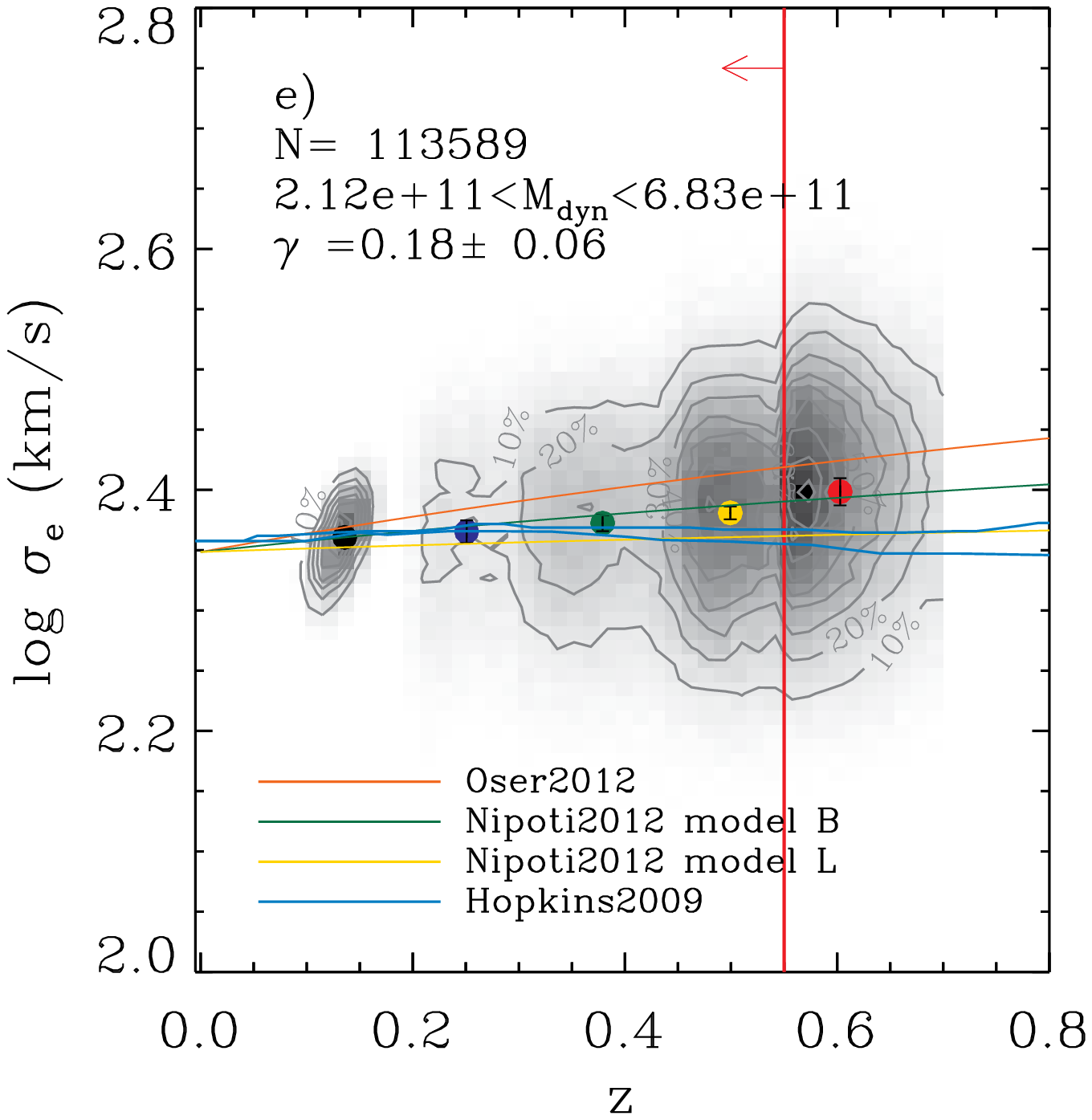}
\includegraphics[trim= 0 0 0 30, clip, width=0.33\textwidth]{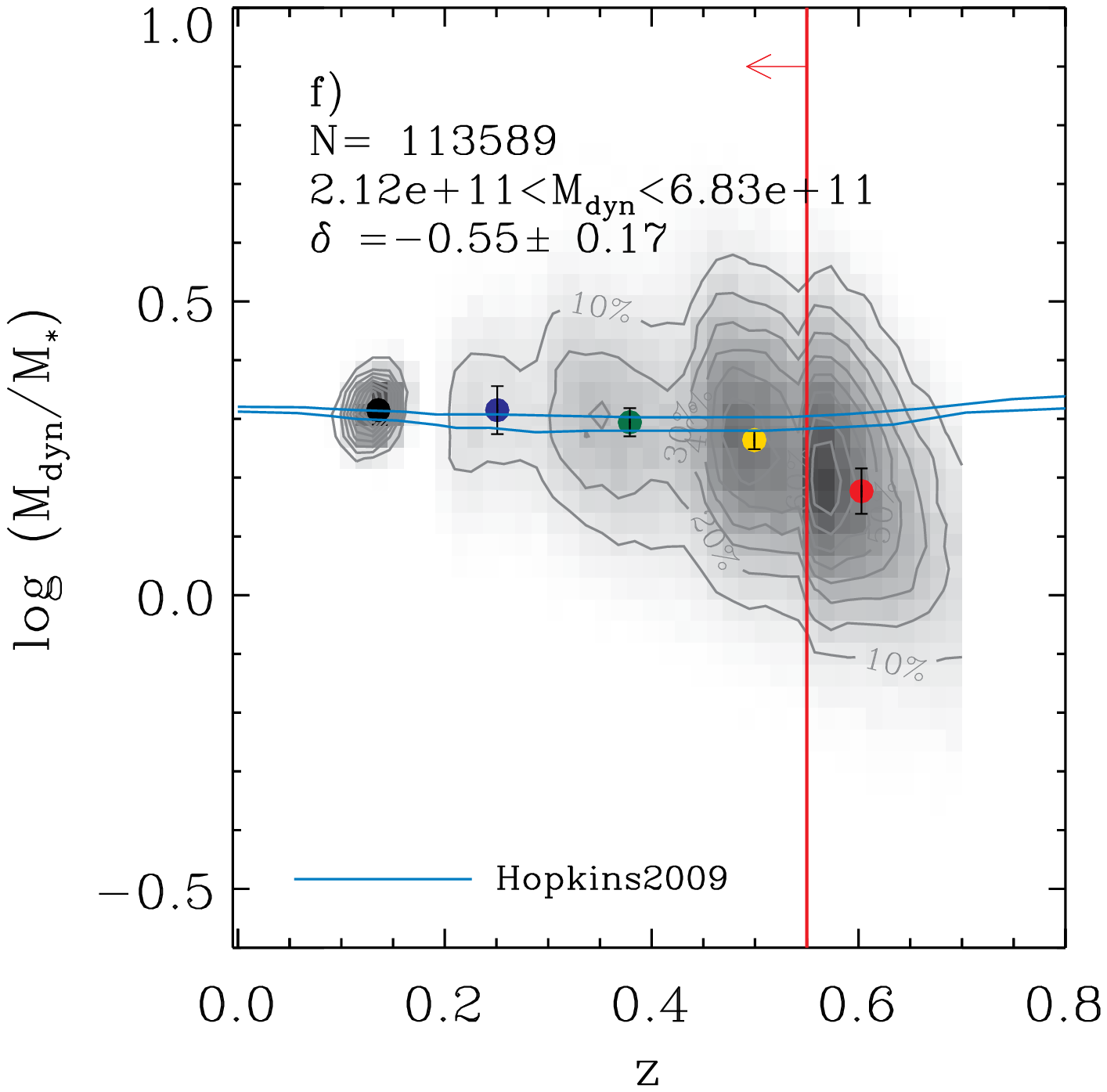}}}
\caption{Same as Figure~\ref{fig:mvir_mstar_sigma_re} with predictions from theoretical models over-plotted as solid colored lines for comparison (see labels for references).}
\label{fig:mvir_mstar_sigma_re_models}
\end{figure*}

In this section we present the redshift evolution of the galaxy
parameters effective radius, velocity dispersion, and dynamical to
stellar mass ratio for our final sample of 256,849 SDSS-III/BOSS
galaxies and 12,089 local early-type SDSS-II galaxies with a typical
stellar mass of $\sim 2.2\times 10^{11}\msun$ and a typical dynamical
mass of $\sim 3.9\times 10^{11}\msun$. The results are presented in
Figure~\ref{fig:mvir_mstar_sigma_re}, where we plot the galaxy
parameters effective radius (left-hand panels), stellar velocity
dispersion (center panels), and dynamical to stellar mass ratio
(right-hand panels) as functions of redshift. Shaded regions and
contours indicate the number density of galaxies (10 equally-spaced
density levels showing the percentage of galaxies compared to the peak
value of each plot), and colored circles are the mean for each
redshift bin. Fixed intervals in stellar mass and dynamical mass are
considered in the top and bottom panels, respectively. They were
selected to be within $\pm 1\sigma$ of the mass distributions of
Figure~\ref{fig:progenitor_bias_mass_distribution}. This allows us to
keep a large number of galaxies with similar mass (186,269 and 189,613
galaxies for \mstar\ and \mvir\ selection for the full local and BOSS
sample, respectively) without being affected by selection effects as a
function of $z$. A finer division in both \mstar\ and \mvir\ would not
change our results.

The solid line in Figure~\ref{fig:mvir_mstar_sigma_re} is a linear fit
to the relation, whereas the dashed line is a fit to the zero point at
constant zero slope and the black dotted line is the linear fit to a
sample without size correction. The fit parameters are summarized in
Table~\ref{tab:fits}.

We fit relationships of the form $X \propto (1+z)^{slope}$ to all
data, but the result does not change significantly by fitting the
means.  We {\it do not account for galaxies in the last redshift bin
  at $z>0.55$} in the fit because of the larger uncertainty of the
radius calibration (see Section~\ref{subsec:size_calibration}).  The
best-fitting values of zero-point, slope, and their associated errors
are derived by performing least-squares linear regressions using the
{\tt MPFIT} package. We additionally consider the case where we only
fit the zero-point (assuming a zero slope) to test the significance of
the derived slopes (dashed lines).  We also assess the latter with a
comparison of the $\chi^{2}$ values of the two fits for free and fixed
slope, accounting for the number of degrees of freedom, using the
F-test statistics.

Figure~\ref{fig:mvir_mstar_sigma_re_models} is a reproduction of
Figure~\ref{fig:mvir_mstar_sigma_re} in which the predictions of
simulations are shown for comparison.  Solid lines in
Figure~\ref{fig:mvir_mstar_sigma_re_models} are model predictions of
\citet{Oser2012}, \citet{Nipoti2012} (here we list a couple of models
with different stellar-to-halo-mass prescriptions as a function of
redshift that those author presented in their work),
\citet{Hopkins2009}, and \citet{Khochfar2006} for the evolution of
galaxy size and velocity dispersion. The predictions of the redshift
evolution of \mvir/\mstar\ are from \citet{Hopkins2009}.

As discussed in the previous sections, the major sources for random
and systematic errors are the size correction (Section~\ref{sec:size})
and the calculation of dynamical mass through the virial estimator
(Section~\ref{subsec:galaxy_mass}). To assess random and systematic
errors in the redshift evolution of the galaxy parameters we perform
Monte Carlo simulations perturbing the slope $a$, the intercept $b$ of
the size correction, as well as the structural dependent quantity
$\beta(n)$ within their errors. For each redshift bin, we produced
distributions of $a$, $b$ and $\beta$ generating random numbers from
their errors and assuming normal distributions. For each of the 200
realizations we then derived the mean slope and the 68\% confidence
intervals for the evolution of \re, \sigmae, \mvir/\mstar\ as a
function of redshift.  The error bars in
Figures~\ref{fig:mvir_mstar_sigma_re} and
\ref{fig:mvir_mstar_sigma_re_models} are estimated through these
simulations hence include both random and systematic errors.

\subsection[]{Evolution of galaxy size}
\label{subsec:mass_size}

The left-hand panel of Figure~\ref{fig:mvir_mstar_sigma_re} shows that
galaxy radius decreases with increasing redshift for both choices of
mass estimator (top and bottom panels) at 1.5$\sigma$ significance.
The F-test between the fits with fixed and free slope yields a
probability $<$ 25\% of the null hypothesis being true (no redshift
evolution)  for both \mstar\ and \mvir\ selected samples, which
supports the significance of the slope derived here.  Qualitatively
similar results are obtained when only using the BOSS sample, although
uncertainties are larger and the significance reduced
(Appendix~\ref{sec:test_boss_alpha}).  The size evolution found in the
present work are consistent within the errors with previous
determinations in the literature, which are mostly based on data at
higher redshifts \citep[e.g.][]{Trujillo2006b, vanDokkum2008,
  Cimatti2008,Saracco2009}, but in particular with \citet{Saglia2010},
who studied a similar $z$ range.  This agreement further validates the
size correction applied here. Without the latter, we would not detect
significant evolution of galaxy sizes (dotted lines in
Figure~\ref{fig:mvir_mstar_sigma_re}) in clear contradiction to
findings in the literature.

We note again that we did not account for the slightly different
mapped rest-frame wavelengths using radii from observed $i$-band
images across all redshifts.  This approach is conservative as even
smaller sizes would obtained from the rest-frame bluer images at
higher redshift \citep{Bernardi2003a,Hyde2009} with the net effect
that we tend to slightly underestimate the size evolution.

In Figure~\ref{fig:mvir_mstar_sigma_re_models} (left panels) we show
the comparison of our results with simulations (solid lines) which
show a very wide range of predictions for the slope $\beta$.  The
evolution we find is consistent with or slightly milder than the
predictions from semi-analytical models or hydrodynamical simulations
(\citealt{Khochfar2006,Naab2009,Hopkins2009,Nipoti2012,Oser2012}). However,
recent work on size evolution suggests that the size evolution at
$z<1$ is much shallower than at high redshifts
\citep{Newman2012,Cimatti2012,Nipoti2012}, which could explain why we
find a milder evolution of the effective radius.

\subsection{Evolution of stellar velocity dispersion}
\label{subsec:mass_sigma}

The central panels of Figure~\ref{fig:mvir_mstar_sigma_re} show the
evolution of \sigmae\ with redshift. We detect a mild but significant
evolution of stellar velocity dispersion, in the sense that \sigmae\
increases with increasing redshift at $>2\sigma$ significance.  Again,
the F-test between the fits with fixed and free slope yields a
probability  $< 1$\% or $<2$\% of the null hypothesis being true
  (no redshift evolution) for \mstar\ and \mvir\ selected samples,
  respectively, which supports the significance of the slope derived
here.  As for the size evolution qualitatively similar results are
obtained when only using the BOSS sample again with somewhat larger
uncertainties and slightly reduced significance
(Appendix~\ref{sec:test_boss_alpha}).
  
Our findings are consistent with previous results from the literature
\citep{Cenarro2009,vanDokkum2009,Saglia2010,vandeSande2011,vandeSande2012},
although the evolution detected here is somewhat milder, possibly
because of the relatively small redshift range mapped in the present
work. Indeed, studying a similar redshift range, \citet{Saglia2010}
find that, depending on the selection criteria and accounting for
progenitor bias, the slope $\gamma$ ranges from $0.59\pm0.10$ to
$0.19\pm0.10$, which is consistent with our results.

In Figure~\ref{fig:mvir_mstar_sigma_re_models} (central panels) we
compare our results with the large range of predictions of $\gamma$
coming from simulations.  The evolution is consistent with the
predictions from the models by \citet{Oser2012} if selected by
dynamical mass (center-bottom panel in
Figure~\ref{fig:mvir_mstar_sigma_re_models}). A milder evolution of
\sigmae, however is predicted by the hydrodynamical simulations of
\citet{Hopkins2009c} and N-body simulations of \citet{Nipoti2012}
which are consistent with center-top panel in
Figure~\ref{fig:mvir_mstar_sigma_re_models}.  

 \citet{Hopkins2009c} suggest that velocity dispersions do not
  evolve significantly with redshift for the redshift range probed
  here; they find a mild evolution at $z>1$, which they explain with
  velocity dispersions being set by the dark matter halos that evolve
  more weakly compared to \re. This absence of evolution in our
  redshift range is in tension with the observational result presented
  here.

\subsection[]{Evolution of the dynamical to stellar mass ratio}
\label{subsec:masses_boss}

The right-hand panels of Figure~\ref{fig:mvir_mstar_sigma_re} display
the evolution of the dynamical to stellar mass ratio \mvir/\mstar\
with redshift.  This ratio decreases with increasing redshift at $>2
\sigma$ significance.  Again, the F-test between the fits with fixed
and free slope yields a probability  $< 2.5$\% and $<1$\% of the
  null hypothesis being true, for \mstar\ and \mvir\ selected samples,
  respectively, which supports the significance of the slope derived
here.  The decrease is driven by the decrease in size, and not
balanced by the very mild increase in stellar velocity dispersion. The
slopes for the evolution are consistent within the errors whether we
select our sample by stellar or dynamical mass.  As for the evolution
of galaxy size and velocity dispersion qualitatively similar results
are obtained when only using the BOSS sample, again with somewhat
larger uncertainties and slightly reduced significance
(Appendix~\ref{sec:test_boss_alpha}).
  
Also in Appendix~\ref{sec:test_boss_alpha}, we discuss the effect of a
redshift-variable parameter $\beta(n)$ for the BOSS sample. We ran
additional simulations with redshift-dependent structural parameter
$n$ and hence a redshift-dependent virial constant $\beta(n)$ based on
the COSMOS/HST measurements. In brief, we find that the results of
this paper are not affected. This ought to be expected as the redshift
evolution of $n$ (and hence $\beta(n)$) is mild as discussed in
Section~\ref{subsecstructural_cosmos}.

Our finding of a decreasing \mvir/\mstar\ ratio with increasing
redshift is consistent with the evolution of \re\ and \sigmae\ from
\citet{Saglia2010}, resulting in a similar trend of decreasing
\mvir/\mstar\ with redshift at a $2\sigma$ level.  We searched for
systematic effects by checking the evolution of \mvir\ and \mstar\
separately.  For fixed \mstar, \mvir\ decreases with $z$
($\delta=-0.23\pm0.12$) whereas for fixed \mvir, \mstar\ increases
with $z$ ($\delta=0.85\pm0.11$).

A change in dynamical to stellar mass ratio can have several physical
explanations. In general, the effects of varying dark matter fraction
and change in the inferred stellar mass due to a variable IMF are
highly degenerate, and it is notoriously difficult to distinguish the
two. In the present study we use an approach in which redshift
evolution is added as a further constraint. As we are probing a well
selected, passively evolving galaxy sample consisting of low-$z$
massive galaxies and their high-$z$ progenitors any variation in
stellar population property  would be minor (in case of galaxy
  mergers, for instance, the variation of the effective IMF would be
  small). As a consequence the decrease of $\mvir/\mstar$ with
redshift is most plausibly caused by a decrease of dark matter
fraction. We emphasize again that we are probing a variation of
stellar kinematics within the effective radius, hence a possible
change of dark matter fraction within $1\re$, even though total masses
are compared. In other words our results imply that the dark matter
fraction in massive galaxies {\em within the half-light radius}
increases with cosmic time.

\subsubsection{Comparison with local and $z\sim0.8$ \mvir}
It is worth noting that the mean ratio between dynamical and stellar
mass is larger than one at all redshifts. This point is crucial, as a
smaller dynamical than stellar mass would be unphysical. As a key
consistency check, we compare our \mvir/\mstar\ values with
derivations for local galaxies based on sophisticated dynamical
modeling by \citet{Thomas2011} as opposed to the simple virial mass
estimator adopted here. \citet{Thomas2011} derive dynamical to stellar
mass ratios of 1.8 (assuming a \citealt{Kroupa2001} IMF) for a sample
of early-type galaxies in the Coma cluster.  The value derived in the
present work for the lowest redshift bin, $z\sim 0.1$, is $\sim 2.1$
(also based on \citealt{Kroupa2001} IMF), is well consistent with this
value, as well as with other published values for the SDSS sample
\citep[$\sim 1.7$,][]{Taylor2010}.

Recent work of \citet{Shetty2014} found that galaxies at
  $z\sim0.7-0.9$, with stellar mass \mstar $> 10^{11}$\msun\ and
  stellar velocity dispersion $\sim 200$\kms, have an average
  normalization of the IMF consistent with a Salpeter slope, similarly
  to recent findings in the local universe
  \citep[e.g.,][]{Cappellari2012}.  In our work we cannot constrain the
  actual normalization of the IMF, but we note that 7\% of our BOSS
  galaxies with stellar velocity dispersion between $200<\sigma_{\rm
    e}<280$\kms\ and errors on the stellar velocity dispersion smaller
  than the typical cut we use in our analysis ($<10$\%), would have
  unphysical \mvir/\mstar\ ratio by using a Salpeter IMF. Those
  galaxies also have a smaller average size, $\sim 3.5$kpc, compared
  to the typical $\sim 4.6$kpc of galaxies with a physical
  \mvir/\mstar\ ratio.  By using a Kroupa IMF, only the 0.5\% of BOSS
  galaxies with $200<\sigma_{\rm e}<280$\kms\ and errors on \sigmae\
  $<10$\%, have an unphysical \mvir/\mstar\ ratio (those galaxies have
  also an average size of $\sim2$kpc).

\subsubsection{Comparison with simulations}

The right panels of Figure~\ref{fig:mvir_mstar_sigma_re_models}
  show the comparison between our results and simulations by
  \citet{Hopkins2009c} for galaxies with $\mstar\sim10^{11}\msun$
  (blue lines).  The latter predict almost no evolution of
  \mvir/\mstar\ with redshift for galaxies at $\mstar\sim10^{11}\msun$
  in our redshift range, in tension with our observational result.
  \citet{Hopkins2009c} predict that \mvir/\mstar\ decreases with
  increasing redshift beyond $z\sim 1$.
  The recent galaxy formation simulations by
  \citet{Hilz2012,Hilz2012b}, instead, are in better agreement with
  our observations. The authors find that galaxy sizes grow
  significantly faster and the profile shapes change more rapidly for
  minor mergers of galaxies embedded in dark matter halos than for
  major mergers. Moreover, the increase in stellar mass is much
  smaller for minor mergers than for major mergers. This growth is
  accompanied by an increase of the dark matter fraction within the
  half-mass radius, driven by the strong size increase probing larger,
  dark matter dominated regions \citep{Hilz2012b}. In this scenario,
  the dark matter fraction in the center of a galaxy is expected to
  increase with cosmic time, in agreement with the observational
  result found in the present study.
 As shown in \citet{Hilz2012b}, major mergers could also result in
  an evolution of \mvir/\mstar, although by a smaller amount (25\%),
  and they would change substantially the internal structure of the
  galaxy. We caution that our data do not constrain any difference
  between minor and major merger, we only study the relative evolution
  of galaxy properties and not absolute quantities.  Minor mergers
  could explain our results but this does not exclude that major
  mergers play a role as well.

\section[]{Discussion}
\label{sec:discussion}

\begin{figure*}
\begin{center}
\includegraphics[width=0.7\textwidth]{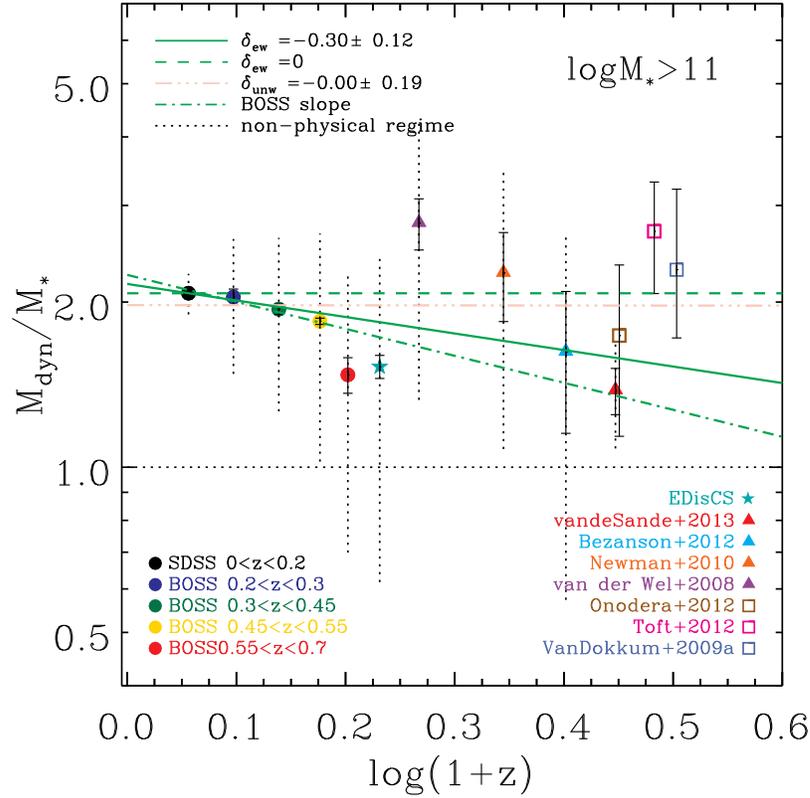}
\end{center}
\caption{Ratio between dynamical and stellar mass as a function of
  redshift extending up to $z\sim 2$ for galaxies with
  $\log$\mstar$/M_{\odot}>11$. Data samples from the literature with
  $z>0.7$ (see labels in the plot) have been added to the sample
  presented in Figure~\ref{fig:mvir_mstar_sigma_re}. The filled
  colored circles are the median \mvir/\mstar\ ratios for the redshift
  bins of Figure~\ref{fig:mvir_mstar_sigma_re}. The filled star and
  triangles are the median \mvir/\mstar\ ratios for each sample from
  the literature. Open squares are individual measurements for
  publications of only one object. The green solid line is  the
    error-weighted linear fit to the median values plotted covering
  the full range in redshift $0.08<z<2.18$, whereas the pink line
    is the unweighted linear fit to the same points.The green dashed
  line is a fit with zero slope for comparison. The green dot-dashed
  line is the linear fit to our sample from
  Figure~\ref{fig:mvir_mstar_sigma_re}. The black dotted horizontal
  line indicates the non-physical regime of $\mvir/\mstar<1$ (below
  the line).  Black dotted vertical lines indicate the standard
  deviation of the distribution at the given redshift.}
\label{fig:comparison_data}
\end{figure*}

In the past years a number of investigations have been dedicated to
studying the dependence of dynamical to stellar mass ratios
\mvir/\mstar\ with galaxy mass in the local universe. There is a clear
concordance that \mvir/\mstar\ increases with galaxy mass. The origin
of this trend remains controversial, however. It is still under debate
whether this phenomenon is driven by dark matter fractions, variations
of the IMF, non-homology of early-type systems or adiabatic
contraction
\citep{Cappellari2006,Hyde2009,Treu2010,Auger2010b,Napolitano2010,Schulz2010,Dutton2011,Thomas2011,Cappellari2012,Conroy2012,Dutton2012a,Dutton2012b,vanDokkum2012,Wegner2012,Conroy2013}.

In this paper we study the evolution of the dynamical to stellar mass
ratio of massive galaxies as a function of cosmic time. The extra
dimension added with look-back time helps to break some of the
degeneracies plaguing local studies, because we analyze a passively
evolving galaxy population in a very small mass range (see
Figure~\ref{fig:progenitor_bias_mass_distribution}).   In this
  case variations of the effective IMF due to mergers would be minor.
We find that the dynamical to stellar mass ratio in massive galaxies
of $\mstar\sim 2 \times 10^{11}\msun$ decreases with increasing
redshift at $> 2\sigma$ significance over the redshift range
$0.1<z<0.55$.

\subsection{Comparison with high-redshift literature data}
\label{subsec:highz}

The relatively modest evolution of $\mvir/\mstar$ over the past seven
billion years found here is well in line with other studies in the
literature generally probing higher redshift and larger look-back
times. The SDSS-III/BOSS data serve well in bridging galaxy properties
from the distant with the local universe. In this section we will put
those two data sets together comparing our results directly with the
data at high redshift.

We collect public catalogs of structural parameters, stellar masses
and stellar velocity dispersions from the EDisCS survey described in
\citet{Saglia2010}, for a sample of 154 cluster and field galaxies (41
field galaxies and 113 cluster galaxies) at median redshift $z\sim
0.7$.  We derive dynamical masses as described in
Section~\ref{subsec:galaxy_mass} using a variable $\beta(n)$ derived
from EDisCS S\'ersic indices and Equation~20 of
\citet{Cappellari2006}.  We rescale sizes in kpc of Table~1 and 2 of
\citet{Saglia2010} to our cosmology and stellar velocity dispersions
are rescaled to \re\ using Equation~2.  \citealt{Saglia2010} stellar
masses, derived using \citet{Bruzual2003} models and a diet-Salpeter
IMF \citep{Bell2001}, are rescaled to a common Kroupa IMF (to match
the IMF used for our local SDSS and BOSS sample), using a $-0.05\;$dex
offset based on Table~2 of \citet{Bernardi2010}.  We select galaxies
with $\log M/M_{\odot}>11$ resulting in 77 objects.

We collect $z>1$ data from \citet{vandeSande2012}. These authors
presented five new kinematic measurements of galaxies at $z>1.5$ and
compiled a catalog of previous data in the range of $0.8<z<2.18$
\citep{vanderWel2008,vanDokkum2009,Newman2010,Onodera2012,Toft2012,Bezanson2013},
for a total of 73 galaxies, 46 of which have $\log
M/M_{\odot}>11$. The dynamical masses in \citet{vandeSande2012} were
derived by using procedures similar to those described in
Section~\ref{subsec:galaxy_mass}, accounting for a variable
$\beta(n)$, hence we only rescale them to our cosmology. Stellar
masses, derived using \citet{Bruzual2003} models and a
\citet{Chabrier2003} IMF, were rescaled to a common Kroupa IMF using a
$+0.05\;$dex offset based on Table~2 of \citet{Bernardi2010}.

\subsubsection{Stellar masses}

Stellar masses derived with different population models may be
different because of the different assumptions of stellar evolution in
the models. Moreover, other assumptions regarding the star formation
history, dust reddening and the assumed IMF all affect the final value
of \mstar.

The variation in \mstar\ is quantified in \citet{Pforr2012} as a
function of population model, using \citet{Maraston2005} and
\citet{Bruzual2003} models, and as a function of the star formation
history and IMF assumed in the models. We use their results for
obtaining a homogeneous sample of \mstar.

As most of the galaxies in the $z>0.7$ sample studied here appear to
be passive \citep{vandeSande2012}, we choose offsets from Table~B4 of
\citet{Pforr2012} derived for mock passive galaxies at $z=2$.  As
fitting setup we select the ``wide BC03'' with reddening included as
adopted in the $z>0.7$ literature stellar masses, which gives an
offset to \citet{Maraston2005} based stellar masses of $0.13\;$dex. We
note that the star-forming mocks show the same offset ($0.12\;$dex)
for the ``wide BC03'' fitting setup, which is important as some of the
galaxies in the sample might not be passive (see for instance the
\citealt{Bezanson2013} sample).  We decrease the stellar masses of the
$z>0.7$ sample by this amount. This offset is consistent with
differences in stellar mass due to stellar population models found for
BOSS galaxies (see Appendix of \citealt{Maraston2012}) and for
galaxies in COSMOS \citep{Ilbert2010}.

It should be noted, however, that stellar masses also depend on the
star formation history adopted for the SED fitting. Stellar masses of
the $z>0.7$ sample were obtained assuming an exponentially decaying
star formation history. However, galaxies at those redshift may be
better modeled with an exponentially {\em increasing} star formation
history \citep{Maraston2010}, which would give higher stellar masses
by $\sim 0.2\;$dex compensating the offset applied here. Ideally the
full sample should be modeled self-consistently, but the photometry of
the $z>0.7$ sample is not available to us. We will therefore discuss
final results based on both with and without the offset of $0.13\;$dex
in stellar mass.

\subsubsection{Evolution of \mvir/\mstar}
Figure~\ref{fig:comparison_data} presents the evolution of the
dynamical to stellar mass ratio as a function of redshift for the
redshift interval $0.08<z<2.18$. The SDSS-II and SDSS-III/BOSS data of
the present study is combined with the high-$z$ samples discussed in
the previous section. The $0.13\;$dex offset between the stellar
masses of the low- and high-$z$ samples has been applied. The symbols
plotted are the median values of both $\mvir/\mstar$ and $z$. Error
bars are standard errors, while the dotted lines indicate the standard
deviation of the distribution at the given redshift.

The high-$z$ sample is consistent with the trend of decreasing
$\mvir/\mstar$, even though the scatter at $z>1$ is large.  The
  BOSS  data at intermediate redshifts clearly drives this
  relationship, because of the large scatter in the data at high
  redshift.  By fitting the data over the full redshift range of
$0.08<z<2.18$ as shown in Figure~\ref{fig:comparison_data} and
  including the error bars in the fit, we find
\mvir/\mstar$\propto(1+z)^{-0.30\pm 0.12}$ (where $\delta_{\rm
    ew}$ is the slope, and ``ew'' stands for error-weighted fit).
This slope is slightly shallower but well consistent with the value
derived in this work from the SDSS-II and SDSS-III/BOSS data alone
(dot-dashed line, see Figure~\ref{fig:mvir_mstar_sigma_re}). Most
importantly, the statistical significance for the presence of a
negative slope is $> 2\sigma$ also in this case. This further
reinforces the evidence for a decrease in $\mvir/\mstar$ with
increasing redshift.

Error weighting the fit could potentially bias our results
  towards the BOSS sample, where the statistic is much larger and
  errors on the mean values are smaller. Therefore, we repeated the
  same procedure above using an equal weighting for all the
  points. This is shown by the pink three-dot-dashed line in
  Figure~\ref{fig:comparison_data} (where $\delta_{\rm unw}$ is the
  slope, ans ``unw'' stands for unweighted fit). In this case we find
  almost no evolution as we could easily expect. Given the large
  uncertainties and the low statistic in the high-z sample applying a
  different weight to the BOSS sample would be preferable.  A larger
  sample of high-redshift data would be helpful to constrain the
  evolution \mvir/\mstar\ ratio over a redshift range wider than that
  of BOSS.

\begin{figure}
\begin{center}
\includegraphics[width=\columnwidth]{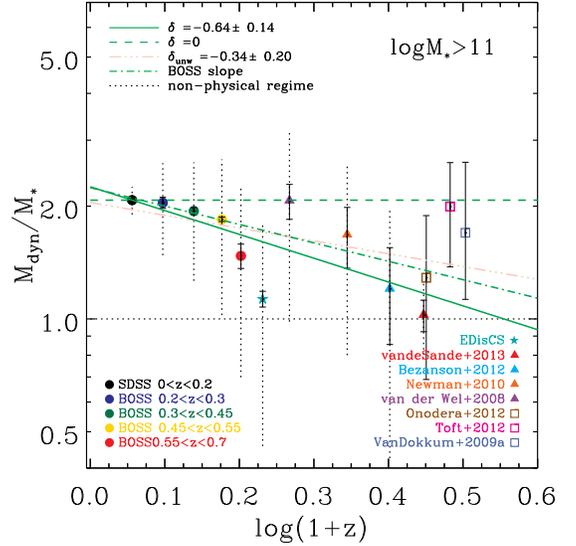}
\end{center}
\caption{Same as Figure~\ref{fig:comparison_data} but without correction for offset in stellar masses due to differences in stellar population modeling (see text for details).}
\label{fig:comparison_data_2}
\end{figure}

Figure~\ref{fig:comparison_data_2} shows the case in which the offset
to the stellar masses has not been applied. As the correction implied
a decrease of stellar masses in the high-$z$ sample, the decrease of
$\mvir/\mstar$ with increasing redshift becomes even steeper and the
statistical significance increases to $>4\sigma$.

\subsection{Decreasing dark matter fraction due to size growth}
The increase of \mvir\ with cosmic time is most plausibly caused by an
increase of dark matter fraction within the effective radius. This
increase can be well understood through size growth, which causes an
increase of the dark matter fraction within an increasing effective
radius \citep{vandeSande2012}.  Indeed simulations show that the
addition of stars in the outskirts of galaxies following the minor
merger scenario can lead to an increased measured dark matter fraction
by $\sim80$\%\ \citep{Johansson2012,Hilz2012} because more area with
larger dark matter fraction is also included within \re\
\citep{Hilz2012b,Hopkins2009c}. Also \citet{Toft2012} studying
galaxies at $z\sim2$ with available kinematics suggest that the low
dark matter fraction of galaxies at $z\sim2$ is in favor of the merger
scenarios which can redistribute dark matter within \re\
\citep{Boylan_Kolchin2005,Oser2012}.

The steady increase of the dark matter fraction in the centers of
massive galaxies with time further implies that massive galaxies in
the local universe must contain some dark matter within their
half-light radii, even if they are baryon dominated. This is
consistent with recent dynamical modeling of nearby galaxies implying
dark matter fractions of $16-28\%$ \citep{Thomas2011,Cappellari2012b}
as well as simulations predicting dark matter fractions of $18-38\%$
\citep[][and references therein]{Naab2007}.

\section{Conclusions}
\label{sec:conclusions}

We study the redshift evolution of the dynamical properties of $\sim
180,000$ galaxies from the SDSS-III/BOSS survey. We examine the
redshift evolution of luminous, massive galaxies ($M\sim 2\times
10^{11}\msun$) at fixed stellar or dynamical mass for the first time
for such a large sample size.

Despite the relatively low $S/N$ of BOSS spectra, it is possible to
measure \sigmas\ for a large sample of galaxies in the range $0.2\leq
z\leq 0.7$ with a typical error $\leq$ 30\%. Stellar velocity
dispersions are adopted from \citet{ThomasD2012}. 

At BOSS redshifts effective radii are barely resolved in the SDSS
imaging, and higher resolution images would be needed, which are not
available for the whole sample.  Therefore, we used a sub-sample of
BOSS galaxies for which HST photometry is available as part of the
COSMOS survey \citep{Masters2011}.  We derived a correction to
physical effective radii derived from SDSS photometry by dividing our
sample in four redshift bins and searching for correlations between
the ratio of the SDSS \re\ and HST/COSMOS \re\ as a function of the
SDSS \re.

We then derive dynamical mass estimates by means of a simple virial
mass estimator based on galaxy effective radius and velocity
dispersions within the effective radius. These total dynamical masses
are compared to the total stellar masses derived by
\citet{Maraston2012} studying the redshift evolution of the galaxy
parameters effective radius, stellar velocity dispersion, and
dynamical to stellar mass ratio $\mvir/\mstar$. We complement the
SDSS-III/BOSS sample with local early-type galaxies from SDSS-II after
matching their mass distributions, so that our study covers the
redshift range $0.1<z<0.55$.

To account for the effects of the so-called progenitor bias we compare
the galaxy ages in each redshift bin and remove those galaxies from
the low-$z$ sample whose ages after evolution to the highest redshift
bin would be lower than a given age threshold. As result we study a
sample of passively evolving galaxies within a relatively narrow mass
range about $M\sim 2\times 10^{11}\msun$ (for a  Kroupa IMF).

We find a moderate size evolution at the $\sim 1.5\sigma$ level, with
galaxy radii decreasing with increasing redshift in agreement with
previous results and model predictions (\citealt{Oser2012}; but better
with \citealt{Nipoti2012}, \citealt{Khochfar2006} or
\citealt{Hopkins2009}). We further observe a mild, but significant
($>2\sigma$), evolution in velocity dispersion of \sigmas\ increasing
with increasing redshift.  The evolution of stellar velocity
dispersion and effective radius together combine to an evolution of
the dynamical to stellar mass ratio, such that $\mvir/\mstar$
increases with decreasing redshift at $> 2\sigma$ significance.  We
emphasize that we are probing a variation of stellar kinematics within
the effective radius, hence this evolution is caused by a change of
dynamical mass within $1\re$, even though total masses are compared.

The major sources for random and systematic errors are the size
correction and the calculation of dynamical mass through the virial
estimator. To assess random and systematic errors in the redshift
evolution of these galaxy parameters we perform Monte Carlo
simulations perturbing the size correction, as well as the structural
dependent constant of proportionality of the virial mass estimator
within their errors. In the appendix we present several additional
tests on the impact of the progenitor bias correction, of aperture
effects, of a possible redshift dependence of the virial constant, and
of a mismatch in mass distribution between the local and the high-$z$
galaxy samples. We show that, while the details and the exact
strengths of the correlations between $\re$, $\sigmas$, and
$\mvir/\mstar$ with redshift vary, the general detection of a redshift
evolution of these parameters is robust against the systematic
uncertainties from these procedures.

Finally, we extend the present study to higher redshifts by combining
our sample with high-redshift literature data
\citep{Saglia2010,vandeSande2012} so that we cover the full redshift
range from $z\sim 0.1$ to $z\sim 2$. The high-$z$ sample  is
  consistent with the trend of decreasing $\mvir/\mstar$, even though
the scatter at $z>1$ is large. By fitting the data over the full
redshift range we find \mvir/\mstar$\propto(1+z)^{-0.30\pm
  0.12}$. This slope is slightly shallower but well consistent with
the value derived in this work from the SDSS-II and SDSS-III/BOSS data
alone. Our results are clearly driven by the BOSS sample in which
  the large number statistics allows us to have smaller uncertainties
  on average quantities as a function of redshift and help to
  identify and quantify redshift-dependent trends.  Most importantly,
the evidence for a decrease of $\mvir/\mstar$ with increasing redshift
is reinforced further at $>2\sigma$ statistical significance, 
  although a larger sample of high-redshift data would be helpful to
  constrain the evolution \mvir/\mstar\ ratio over a redshift range
  wider than that of BOSS.

We discuss that the increase of \mvir\ with cosmic time is most
plausibly caused by an increase of dark matter fraction within the
effective radius. This evolution can be well understood through the
size growth, which causes an increase of the dark matter fraction
within an increasing effective radius as also predicted by galaxy
formation simulations based on minor merger driven mass growth.  Major
mergers could also result in an evolution of \mvir/\mstar, but of
smaller amount; however, with our data we cannot constrain any
difference between minor and major mergers.
Finally it is interesting to note that the steady increase of
the dark matter fraction in the centers of massive galaxies with time
further implies that massive galaxies in the local universe must
contain some dark matter within their half-light radii, even if they
are baryon dominated.

\acknowledgments

 We acknowledge the anonymous referee for valuable comments that
  led to an improved presentation.
AB, DT, CM, OS, KM, JP, RT, JJ, and RN acknowledge STFC rolling grant
ST/I001204/1 ``Survey Cosmology and Astrophysics'' for support.  AB is
indebted with Michele Cirasuolo, Lodovico Coccato, Enrico Maria
Corsini, Marc Sarzi, Sadegh Khochfar, Thorsten Naab, Nicola
Napolitano, Stefanie Phleps, Jens Thomas, David Wilman for many useful
discussions and suggestions.

KLM acknowledges funding from The Leverhulme Trust as a 2010 Early
Career Fellow.

Funding for SDSS-III has been provided by the Alfred P. Sloan Foundation, the Participating Institutions, the National Science Foundation, and the U.S. Department of Energy Office of Science. The SDSS-III web site is http://www.sdss3.org/.

SDSS-III is managed by the Astrophysical Research Consortium for the Participating Institutions of the SDSS-III Collaboration including the University of Arizona, the Brazilian Participation Group, Brookhaven National Laboratory, Carnegie Mellon University, University of Florida, the French Participation Group, the German Participation Group, Harvard University, the Instituto de Astrofisica de Canarias, the Michigan State/Notre Dame/JINA Participation Group, Johns Hopkins University, Lawrence Berkeley National Laboratory, Max Planck Institute for Astrophysics, Max Planck Institute for Extraterrestrial Physics, New Mexico State University, New York University, Ohio State University, Pennsylvania State University, University of Portsmouth, Princeton University, the Spanish Participation Group, University of Tokyo, University of Utah, Vanderbilt University, University of Virginia, University of Washington, and Yale University.

This research has made use of the NASA/ IPAC Infrared Science Archive,
which is operated by the Jet Propulsion Laboratory, California
Institute of Technology, under contract with the National Aeronautics
and Space Administration.



\appendix

\section{A.~~Stellar masses from one effective radius aperture magnitudes}
\label{sec:ap_stellar_masses}

The dynamical mass obtained using the virial mass estimator (see
Section~\ref{subsec:galaxy_mass}) is based on stellar kinematics
within an aperture of 1 effective radius and scaled to {\em total}
dynamical mass via equation~\ref{eq:vir}. This quantity is compared
with the {\em total} stellar mass from \citet{Maraston2012} based on
{\tt cmodelMag} magnitudes. Hence both dynamical and stellar masses
are {\em total} masses, which ensures a consistent comparison.

Still, the total dynamical mass is derived from observations within
the effective radius, while the stellar mass comes from the total
stellar light. We explore therefore the possible presence of a
systematic effect from the different apertures in which kinematics and
stellar populations have been measured. In this test we compare
\mstar\ derived from {\tt modelmag} (rescaled to {\tt cmodelmag}) and
\mstar\ from aperture magnitudes within \re\ (rescaled to {\tt
  cmodelmag}), for a sub-sample of 1,000 galaxies randomly selected
among the BOSS sample to assess differences between the two
quantities.  We derive magnitudes within the two apertures closest to
the effective radius of each galaxy in $i$-band (the same \re\ we used
for \mvir\ determination) and we interpolate magnitudes to derive the
value we would have at \re\ in each band.
Aperture magnitudes are then rescaled to {\tt cmodelmag} in $i$-band
to make sure that any difference on the resulting stellar mass comes
from differences in the $M/L$ ratio and SED shape.

In detail, we collect our reference (circularized) $i$-band radii
following Section~\ref{subsec:size_sdss}.  We estimate aperture
magnitudes following the step described at
http://www.sdss3.org/dr8/algorithms/magnitudes.php\_photo\_profile
through a casjob query.  We first derive {\tt nprof}, from the {\tt
  PhotoObjAll} table, which gives us the number of annuli (concentric
circles) for which there is a measurable signal in $u, g, r, i, z-$
bands.  From the {\tt photoProfile} table, where azimuthally-averaged
radial profiles of SDSS photo objects are listed, we create circular
aperture magnitudes from {\tt profMean} (the mean surface brightness
within the annuli) and their errors ({\tt profErr}) which are both in
$nanomaggies/arcsec^2$.  Quantities are then converted from nanomaggie
following
http://www.sdss3.org/dr8/algorithms/magnitudes.php\_nmgy. The {\tt
  bin} keyword gives the annuli from which the profile was derived
(from 0 to 14) and {\tt band} gives the selected $u, g, r, i, z-$ band
(from 0$-$4).  We integrate {\tt profMean} values within each of the two
annuli close to the circularized $i$-band effective radii.  In our
query we impose that magnitudes are only calculated when fluxes within
the aperture in nanomaggies are positive. We also tested whether the
number of available bins for each galaxy is smaller or equal to the
requested aperture.
The final sample for which we are able to derive aperture
magnitudes reduces to 604 galaxies due to the quality of the $u$-band data.
Magnitudes in the two annuli close to \re\ are then linearly
interpolated to derive the value at \re.  

\begin{figure}
\begin{center}
\includegraphics[width=0.6\textwidth]{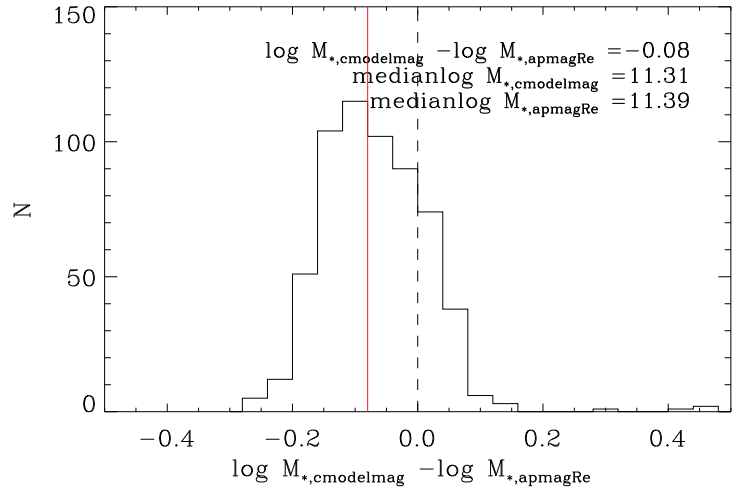}
\end{center}
\caption{Comparison between stellar masses derived from {\tt modelmag}
  and aperture magnitudes within $1\re$ both rescaled to $i$-band {\tt
    cmodelmag} to give total mass. The median difference between the
  two masses is $\sim -0.08\;$dex (25\%, continuous-red
  line). Galaxies are redder inside the effective radius leading to
  higher $M/L$ ratios and hence larger total mass (when scaled to
  total light).}
\label{ap_cmodel_plot}
\end{figure}

The histogram of the resulting difference between \mstar\ obtained
using {\tt modelmag} rescaled to {\tt cmodelmag} and 1\re\ aperture
magnitudes rescaled to {\tt cmodelmag} is shown in
Figure~\ref{ap_cmodel_plot}. The difference is $\sim -0.08\;$dex in
log (25\%).  The slightly redder populations in the more central
aperture photometry (1\re\ aperture magnitude colors are redder than
{\tt modelmag} colors) leads to slightly higher $M/L$ ratios and hence
slightly higher masses. However, this result suggests that the
systematic offset is small. Our \mvir/\mstar\ ratios are always above
0.08 dex, so by rescaling \citet{Maraston2012} stellar masses there is
no risk that \mstar\ exceeds \mvir, which would be unphysical.
We also note that the difference between \mstar\ obtained using {\tt
  modelmag} rescaled to {\tt cmodelmag} and 1\re\ aperture magnitudes
rescaled to {\tt cmodelmag} does not depend on redshift (the redshift
variation is $\sim 0.01$ dex, of the order of the uncertainties on
\mstar). Therefore, our analysis, focused on redshift variations, is
not affected.

\section{B.~~Impact of unresolved multiple systems in the SDSS images}
\label{sec:multiples}

 The presence of unresolved multiple systems in the SDSS images could
potentially introduce spurious trends in the measured evolution of
sizes, dynamical masses and stellar masses.  In this Appendix we study
their influence in our size calibration and in the stellar mass estimation.

\citet{Masters2011} found that the 23\% of the COSMOS/BOSS sub-sample
shows two components in the COSMOS images, which are unresolved in the
SDSS images. We further investigate this, considering only objects
showing a size overestimation after our size calibration.

\begin{figure}
\begin{center}
\includegraphics[width=0.33\textwidth]{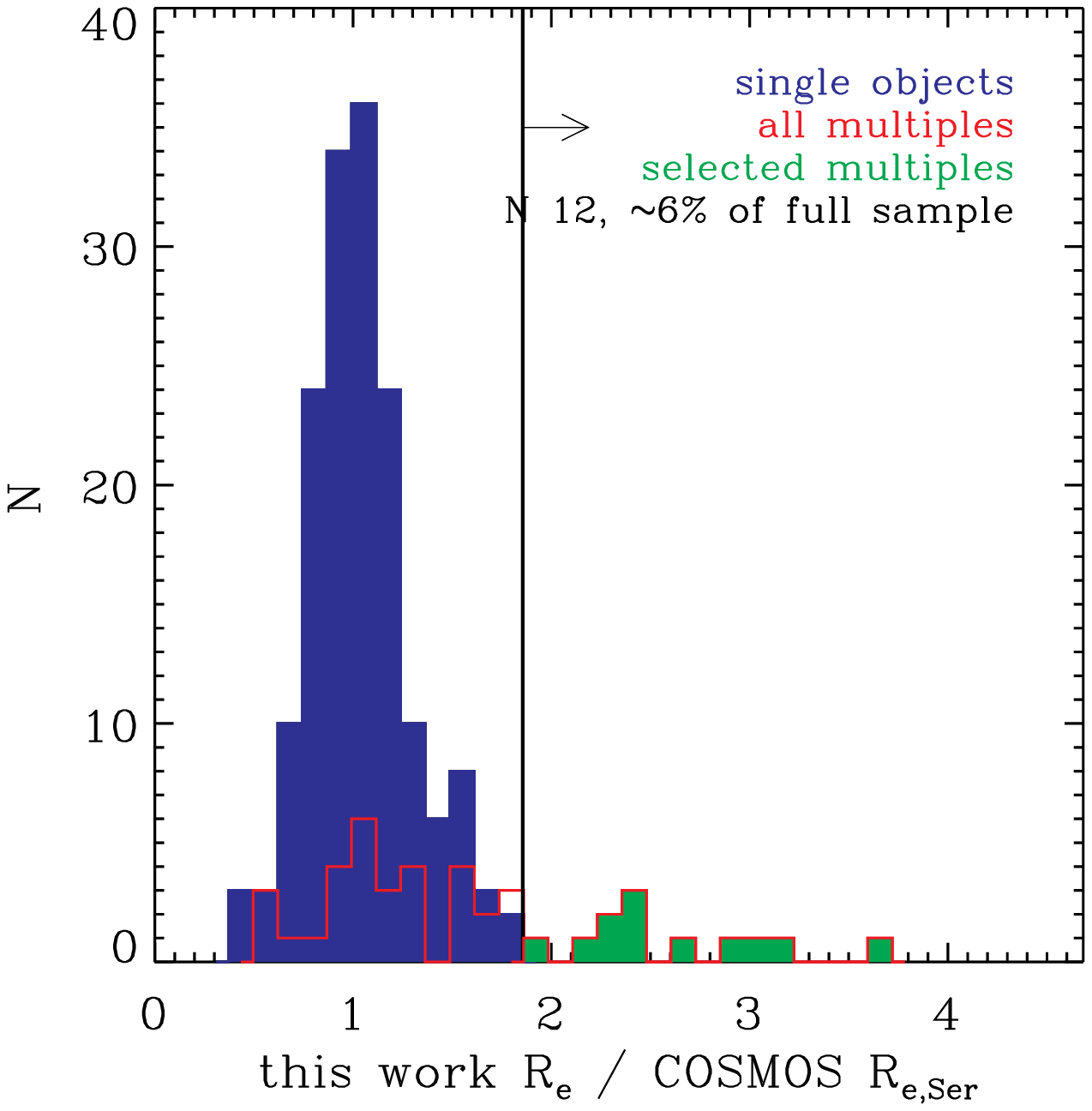}
\includegraphics[width=0.33\textwidth]{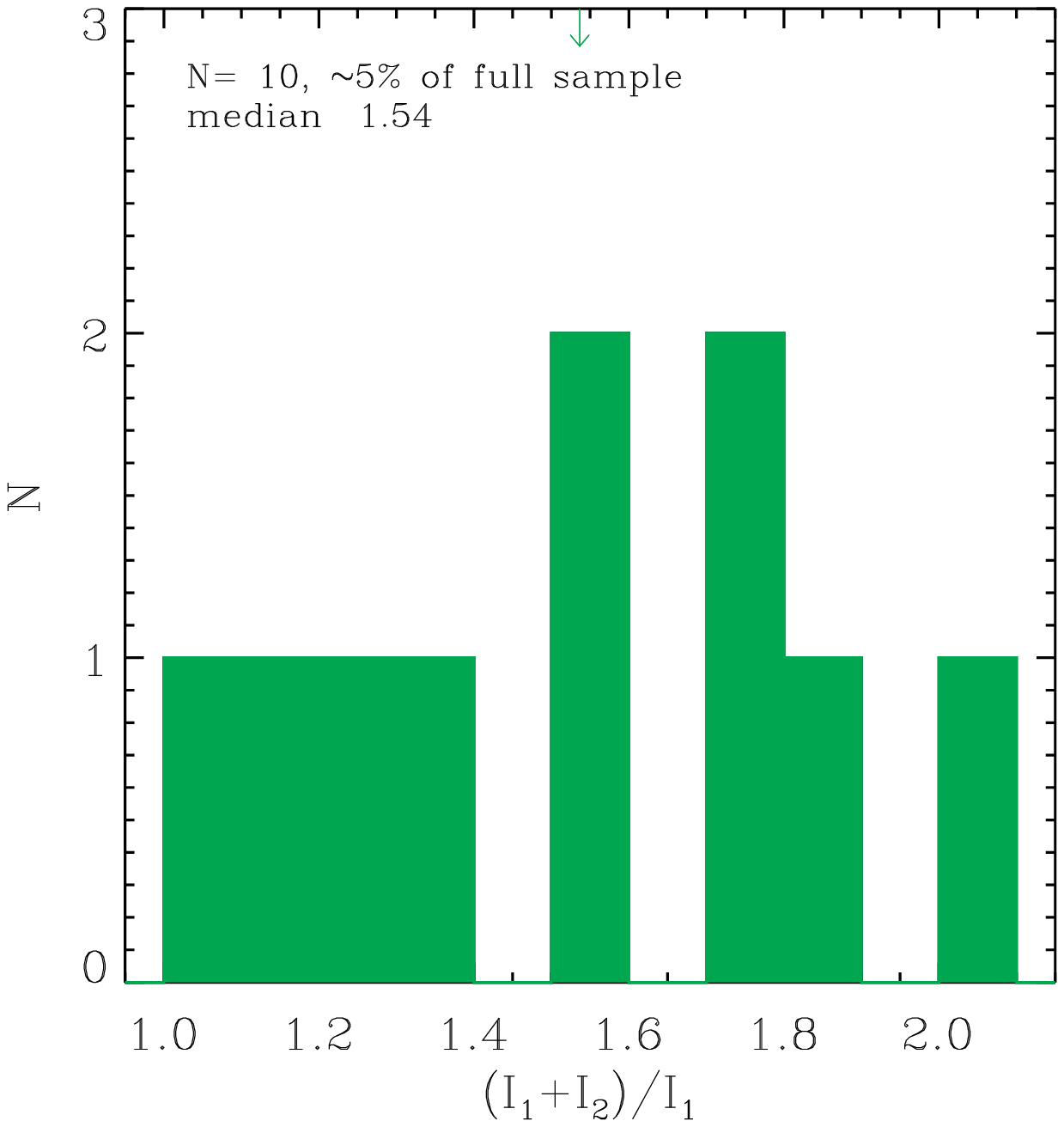}
\includegraphics[width=0.33\textwidth]{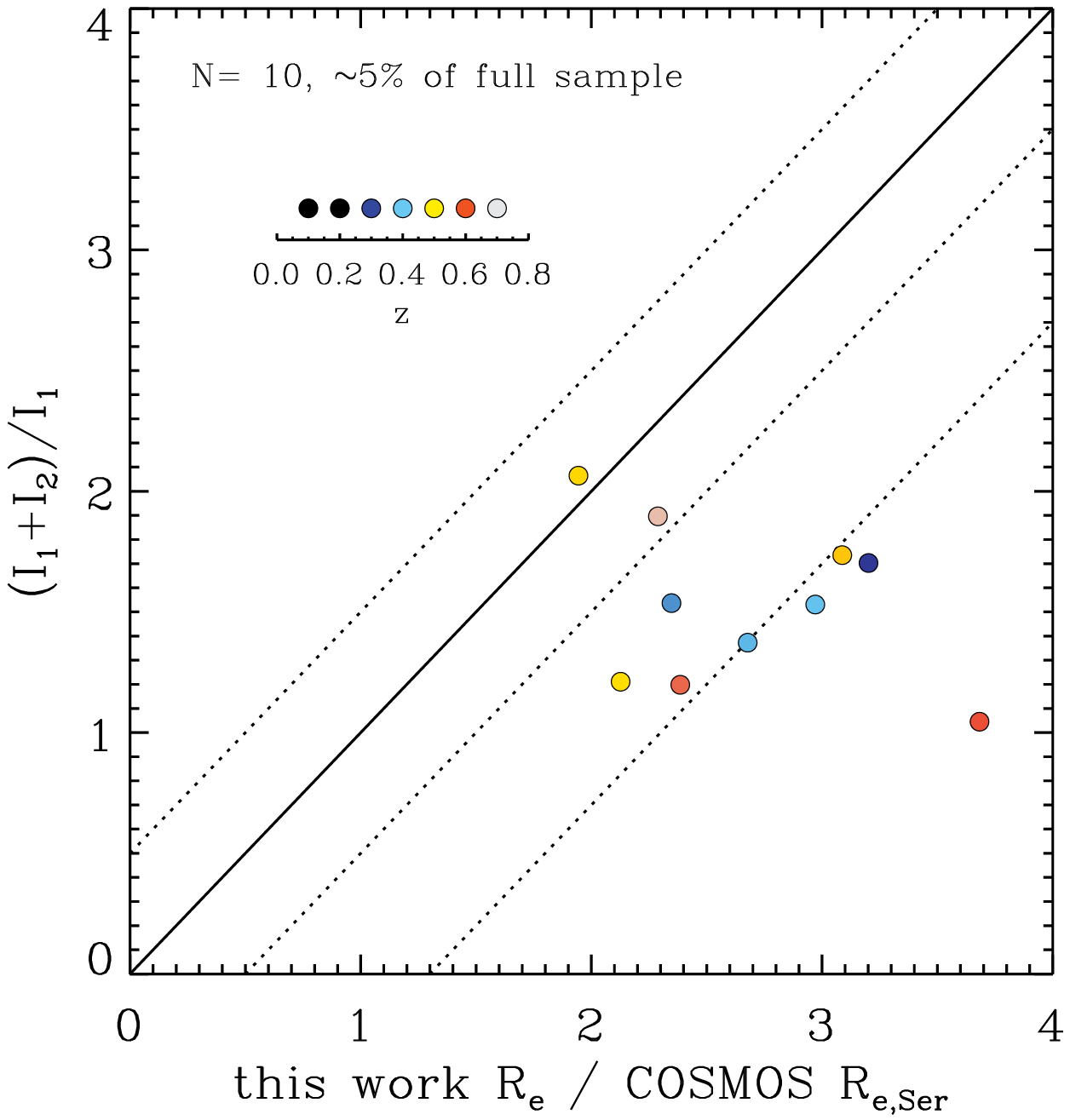}
\end{center}
\caption{{\em Left Panel:} Distribution of the ratio between our
    corrected SDSS \re\ and COSMOS \ensuremath{R_{\rm e, Ser}} (blue
    filled histogram) and distribution of unresolved multiple systems
    (red empty histogram). The green filled histogram shows the sample
    of unresolved multiple systems which were used to study the impact
    of these objects on our results. {\em Central Panel:} Distribution
    of the ratio between the fluxes (masses) of the two components
    which are resolved in the COSMOS photometry. {\em Right Panel:}
    Comparison between the size excess and the flux (mass) excess in
    our sample of 10 multiple systems coded as a function of
    redshift. By using sizes from SDSS we tend to overestimate sizes
    of a factor 2 to 3 and fluxes of a factor $\sim$1.54. The
    continuous line corresponds to the one-to-one relation and dotted
    lines show the intercepts $\pm$0.5 and 1.3x.}
\label{fig:multiples_re_histo}
\end{figure}

Figure~\ref{fig:multiples_re_histo}, left panel, shows the
distribution of the ratio between corrected sizes and COSMOS sizes for
206 objects part of the COSMOS/BOSS sub-sample in the redshift range
$0.2\la z\la 0.7$.  Those include both single objects (163 galaxies,
blue filled histogram) and unresolved multiple systems (43 galaxies,
red empty histogram).  Our size correction allows to adjust most of
the unresolved multiple systems, resulting in a distribution similar
to that of the single objects but with a tail at larger ratios.  
A closer look at the catalog of \citet{Masters2011} (available at
http://www.icg.port.ac.uk/$\sim$mastersk/BOSSmorphologies/) shows that
when the secondary object in the system is very faint, its effect on
the radius measured by the SDSS pipeline seems to be reduced, resulting in sizes
not massively overestimated.

We estimate the further correction we have to apply to unresolved
multiple systems, considering the 12 objects ($\sim 6$\% of our
COSMOS/BOSS sub-sample) which are in the tail, as shown by the
green histogram in the central panel of Figure~\ref{fig:multiples_re_histo}.
Although the effect of the multiple systems is likely to be redshift
dependent, the small sample we are considering here does not allow a
division in redshift bins.  Moreover, due to the small statistic, we
make use the full distribution to derive the correction factor for
sizes, which ranges between a factor 2 and 3.

Similarly to sizes, stellar masses of unresolved multiple systems
could be biased too. We use the same 12 galaxies with overestimated
corrected sizes to assess this effect.
Under the assumption that our galaxies are passively evolving, we can
estimate the bias in the stellar mass from the ratio of the
fluxes of the two components resolved by the COSMOS imaging.
If galaxies were not passively evolving the luminosity would be
diluted but our conclusions would not change appreciably.
Fluxes were derived from the {\tt MAG\_AUTO} listed in the Zurich
Structure \& Morphology Catalog v1.0 available at
http://irsa.ipac.caltech.edu/data/COSMOS/datasets.html and described in
Section~\ref{subsec:size_cosmos}.  Given that we do not have accurate
information about the redshifts of the two systems we used as estimate
the parent magnitudes and not absolute magnitudes.
Two objects are discarded because, by visual inspection, their
separation looked large enough for each of them to be considered two
separated sources, suggesting a smaller bias in the SDSS magnitudes
used in the stellar mass calculation (separation $> 2\times {\rm
  FWHM_{seeing}}$).  This retains 10 objects ($\sim$ 5\% of the total
sample of COSMOS/BOSS galaxies).  The distribution of the relative
flux ratio of the objects which are resolved in the COSMOS images,
$(I_1+I_2)/I_1$, is shown in the central panel of
Figure~\ref{fig:multiples_re_histo}, where $I_1$ is the flux of the
primary brightest object in COSMOS, and $I_2$ is the flux of the
secondary objects within 2 times the typical FWHM of the SDSS seeing
($\sim $1\farcs1, see Section~\ref{subsec:size_sdss} for details).
The average correction factor for fluxes (or masses) is about 1.54.

Although stellar velocity dispersions, to some extent, can be biased
too, the estimation of this effect is not trivial and cannot be done
without additional spectroscopy, therefore we disregard any possible
biases to stellar velocity dispersions due to unresolved multiple in
our analysis.

Right panel of Figure~\ref{fig:multiples_re_histo} compares the flux
(i.e., mass) correction with the size correction for final sample of
10 unresolved multiple systems used here. Typical size corrections
range between a factor 2 and 3, whereas the flux (mass) correction is
much smaller ($\sim 1.54$) as shown by the dotted lines parallel to
the continuous one-to-one relation in a manner which does not seem to
change with redshift (but the statistic is too small to make strong
statements).

We finally apply the redshift independent size and mass corrections to
a randomly selected 6\% sub-sample of BOSS galaxies and analyze the
effects on our results.  The corrections were also included in our
Monte Carlo estimation of systematic errors.  The resulted combined
effect on sizes and masses is negligible and our results do not change
substantially considering this additional correction or not (if not
improve them of about $<2$\% ).
Figures~\ref{fig:mvir_mstar_sigma_re},
\ref{fig:mvir_mstar_sigma_re_models}, \ref{fig:comparison_data} and
\ref{fig:comparison_data_2} and Table~2 include this additional
correction.

\section{C.~~Tests on evolution: BOSS sample, and use of constant or
  redshift dependent $\beta$}
\label{sec:test_boss_alpha}

In this Appendix we show that our results do not change when we
analyze scaling relations only using the BOSS sample.  Table~3 (case
in which $\beta$ is assumed constant with redshift) shows that the
trends we find in Section~\ref{sec:redshift_ev_all} are maintained,
although the significance of the results is reduced. Errors from the
Monte Carlo simulations for the BOSS sample are $50-60$\% larger than
using the combined sample of local and BOSS data (see
Figure~\ref{fig:mvir_mstar_sigma_re} and Table~2).  The significance
is particularly reduced for the size evolution.  We conclude that the
addition of the local early-type sample does not affect our results
but helps constraining the fits. Results are summarized in Table~3.

S\'ersic indices from HST photometry are available as part of the
  BOSS/COSMOS sub-sample (see Section~\ref{subsecstructural_cosmos}).

\begin{figure*}
\vbox{
\hbox{
\includegraphics[width=0.33\textwidth]{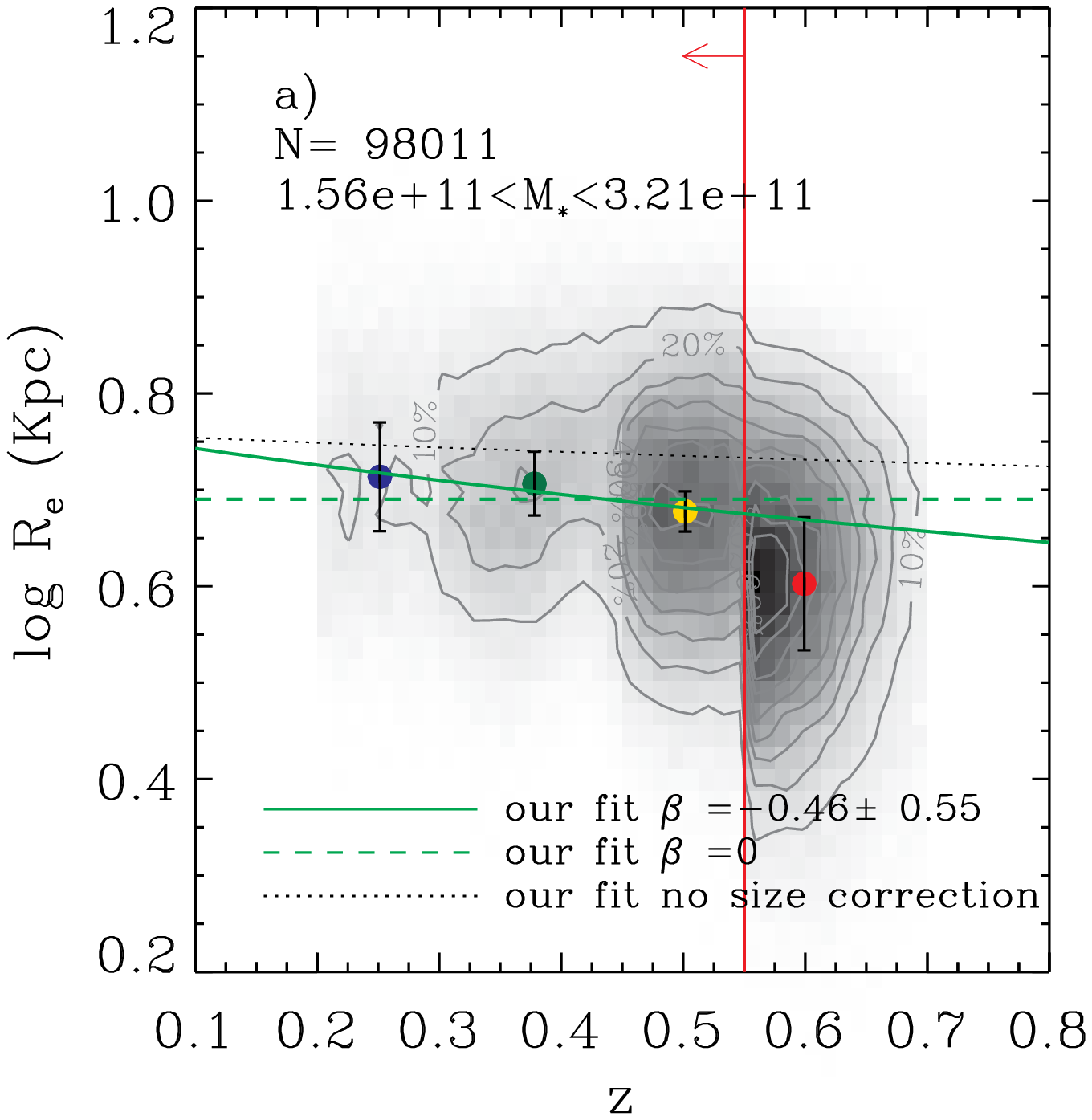}
\includegraphics[width=0.33\textwidth]{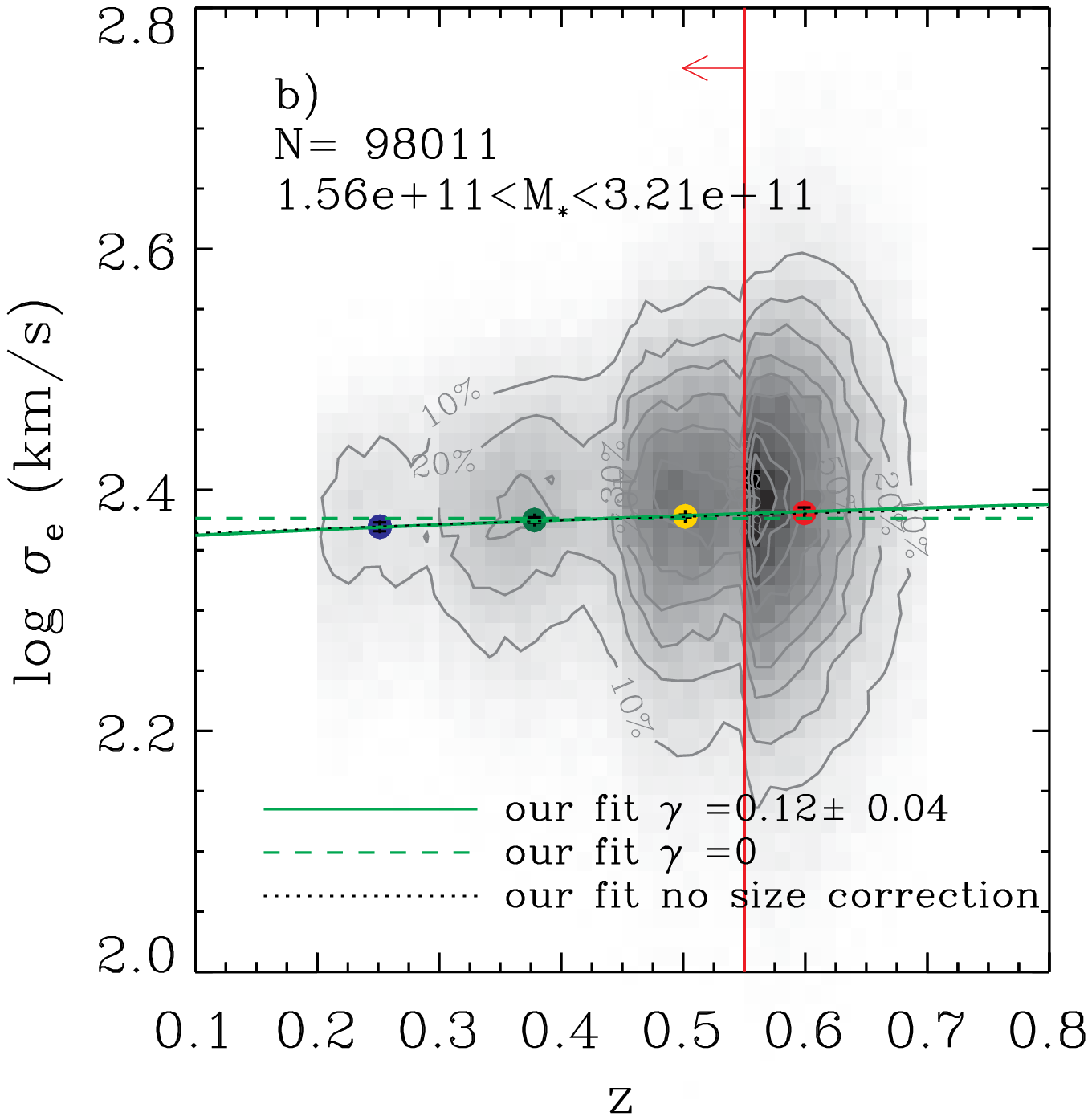}
\includegraphics[width=0.33\textwidth]{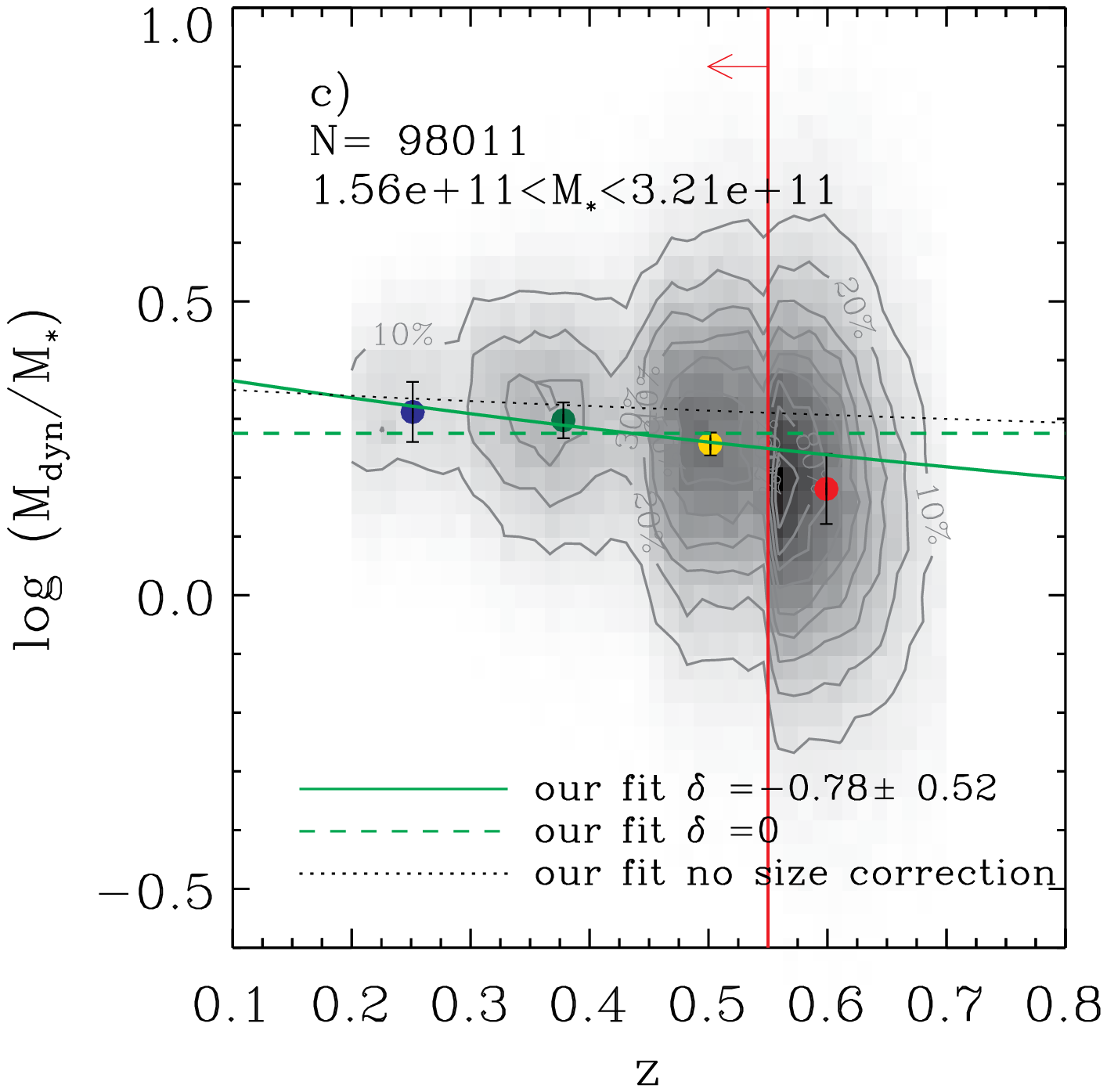}}}
\vbox{
\hbox{
\includegraphics[width=0.33\textwidth]{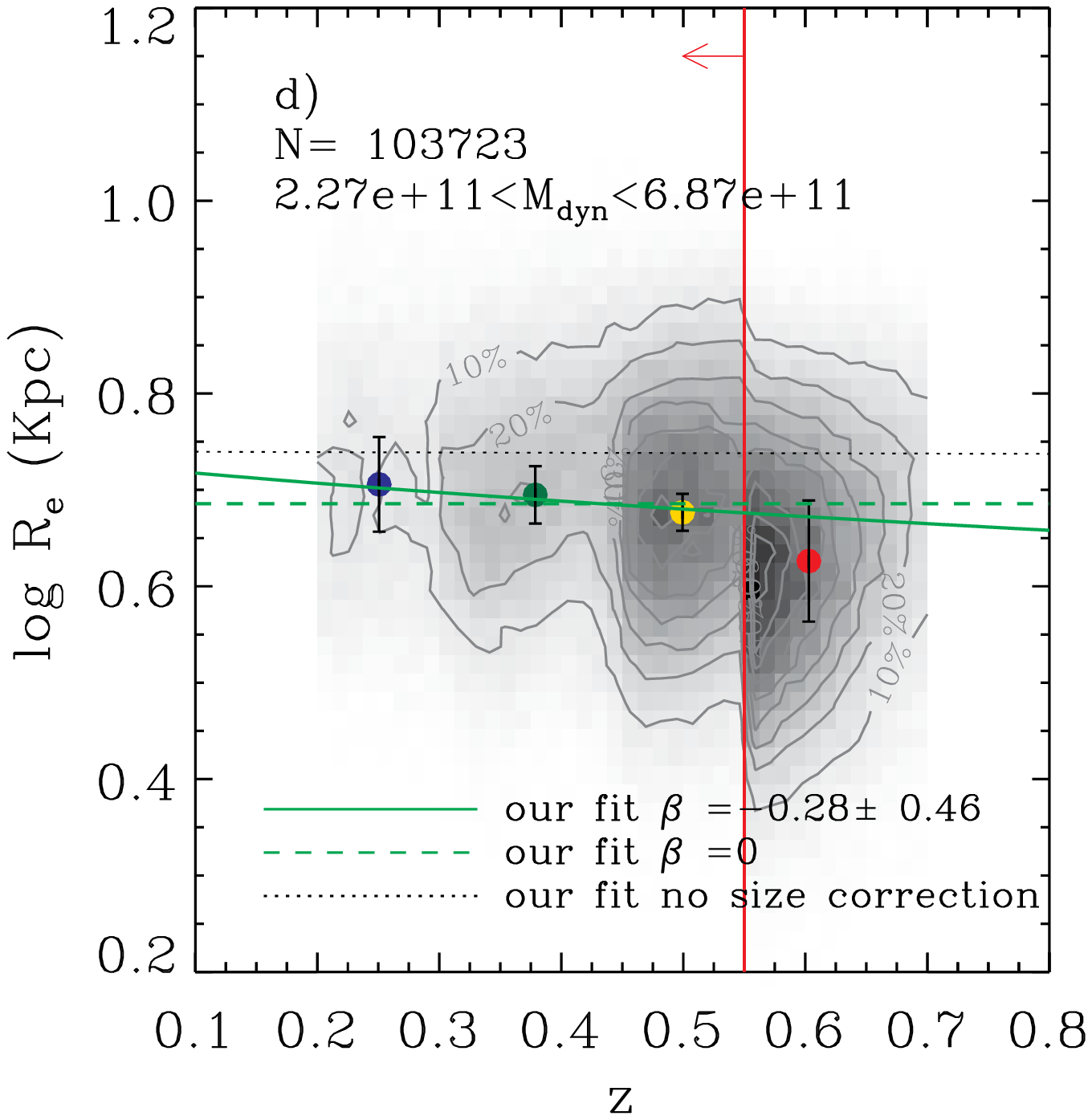}
\includegraphics[width=0.33\textwidth]{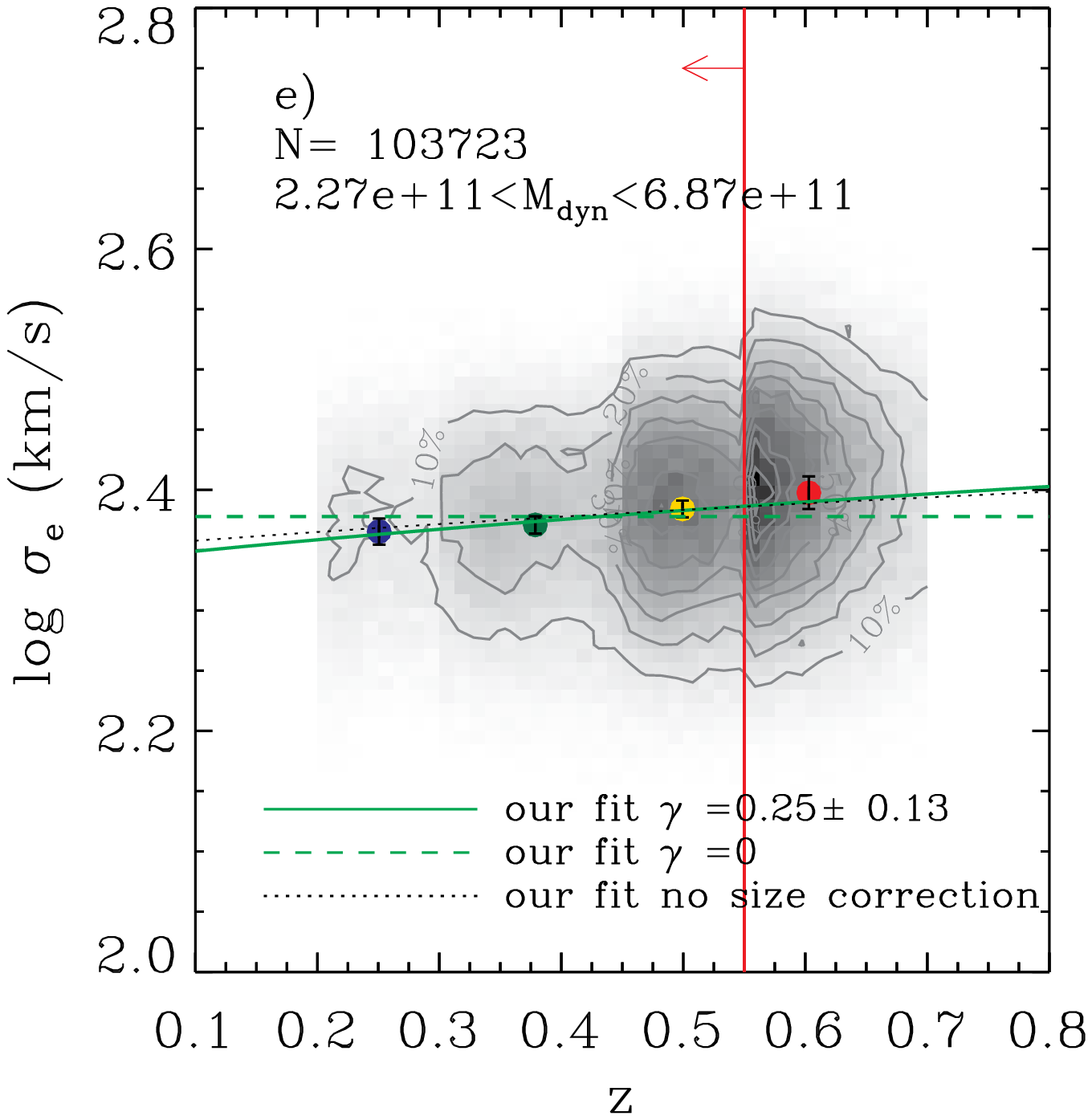}
\includegraphics[trim= 0 0 0 30, clip, width=0.33\textwidth]{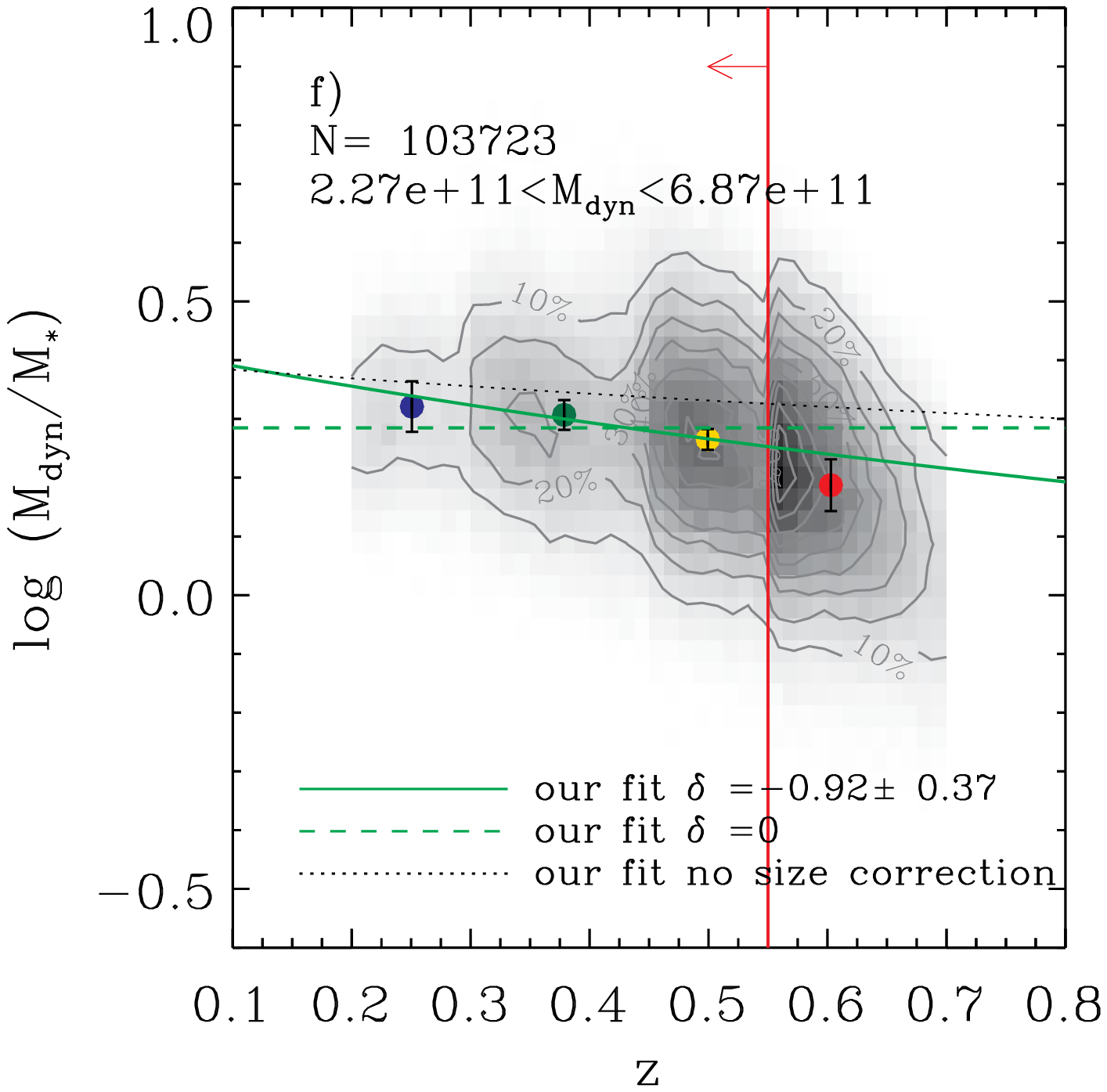}}}
\caption{Same as Figure~\ref{fig:mvir_mstar_sigma_re} but for the
  SDSS-III/BOSS sample only and with a {\em redshift dependent} $\beta$ (see Equation~\ref{eq:vir}).}
\label{fig:mvir_mstar_sigma_re_alphan_BOSS}
\end{figure*}

  This allows us to assess the effect of a redshift-dependent
  $\beta(n)$ parameter, through our Monte Carlo simulations.  The
  results of this test are shown in
  Figure~\ref{fig:mvir_mstar_sigma_re_alphan_BOSS} and summarized in
  Table~3 (columns with $z$-variable $\beta$).  By using a redshift
  dependent $\beta$ on Equation~\ref{eq:vir} for \mvir, the evolution
  of the \mvir/\mstar\ ratio becomes more significant than using a
  constant $\beta$ (more conservative results). The difference on the
  slopes, however, is very small ($\sim 10$\%) suggesting that
  $\beta(n)$ does not strongly evolve in the BOSS redshift range and
  any possible variation would not affect our results.  Table~3
  summarizes the comparison between a redshift dependent or a constant
  $\beta$.

  The analysis above can be done only for the COSMOS/BOSS sample,
  because structural parameters are measured in a similar way and on
  the same images. 

  We use the photometric catalog of \citet{Simard2011} for SDSS
    local galaxies to compare structural parameters (i.e., S\'ersic
    indices) of the local and BOSS/COSMOS sample.  \citet{Simard2011}
    presented new structural parameters from both single S\'ersic fits
    and bulge-to-disk decomposition on the full DR7 sample in $g$ and
    $r$-band.  For our tests we chose data from single S\'ersic fit to
    be consistent with COSMOS (in particular around the transition
    redshift $z\sim0.2$).

    The \citet{Simard2011} and COSMOS catalogs have a limited number
    of objects in common. For $n<4$ we found a fair agreement between
    structural parameters in the two catalog, whereas for $n>4$,
    \citet{Simard2011} S\'ersic indices are larger than COSMOS ones.
    By using median \citet{Simard2011} S\'ersic indices for our
    early-types sample we reproduce the commonly used $\beta(n)\sim5$
    for local galaxies.  S\'ersic indices of the two samples probably
    differ because they were derived with different images and
    approaches and there is no clear correction to be applied to this.

    For this reason in our analysis we do not combine information
    about the S\'ersic indices from \citet{Simard2011} and
    COSMOS. Instead we adopt the redshift independent $\beta$ of the
    BOSS galaxies also for the SDSS sample. 

An explanation of these differences is beyond the scope
of this paper.

\begin{table*}
\label{tab:results_ratio_sigma_re_noerror_weight_simu}
\begin{scriptsize}
\begin{center}
  \caption{Fitting parameters for the BOSS sample between
    $0.2<z<0.55$, in which the progenitor-bias correction is
    applied.}
\begin{tabular}{  l | c c | c c | c c | c c}
\hline
\hline
\noalign{\smallskip}
\multicolumn{1}{c}{} &
\multicolumn{4}{c}{\mstar} &
\multicolumn{4}{c}{\mvir}  \\
\hline
\noalign{\smallskip}
\multicolumn{1}{c|}{Parameter} &
\multicolumn{2}{c|}{constant $\beta$} &
\multicolumn{2}{c|}{$z$-variable $\beta$} &
\multicolumn{2}{c|}{constant $\beta$}  &
\multicolumn{2}{c}{$z$-variable $\beta$}   \\
\hline
\noalign{\smallskip}
\multicolumn{1}{c|}{} &
\multicolumn{1}{c}{slope} &
\multicolumn{1}{c|}{z=0} &
\multicolumn{1}{c}{slope} &
\multicolumn{1}{c|}{z=0} &
\multicolumn{1}{c}{slope} &
\multicolumn{1}{c|}{z=0} &
\multicolumn{1}{c}{slope} &
\multicolumn{1}{c}{z=0} \\
\hline
\re &$-0.46\pm 0.56$ &$0.76\pm 0.09$ & $-0.46\pm 0.55$ & $0.76\pm 0.09$& $-0.32\pm 0.47$ &$0.73\pm 0.08$& $-0.28\pm 0.46$ & $0.73\pm 0.08$ \\
\sigmae &$ 0.12\pm 0.04$ & $2.36\pm 0.006$ & $ 0.12\pm 0.04$ & $2.36\pm 0.006$& $ 0.20\pm 0.13$ & $2.35\pm 0.02$& $ 0.25\pm 0.13$ & $2.34\pm 0.02$ \\
$M_{\rm dyn}/M_{\star} $ &$-0.53\pm 0.48$ & $0.36\pm 0.08$ &$-0.78\pm0.52$ & $0.40\pm 0.08$& $-0.77\pm 0.35$ & $0.41\pm 0.06$&$-0.92\pm0.37$ & $0.43\pm 0.06$ \\ 
\hline
\noalign{\smallskip}
\label{tab:betaevol}
\end{tabular}
\begin{minipage}{\textwidth}

  {\sc Notes.} --- {Uncertainties on each parameter are
    $1\sigma$ errors derived from Monte Carlo simulations}.  The relation we fitted for \re\ is $\log R_{\rm
      e} =\log R_{\rm e,0} +\beta (1+z) $, for \sigmae\ is $\log
    \sigma_{\rm e} =\log \sigma_{\rm e,0} +\gamma (1+z)$, and for
    $M_{\rm dyn}/M_{\star}$ is $\log (M_{\rm dyn}/M_{\star}) = \log
    (M_{\rm dyn}/M_{\star})_0+\delta(1+z)$ .
\end{minipage}
\end{center}
\end{scriptsize}
\end{table*}

\section{D.~~Tests on evolution: local early-types \& BOSS samples
  without progenitor bias correction}
\label{sec:noPB_show}

\begin{figure*}
\vbox{
\hbox{
\includegraphics[width=0.33\textwidth]{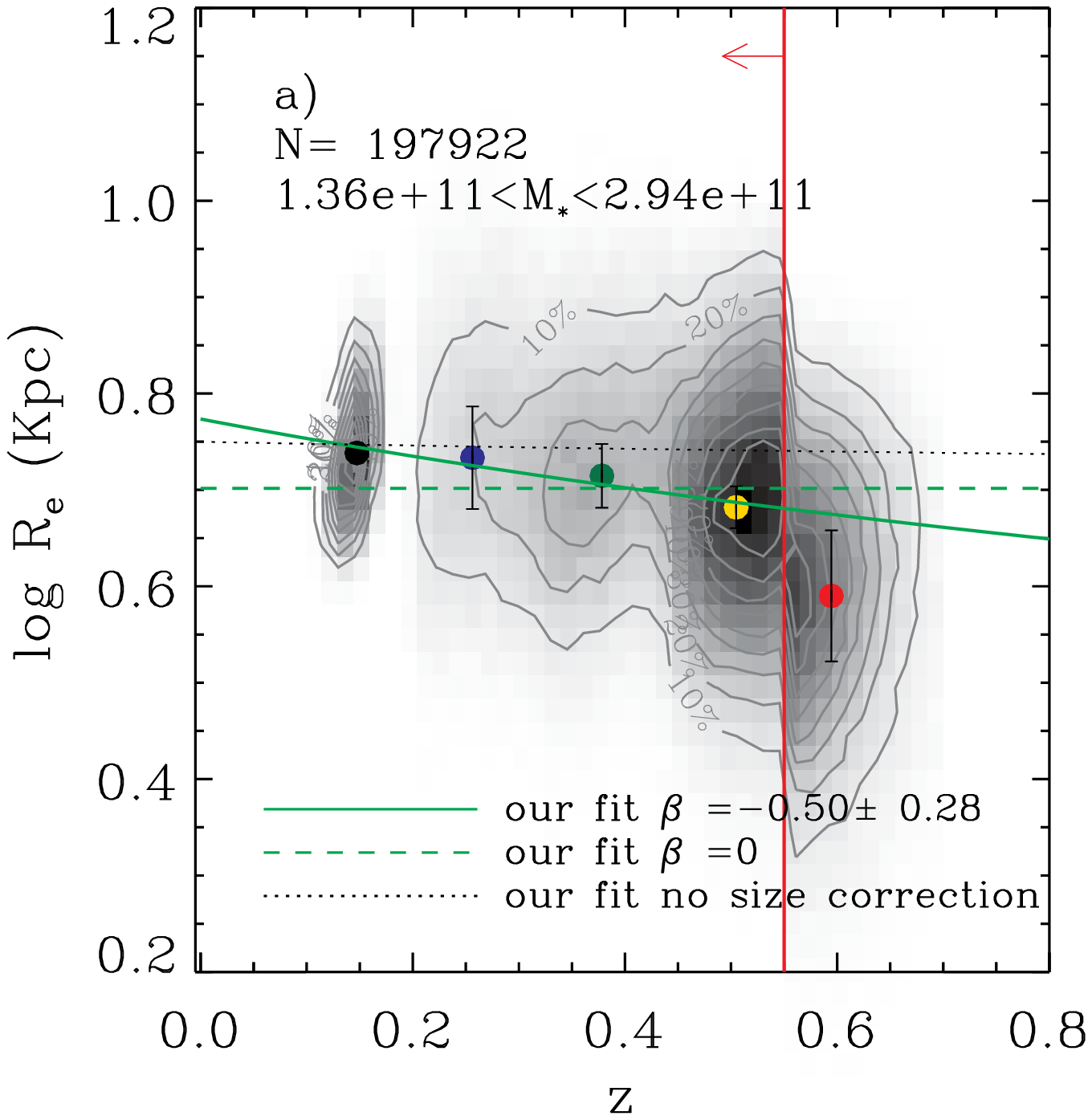}
\includegraphics[width=0.33\textwidth]{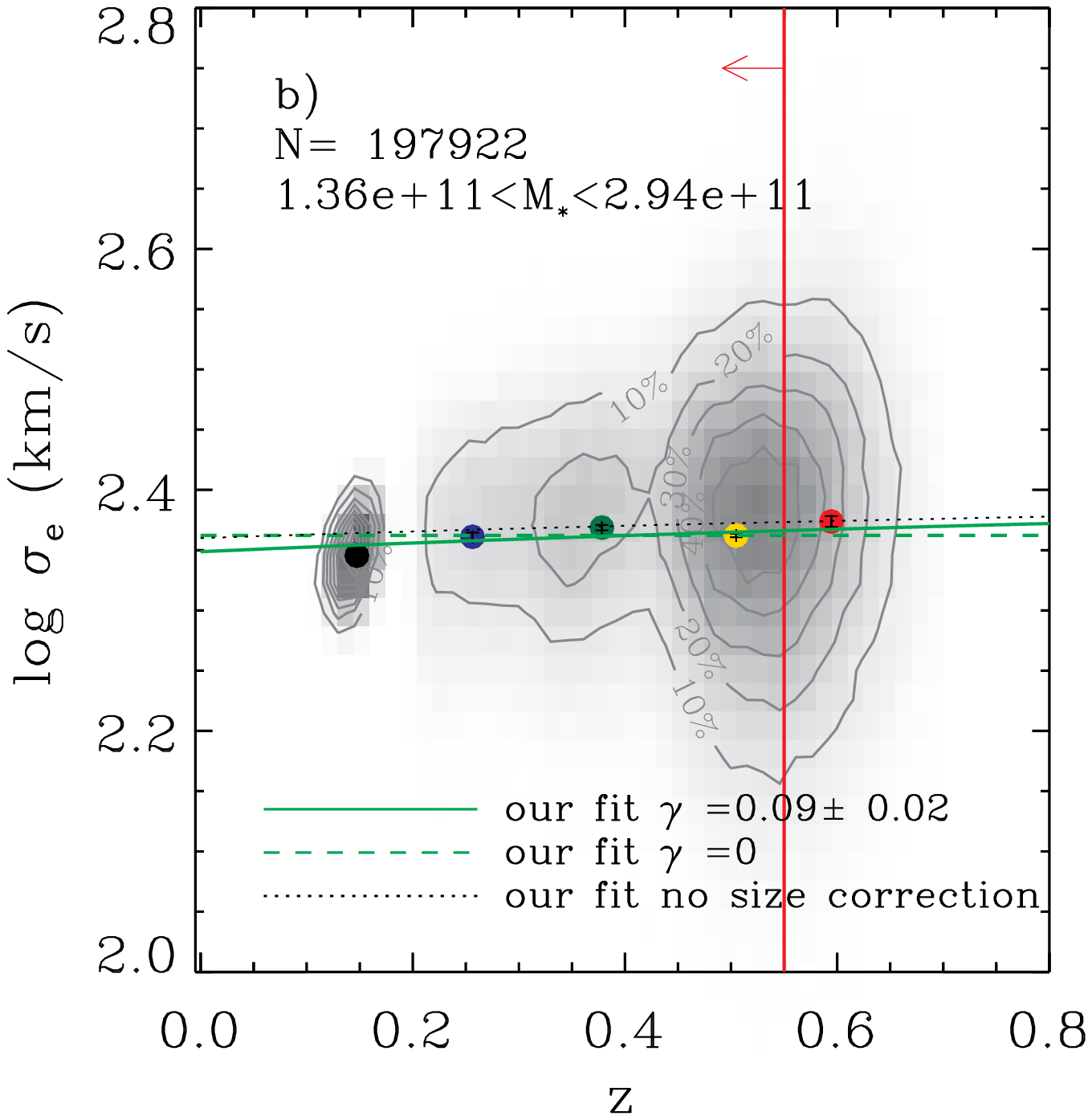}
\includegraphics[width=0.33\textwidth]{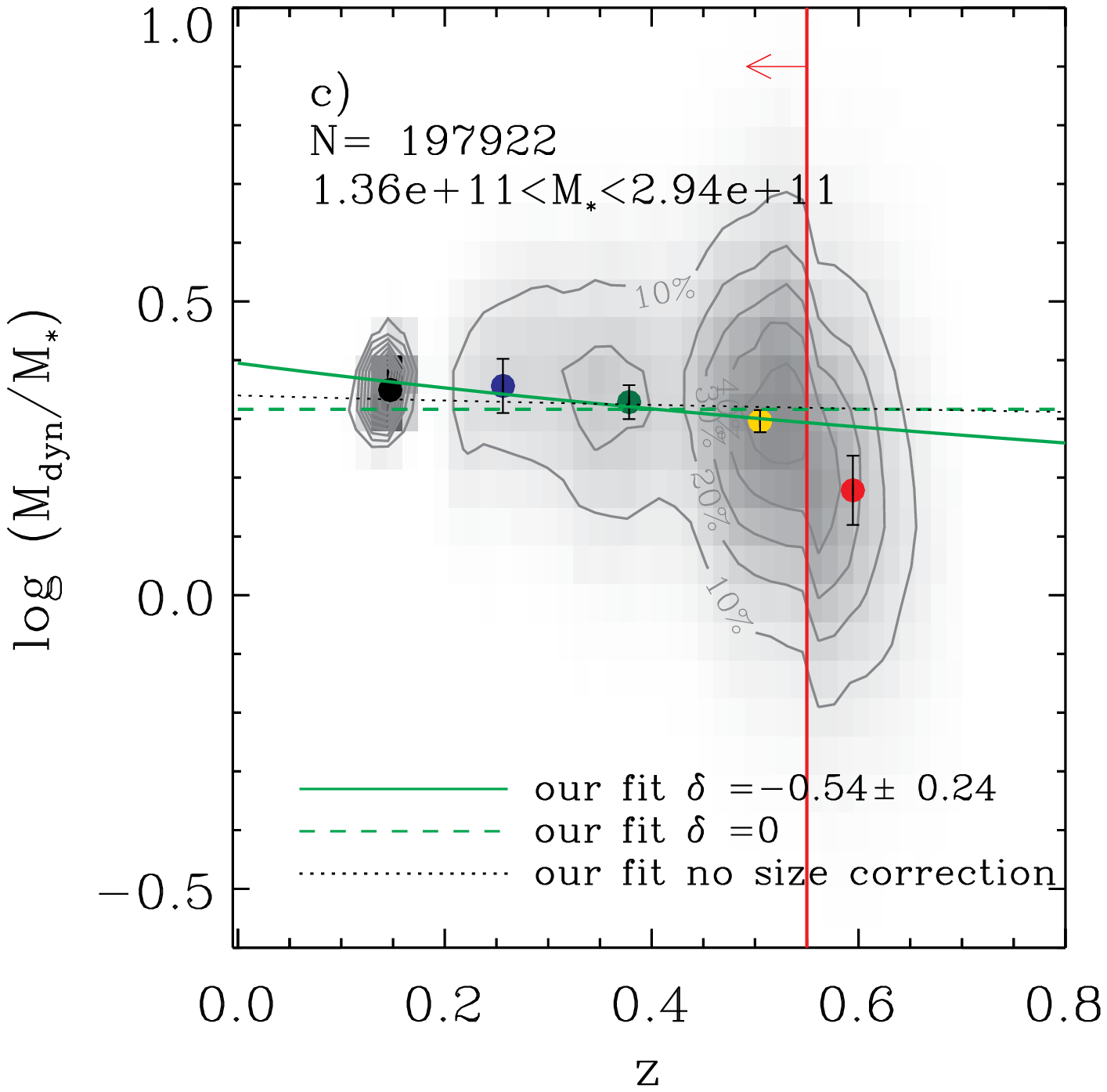}}}
\vbox{
\hbox{
\includegraphics[width=0.33\textwidth, trim= 0 0 0 30, clip]{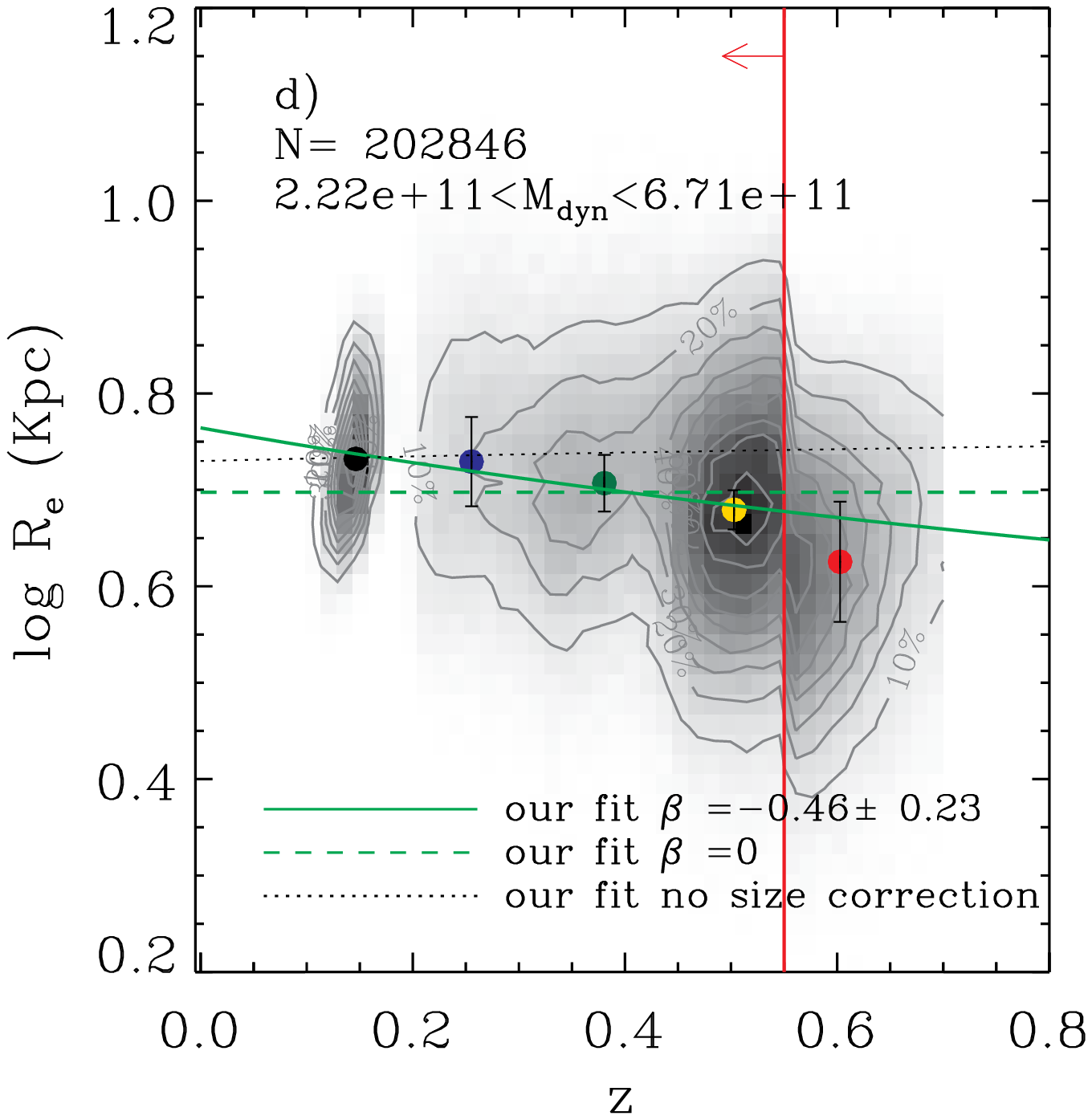}
\includegraphics[width=0.33\textwidth]{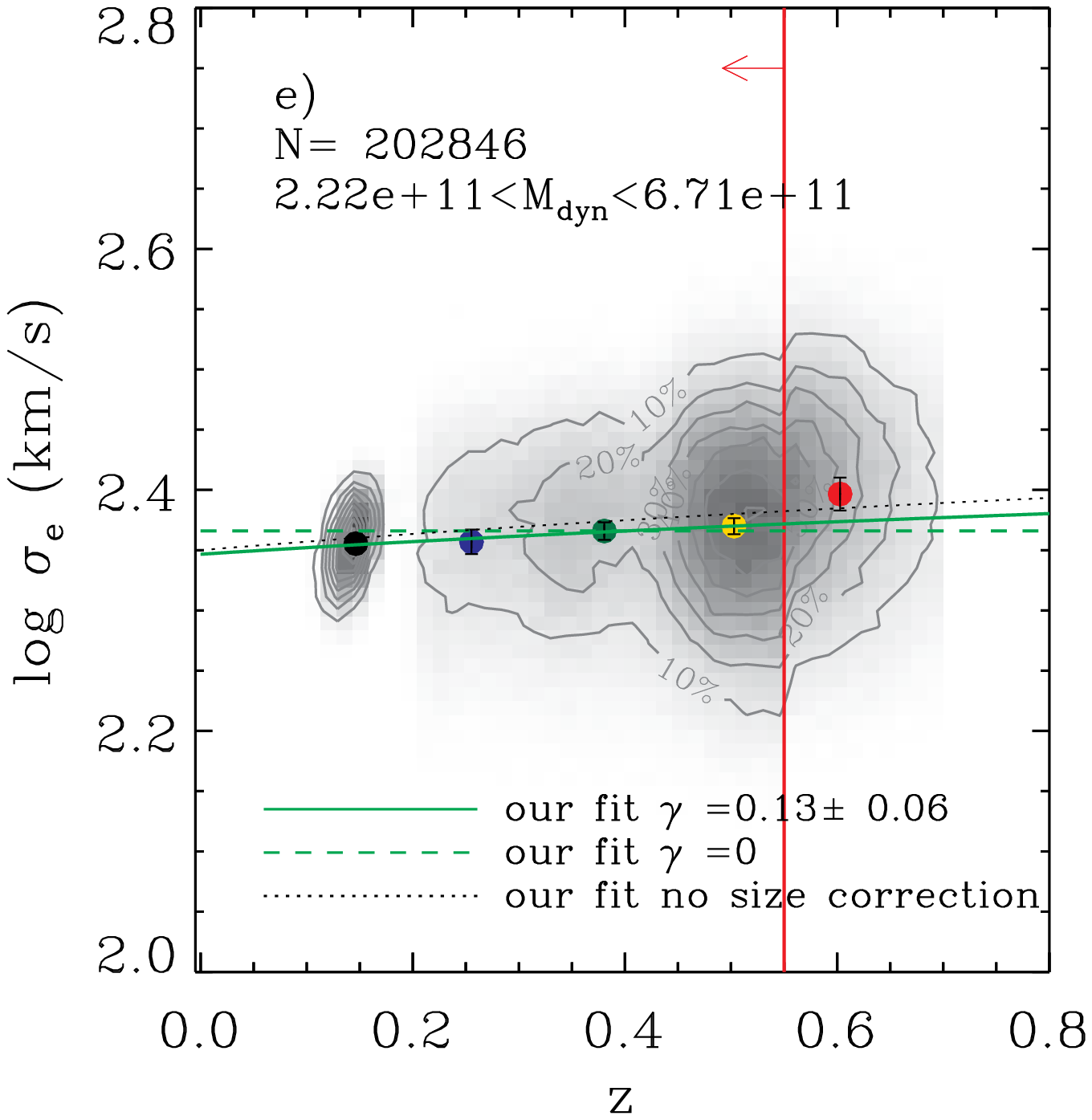}
\includegraphics[trim= 0 0 0 30, clip,width=0.33\textwidth]{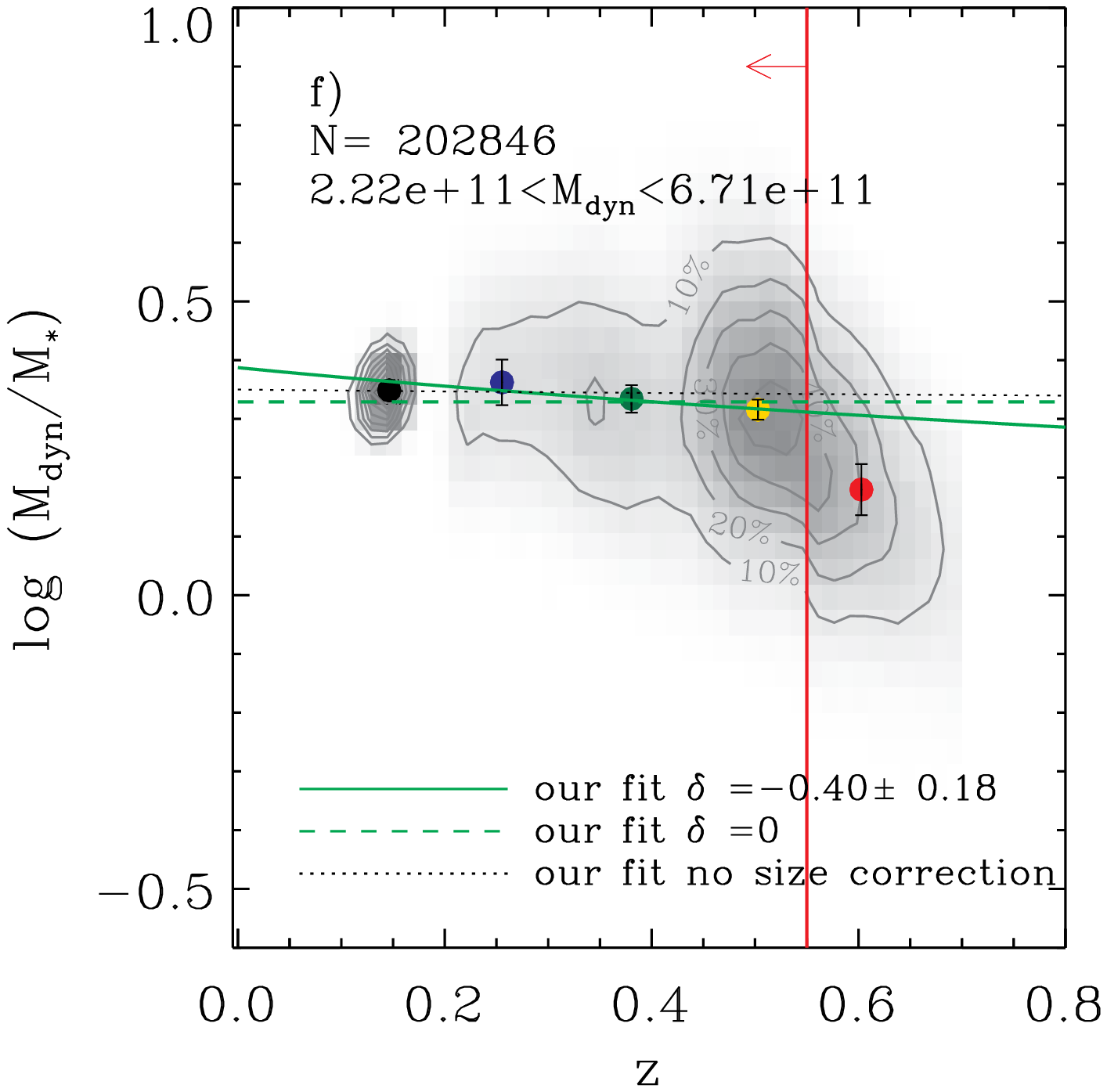}}}
\caption{Same as Figure~\ref{fig:mvir_mstar_sigma_re} but no correction for progenitor bias.}
\label{fig:mvir_mstar_sigma_re_alpha5_noerror_ETGs_noPB}
\end{figure*}

In Section~\ref{sec:progenitor} we discuss the correction for
progenitor bias applied in this work. To study the impact of this
correction on our analysis, we have performed a re-analysis for a
sample without progenitor bias
correction. Figure~\ref{fig:mvir_mstar_sigma_re_alpha5_noerror_ETGs_noPB}
shows the redshift evolution of the galaxy parameters size, stellar
velocity dispersion and \mvir/\mstar\ without progenitor bias
correction. The fit parameters are given in
Table~\ref{tab:fitsnobias}. Comparing to
Figure~\ref{fig:mvir_mstar_sigma_re} and Table~\ref{tab:fits} in the
main text it can be seen that generally the results are fairly stable
against the progenitor-bias correction, significant evolution with
redshift is still detected for all three parameters. We infer that the
main conclusions of this paper do not critically depend on the
progenitor bias correction.

\begin{table*}
\begin{scriptsize}
\begin{center}
\begin{minipage}{0.5\textwidth}
\caption{Fitting parameters for the redshift evolution of galaxy parameters
  between $0.1\leq z \leq0.55$ without progenitor bias correction}
\begin{tabular}{c r r r r}
\hline
\hline
\noalign{\smallskip}
\multicolumn{1}{c}{} &
\multicolumn{2}{c}{\mstar} &
\multicolumn{2}{c}{\mvir}  \\
\hline
\noalign{\smallskip}
\multicolumn{1}{c}{Parameter} &
\multicolumn{1}{c}{slope} &
\multicolumn{1}{c}{zero point} &
\multicolumn{1}{c}{slope} &
\multicolumn{1}{c}{zero point} \\
\hline
\re                              &$-0.50 \pm 0.28 $ & $0.78 \pm  0.04 $  &$ -0.46 \pm  0.23  $ & $ 0.76  \pm 0.03 $  \\
\sigmae                      &$  0.09  \pm 0.02 $ & $2.36 \pm  0.00$ &$ 0.13   \pm 0.06  $ & $ 2.35  \pm 0.01 $  \\
$M_{\rm dyn}/M_{\star} $ &$ -0.54 \pm 0.24 $ & $0.40 \pm  0.03 $  &$-0.40  \pm  0.18  $ & $ 0.38 \pm  0.02 $ \\ 

\hline
\noalign{\smallskip}
\label{tab:fitsnobias}
\end{tabular}
\end{minipage}
\begin{minipage}{\textwidth}
 {\sc Notes.} --- Uncertainties on each parameter are $1\sigma$
  errors derived from Monte Carlo Simulations.  The relation
  we fitted for \re\ is $\log R_{\rm e} =\log R_{\rm e,0} +\beta (1+z)
  $, for \sigmae\ is $\log \sigma_{\rm e} =\log \sigma_{\rm e,0}
  +\gamma (1+z)$, and for $M_{\rm dyn}/M_{\star}$ is $\log (M_{\rm
    dyn}/M_{\star}) = \log (M_{\rm
    dyn}/M_{\star})_0+\delta(1+z)$. 
\end{minipage}
\end{center}
\end{scriptsize}
\end{table*}

\section{E.~~Tests on evolution: stellar mass distribution of 
local early-types \& BOSS galaxies}
\label{sec:mass_distrib_local}

\begin{figure*}
\begin{center}
\includegraphics[width=0.90\textwidth]{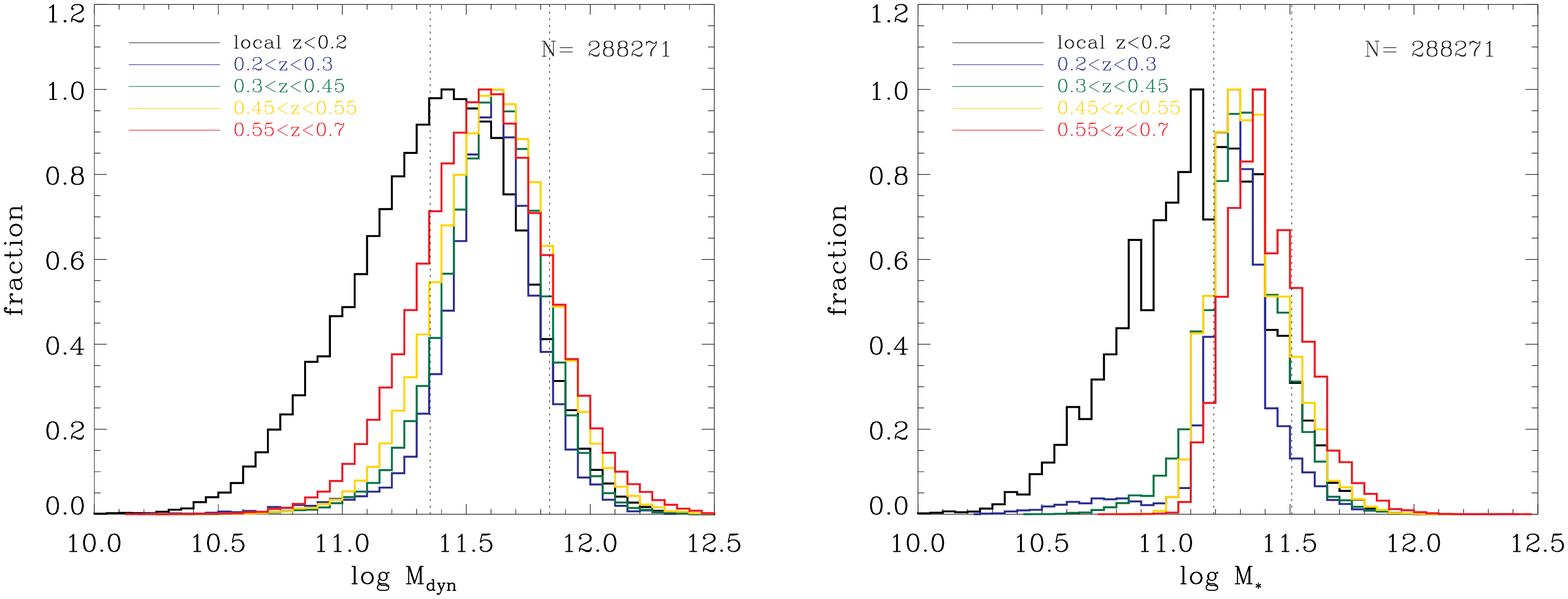}
\includegraphics[width=0.45\textwidth]{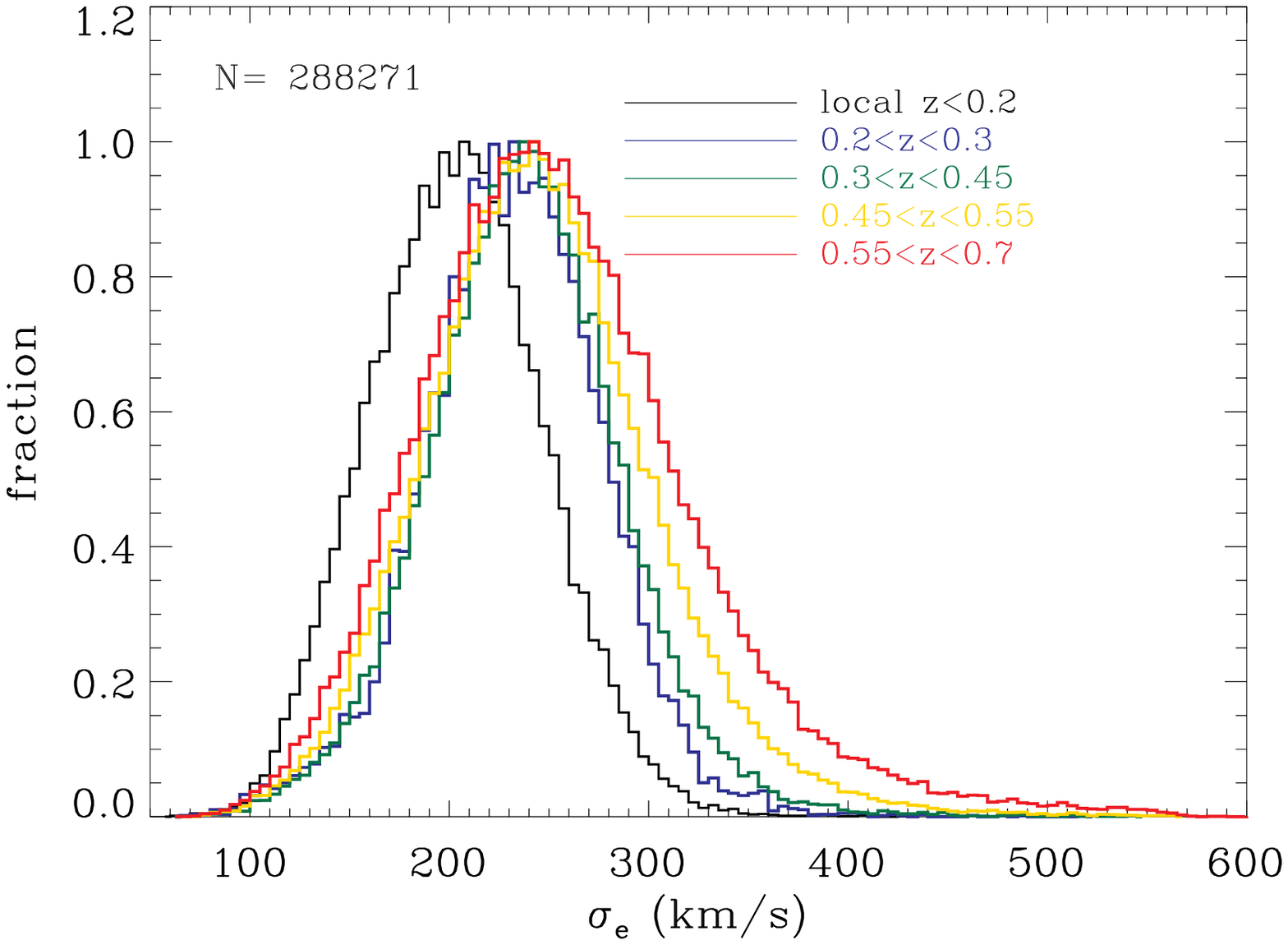}
\includegraphics[width=0.45\textwidth]{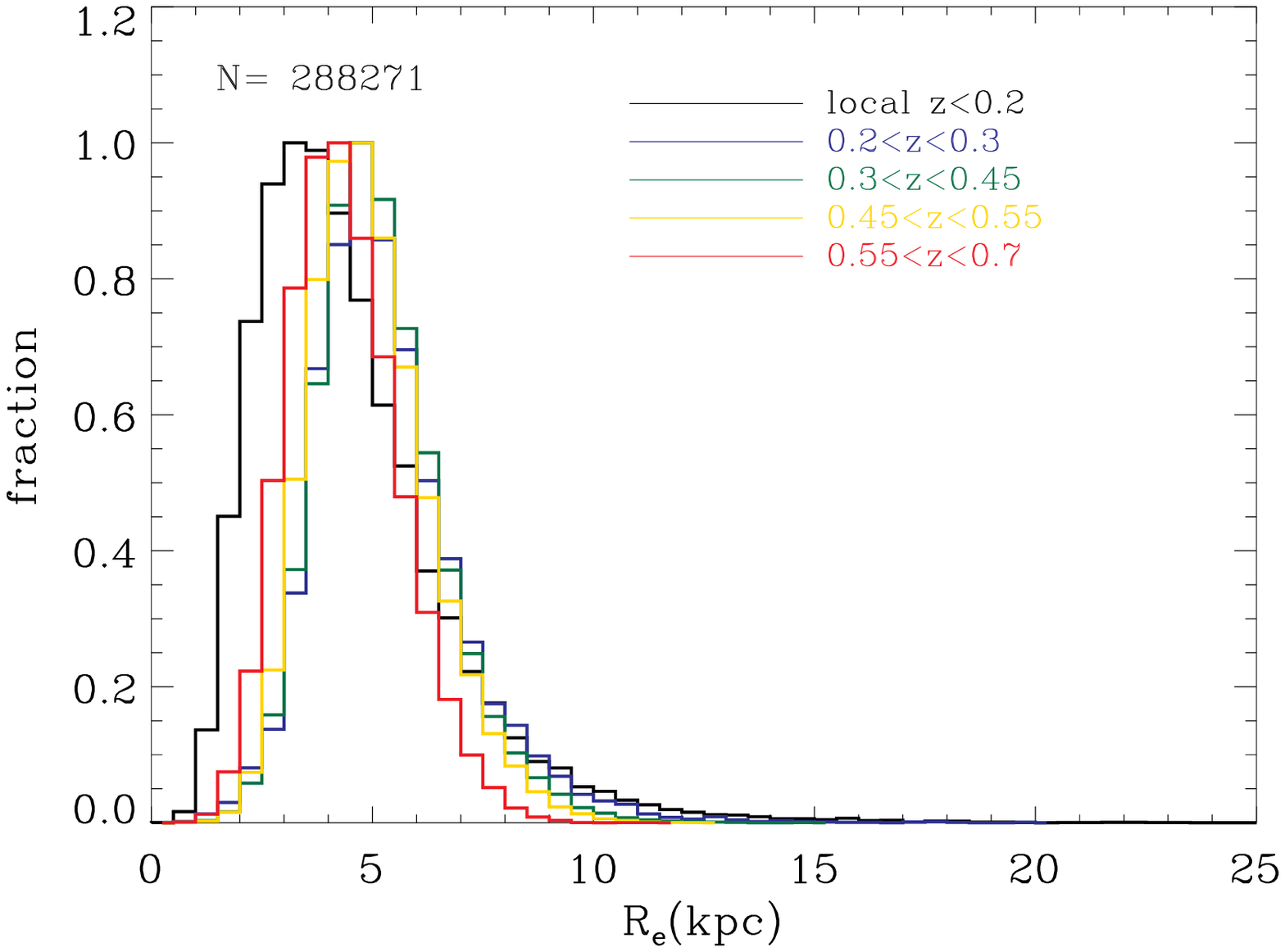} 
\end{center}
\caption{Same as Figure~\ref{fig:progenitor_bias_mass_distribution} but no homogenization between the mass distributions of the local sample from SDSS-II and the sample from SDSS-III/BOSS.}
\label{fig:updated_mass_distrib}
\end{figure*}

\begin{figure*}
\vbox{
\hbox{
\includegraphics[width=0.33\textwidth]{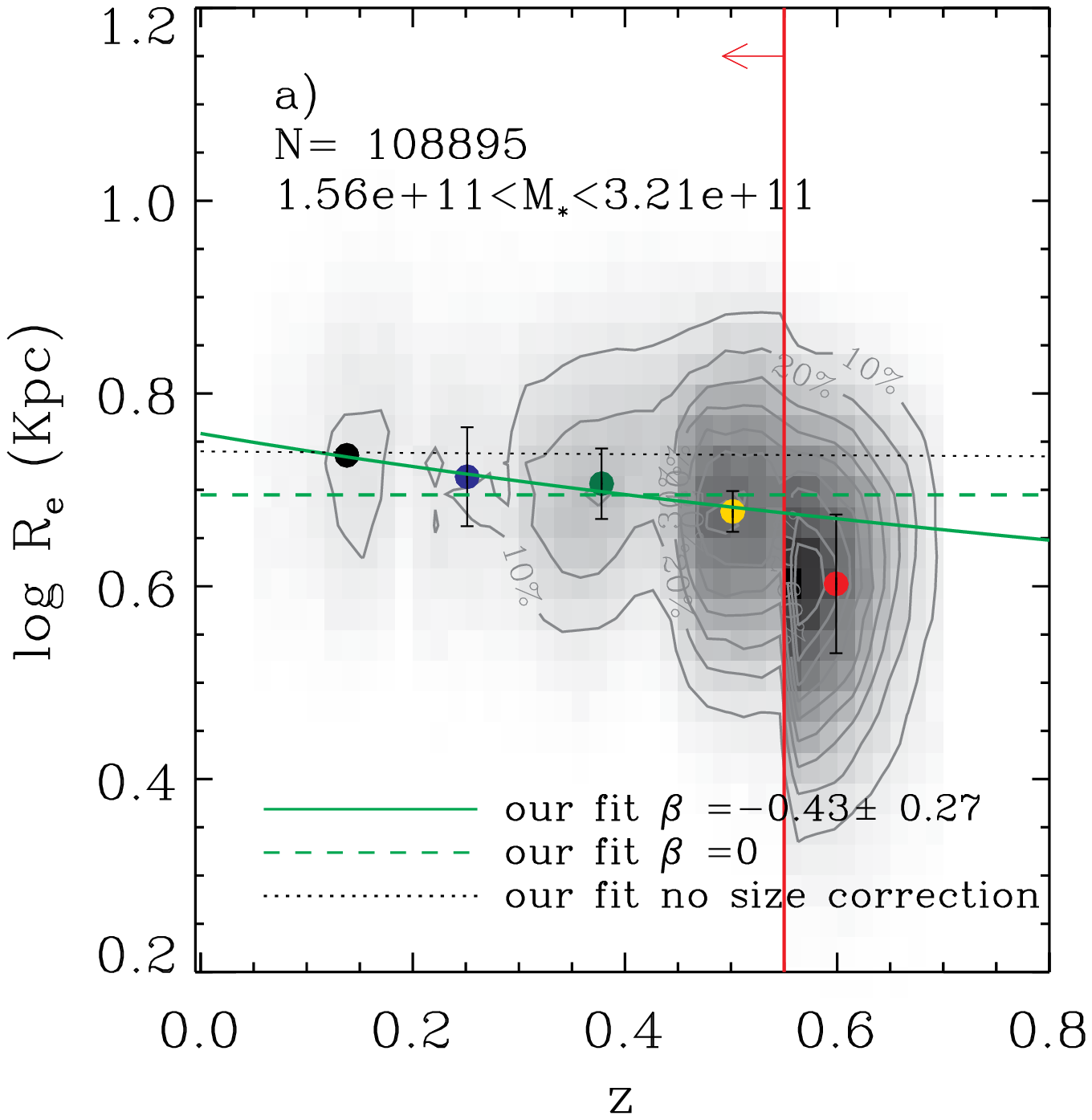}
\includegraphics[width=0.33\textwidth]{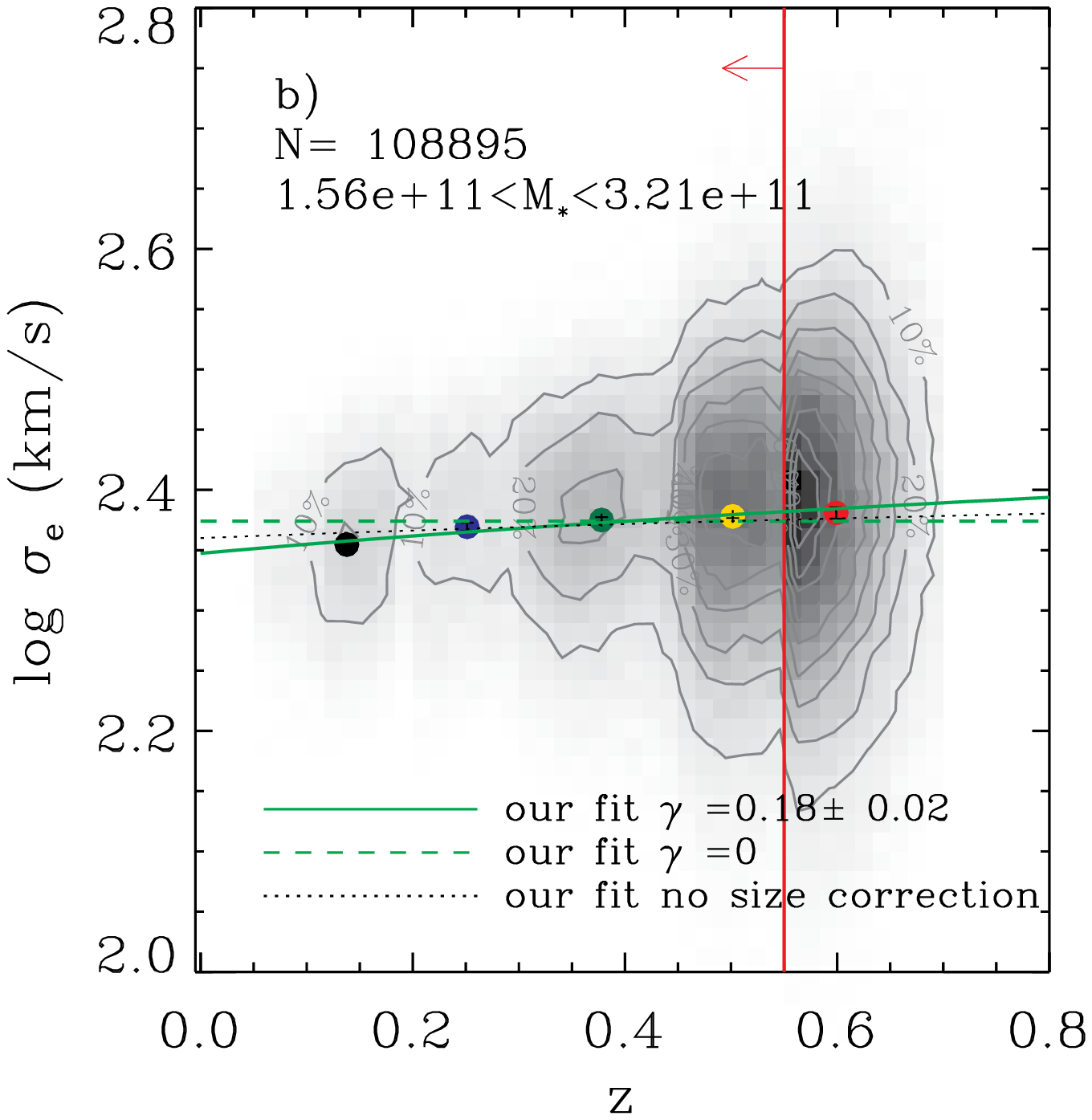} 
\includegraphics[width=0.33\textwidth]{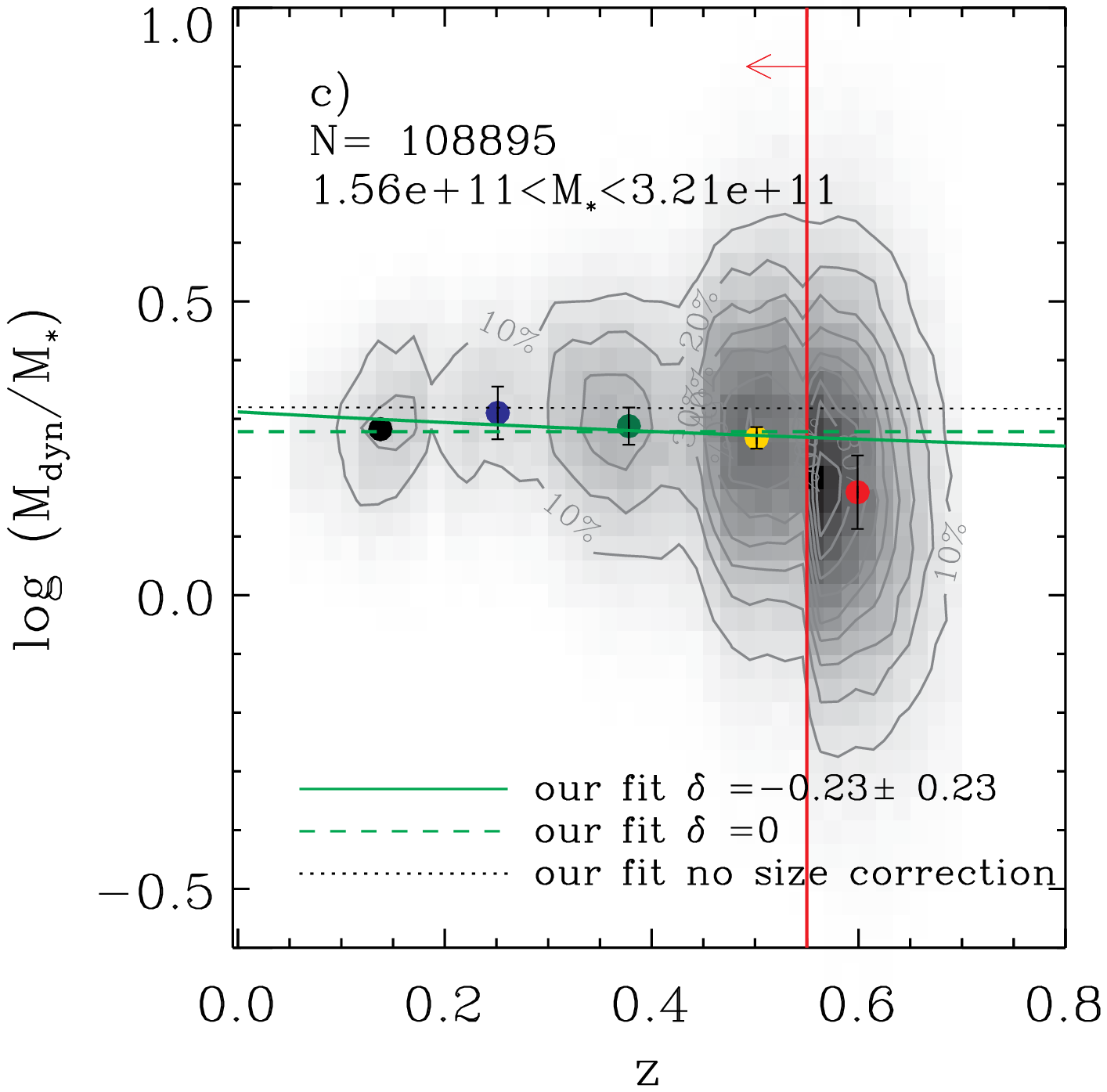}}}
\vbox{
\hbox{
\includegraphics[trim= 0 0 0 30, clip,width=0.33\textwidth]{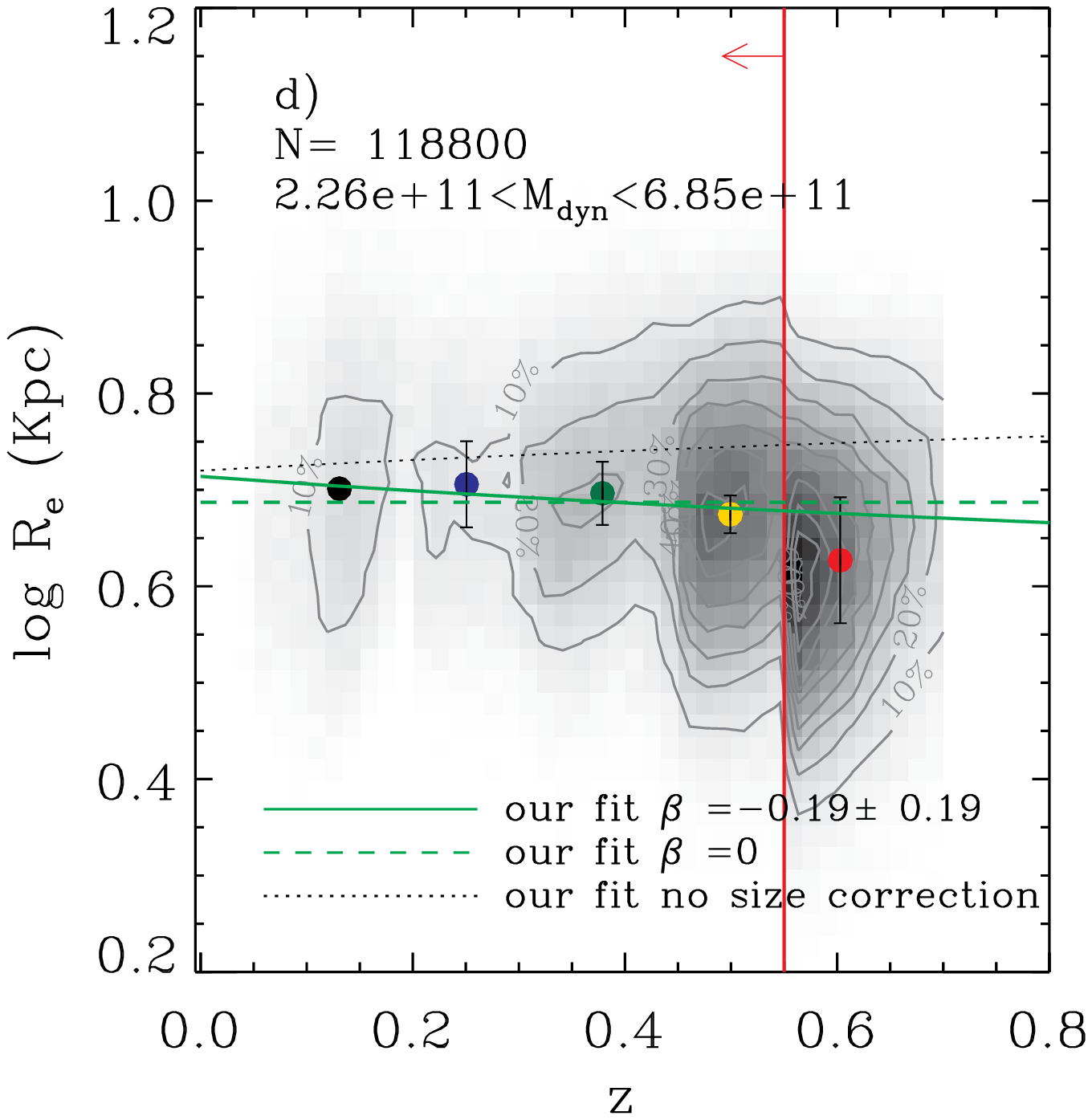}
\includegraphics[width=0.33\textwidth]{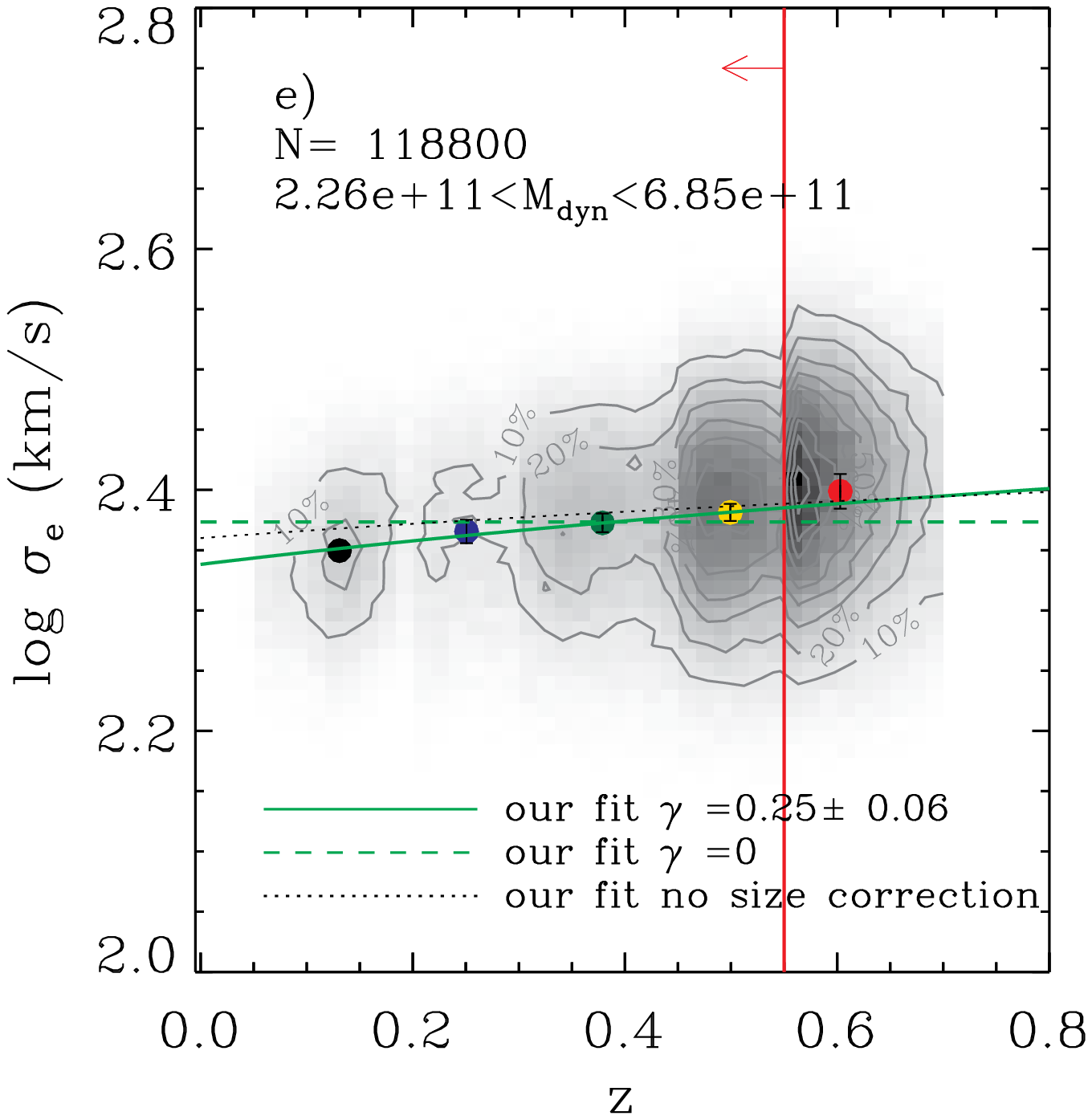}
\includegraphics[trim= 0 0 0 30, clip ,width=0.33\textwidth]{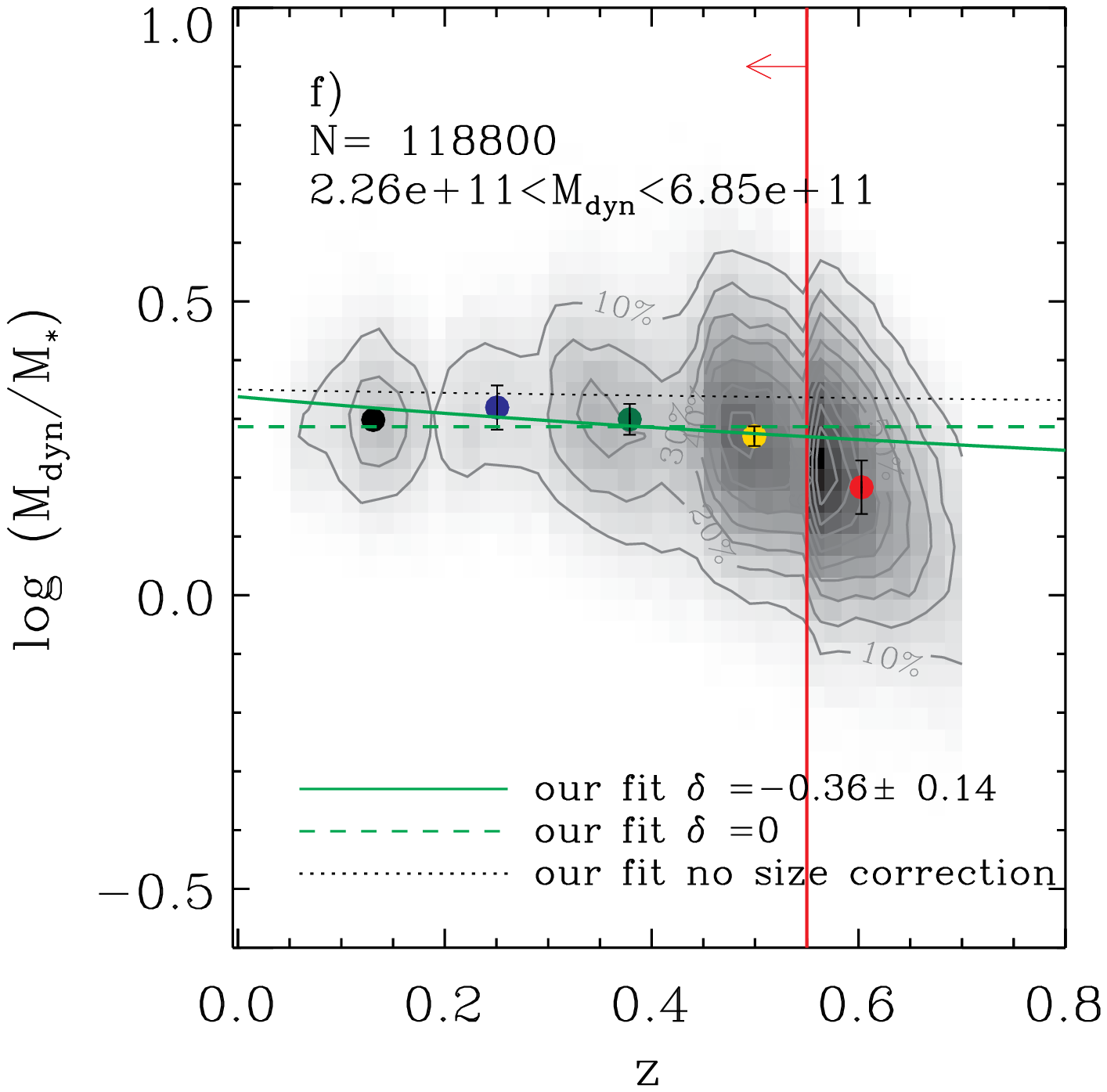}}}
\caption{Same as Figure~\ref{fig:mvir_mstar_sigma_re} but no homogenization between the mass distributions of the local sample from SDSS-II and the sample from SDSS-III/BOSS.}
\label{fig:mvir_mstar_sigma_re_localdistrub_new}
\end{figure*}

As described in Section~\ref{sec:localETGs} we homogenize the stellar
mass distributions between the local sample from SDSS-II and the
high-$z$ sample from SDSS-III/BOSS by selecting a local sub-sample
that matches the mass distribution of the BOSS sample. In this
Appendix we present a re-analysis in which we do not apply this
homogenization. The distributions of stellar masses, dynamical masses,
stellar velocity dispersions, and effective radii are shown in
Figure~\ref{fig:updated_mass_distrib}. Figure~\ref{fig:mvir_mstar_sigma_re_localdistrub_new}
shows the redshift evolution of the galaxy parameters size, stellar
velocity dispersion and \mvir/\mstar, the fit parameters are
summarized in Table~\ref{tab:fitsnomassdistrib}. It can be seen that
our finding of redshift evolution remains intact for all three
parameters, and the significance of the slopes only changes slightly.

\begin{table*}
\label{tab:results_ratio_sigma_re}
\begin{scriptsize}
\begin{center}
\begin{minipage}{0.5\textwidth}
\caption{Fitting parameters for the redshift evolution of galaxy parameters
  between $0.1\leq z \leq0.55$ without matching mass distributions between the local SDSS-II and the SDSS-III/BOSS samples}
\begin{tabular}{c r r r r}
\hline
\hline
\noalign{\smallskip}
\multicolumn{1}{c}{} &
\multicolumn{2}{c}{\mstar} &
\multicolumn{2}{c}{\mvir}  \\
\hline
\noalign{\smallskip}
\multicolumn{1}{c}{Parameter} &
\multicolumn{1}{c}{slope} &
\multicolumn{1}{c}{zero point} &
\multicolumn{1}{c}{slope} &
\multicolumn{1}{c}{zero point} \\
\hline
\re                  &$-0.43\pm 0.27$ & $0.76\pm 0.04$  & $-0.19\pm 0.19$ &$0.72\pm 0.03$  \\
\sigmae              &$ 0.18\pm 0.02$ & $2.35\pm 0.00$ & $ 0.25\pm 0.06$ & $2.34\pm 0.01$ \\
$M_{\rm dyn}/M_{\star} $ &$-0.23\pm 0.23$ & $0.32\pm 0.03$  & $-0.36\pm 0.14$ & $0.34\pm 0.02$\\ 
\hline
\noalign{\smallskip}
\label{tab:fitsnomassdistrib}
\end{tabular}
\end{minipage}
\begin{minipage}{\textwidth}

  {\sc Notes.} --- Uncertainties on each parameter are $1\sigma$
  errors derived from Monte Carlo simulations.  The relation we fitted
  for \re\ is $\log R_{\rm e} =\log R_{\rm e,0} +\beta (1+z) $, for
  \sigmae\ is $\log \sigma_{\rm e} =\log \sigma_{\rm e,0} +\gamma
  (1+z)$, and for $M_{\rm dyn}/M_{\star}$ is $\log (M_{\rm
    dyn}/M_{\star}) = \log (M_{\rm dyn}/M_{\star})_0+\delta(1+z)$.

\end{minipage}
\end{center}
\end{scriptsize}
\end{table*}


\end{document}